\documentclass[trackchanges, twocolumn]{aastex701}

\usepackage{float}
\usepackage{comment}
\usepackage[T1]{fontenc}
\usepackage{amsmath}

\usepackage{makecell}
\usepackage{url}

\newcommand\frappe{\texttt{FRAP}}

\usepackage{fontawesome}
\definecolor{linkcolor}{HTML}{2AAD2E}

\begin{document}

\title{ Direct Retrieval of Protoplanetary Disk Dust Properties using Auto-differentiable Gaussian Processes and Its Application to the HD 169142 Disk }

\author[0000-0001-8002-8473,sname='Yoshida']{Tomohiro C. Yoshida}
\affiliation{Dipartimento di Fisica, Universit\`{a} degli Studi di Milano, Via Celoria 16, 20133 Milano, Italy}
\affiliation{National Astronomical Observatory of Japan, 2-21-1 Osawa, Mitaka, Tokyo 181-8588, Japan}
\email[show]{tomohiroyoshida.astro@gmail.com}

\author[0000-0003-1283-6262,sname='Mac\'{i}as']{Enrique Mac\'{i}as}
\email{enrique.macias@eso.org}
\affiliation{European Southern Observatory (ESO), Karl-Schwarzschild-Str. 2, Garching bei München, Germany}

\author[0000-0003-4853-5736,sname='Rosotti']{Giovanni Rosotti}
\email{giovanni.rosotti@unimi.it}
\affiliation{Dipartimento di Fisica, Universit\`{a} degli Studi di Milano, Via Celoria 16, 20133 Milano, Italy}

\author[0000-0003-4689-2684]{Stefano Facchini}
\email{stefano.facchini@unimi.it}
\affiliation{Dipartimento di Fisica, Universit\`{a} degli Studi di Milano, Via Celoria 16, 20133 Milano, Italy}

\author[0009-0005-5187-6074,sname='Viscardi']{Elena Viscardi}
\email{elena.viscardi@eso.org}
\affiliation{European Southern Observatory (ESO), Karl-Schwarzschild-Str. 2, Garching bei München, Germany}

\author[0000-0003-1958-6673]{Kiyoaki Doi}
\email{doi.kiyoaki.astro@gmail.com}
\affiliation{Max-Planck Institute for Astronomy (MPIA), Königstuhl 17, 69117 Heidelberg, Germany}


\begin{abstract}
Retrieving dust properties in protoplanetary disks, including the surface density distribution, temperature, and grain size distribution, is a fundamental task in observational studies of planet formation.
While multi-wavelength analysis of the spectral energy distribution (SED) using interferometers such as the Atacama Large Millimeter/submillimeter Array (ALMA) is a powerful diagnostic tool, traditional methods are often hindered by strong biases arising from a limited imaging beamsize.
In this paper, we present a new retrieval framework for dust disk properties designed to overcome this challenge.
We assume that the underlying physical structures are expressed as sample paths from Gaussian processes, compute the radial intensity distributions at observed wavelengths, and produce one-dimensional visibility models.
The models are compared with the observed data and the posterior distributions are sampled via the Markov-Chain Monte-Carlo method.
The whole procedure is implemented in JAX, which enables end-to-end auto-differentiation and significantly accelerate{s} the inference. 
We validate our methodology using mock datasets, and find that the results are not strongly biased and better reproduce the input profiles.
We also demonstrate its capabilities through an application to ALMA Band 3, 6, and 9 observations of the HD 169142 disk, revealing a new complex structure.
Our developed code is publicly available as a Python module, \frappe\ (Flexible Radial Analysis of Protoplanetary disks).
This framework provides a next-generation infrastructure for disk SED modeling, enabling high-precision studies of the physical environments in which planets form.
\href{https://github.com/tomyoshida/frap}{\color{linkcolor}\faGithub} \href{https://frap.readthedocs.io/en/latest/index.html}{\color{linkcolor}\faBook}
\end{abstract}


\section{Introduction}
Retrieving physical quantities in protoplanetary disks is vital for understanding the mechanisms of planet formation. Specifically, characterizing the radial distribution and grain growth of dust grains is a central objective in disk studies.
Recent advancements in instrumentation, most notably the Atacama Large Millimeter/submillimeter Array (ALMA), have enabled the characterization of the dust disk properties at high angular resolution with multiple frequencies \citep[e.g.,][]{2019ApJ...881..159M, 2019ApJ...883...71C, 2022A&A...664A.137G, 2021A&A...648A..33M, 2021ApJS..257...14S, 2022ApJ...930...56U,  2023ApJ...954..110O, 2023ApJ...953...96Z, 2024ApJ...971..129C, 2024A&A...686A.298G, 2024ApJ...974..306S, 2025MNRAS.538.2358S, 2025A&A...702A..56Z}.

In typical studies, the spectral energy distribution (SED) is measured as a function of radius from the central star. These data are then fitted with specific SED models to derive key parameters such as the dust surface density, temperature, and grain size distribution. While this approach is straightforward, it is subject to several critical limitations.

First, many analyses assume that the observed intensity directly represents the intrinsic intensity. While this approximation may hold for well-resolved sources, it becomes problematic in marginally resolved cases where beam convolution effects are significant \citep[e.g.,][]{2024A&A...688A.204L}.
Second, conventional methods often require images or radial profiles to be processed at the same spatial resolution. This requirement is sometimes technically challenging; for instance, long wavelength observations typically achieve significantly lower spatial resolution than higher wavelengths, despite their importance.
Consequently, datasets are often smoothed to match the coarsest resolution, which can lead to misinterpretation of radial substructures.
It is notable that a multi-stage fitting strategy to mitigate some of these resolution-related issues has been recently proposed \citep{2025A&A...695A.147V, 2025A&A...702A..56Z}.
Thirdly, a standard imaging algorithm, CLEAN, can introduce systematic uncertainties due to the sidelobes, potentially leading to underestimated errors or biases in the retrieved physical quantities.
A more robust approach would be to fit the model directly to the interferometric visibilities, the fundamental observables of the interferometer.
Finally, previous methods typically treat observational noise as being spatially or radially independent. In reality, flux scaling uncertainties, which often dominate the error budget, are radially correlated and must be properly accounted for.

In this paper, we propose a new retrieval framework for dust disk properties that addresses these limitations. Our approach employs Gaussian Processes (GP) to construct the underlying physical structures and utilizes auto-differentiation for accelerated posterior sampling from the visibility data.
We have developed an open-source Python module, Flexible Radial Analysis of Protoplanetary disks (\frappe), to implement the methodology presented herein.

The paper is organized as follows. In Section \ref{sec:mod}, we describe our modeling framework. In Section \ref{sec:injrec}, we validate \frappe\ using mock datasets. We then apply this method to real ALMA observations of the HD 169142 disk in Section \ref{sec:hd169}. In Section \ref{sec:disc}, we discuss the performance of the method and the scientific implications for the HD 169142 disk. Finally, we summarize our findings in Section \ref{sec:sum}.

\section{Method} \label{sec:mod}

The primary objective of \frappe\ is to retrieve the radial distribution of dust properties, such as the dust surface density, grain size distribution, and temperature, from interferometric visibilities directly.
Throughout this section, we assume that the disk is axisymmetric and that the observed visibilities have been deprojected and have real part values as a function of the deprojected uv-distance, or radial visibility profiles.
We describe the practical process for this manipulation in Section \ref{sec:datared}.

The architecture of \frappe\ consists of three main stages.
First, the radial profiles of the physical quantities are represented as samples from a non-parametric Gaussian Process (GP).
Second, the spectral energy distribution (SED) at each radius is computed using a reduced radiative transfer model, which generates radial intensity profiles at each wavelength. Finally, these intensity profiles are Hankel-transformed to obtain the corresponding radial visibility profiles.

The entire modeling framework is implemented in \texttt{JAX} \citep{jax2018github}, a Python library optimized for high-performance machine learning and numerical computing.
This choice enables end-to-end automatic differentiation of the model; for example, the gradient of a specific visibility value with respect to the dust surface density at a given radius can be computed efficiently.
This auto-differentiation capability significantly accelerates the sampling of the posterior distribution via gradient-based Markov Chain Monte Carlo (MCMC) methods.
Specifically, it enables the retrieval of dust properties with the high degree of flexibility afforded by GPs, which would otherwise be computationally prohibitive.
In our practical implementation, we utilize \texttt{numpyro} \citep{bingham2019pyro, 2019arXiv191211554P} to perform the posterior sampling.

\subsection{Radial Profile Modeling with Gaussian Processes} \label{sec:gp}
We employ a GP as a prior for the radial profile of a given parameter $X(r)$, where $r$ denotes the radius from the disk center.
A GP is a multivariate normal distribution characterized by a covariance (kernel) matrix.
In this study, we adopt the radial basis function (RBF) kernel for the parameter $X$:
\begin{equation}
    K_X(r_1, r_2) = \exp\left( -\frac{(r_1-r_2)^2}{2l_X^2} \right),
\end{equation}
where $r_{1}$ and $r_{2}$ represent radial positions and $l_X$ is the characteristic length scale.

To treat parameters with upper and lower boundaries, we re-parameterize them.
Specifically, we apply a sigmoid-like transformation such that $X(r)$ is mapped from its physical range $[X_{\rm min}, X_{\rm max}]$ to a latent space:
\begin{equation}
    x(r) = \log\left\{ 1 - \frac{X(r) - X_{\rm min}}{X_{\rm max} - X_{\rm min}} \right\},
\end{equation}
where $X_{\rm min}$ and $X_{\rm max}$ are the prescribed minimum and maximum bounds for the parameter, respectively.

\begin{figure}
\centering
\includegraphics[width=0.9\linewidth]{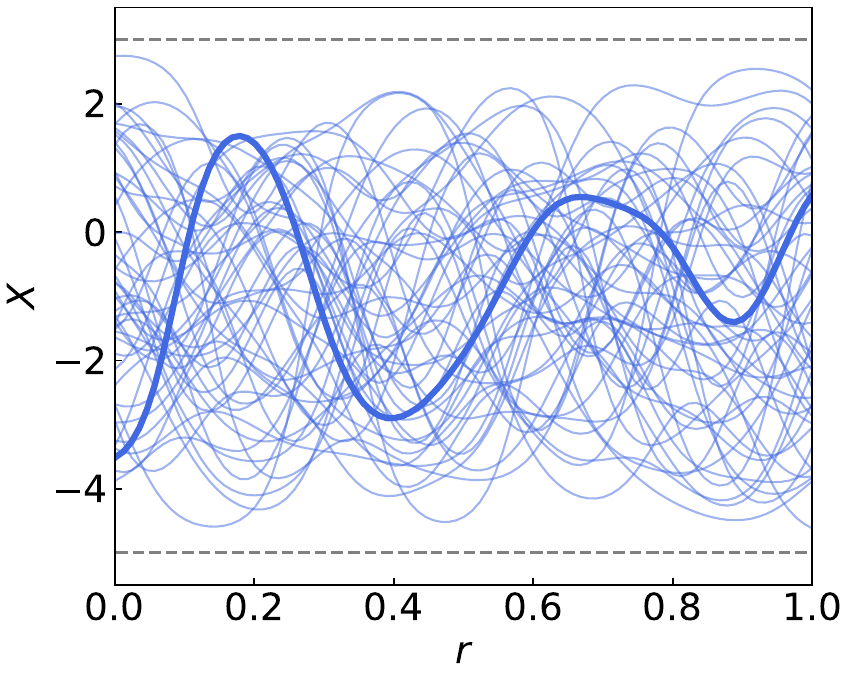}
\caption{Illustrative radial profiles of an arbitrary quantity $X$ sampled from the Gaussian Process prior after applying the sigmoid transformation. We assume an RBF kernel with a length scale of $l_X = 0.1$. A single representative sample is shown as a solid curve, while the ensemble of 20 samples demonstrates the prior's flexibility. The horizontal dashed lines indicate the boundaries $X_{\rm max}$ and $X_{\rm min}$.}
\label{fig:prior}
\end{figure}

Figure \ref{fig:prior} displays 20 sample radial profiles generated from this GP prior with $l_X = 0.1$ and the range defined by $(X_{\rm min}, X_{\rm max}) = (-5, 3)$ in arbitrary units to demonstrate the flexibility of GPs.
These flexible curves serve as the prior distributions for the fitting process in \frappe.

\subsection{Reduced Radiative Transfer Model} \label{sec:rt}

The Gaussian process framework described in Section \ref{sec:gp} provides the non-parametric priors for the physical quantities we aim to retrieve. For the sake of clarity, we consider a model characterized by three free radial parameters in this and next sections: the dust surface density $\Sigma(r)$, the dust temperature $T(r)$, and the maximum dust grain size $a_{\rm max}(r)$. The dust grain size distribution $n(a)$ is assumed to follow a power law {with an exponent $q$}:
\begin{equation}
\frac{dn}{da} \propto
      \begin{cases}
      a^{-q} & (a_{\rm min} < a < a_{\rm max}) \\
      0 & (\text{otherwise}),
      \end{cases}
\end{equation}
{where we fix the minimum grain size to $a_{\rm min} = 0.1~\mu\text{m}$.}

Once the dust size distribution is defined, the absorption and effective scattering opacity profiles at each frequency $\nu$, $\kappa_{\rm abs}(\nu, r)$ and $\kappa_{\rm sca,\ eff}(\nu, r)$, are calculated.
While we adopt the DSHARP opacity model \citep{2018ApJ...869L..45B} in this and next sections, \frappe\ can accommodate any user-defined opacity model that maps physical quantities (e.g., $a_{\rm max}$) to optical properties.
From these opacities, we derive the radial profiles of the scattering albedo,
\begin{equation}
    \omega(\nu, r) = \frac{\kappa_{\rm sca,\ eff}}{\kappa_{\rm sca,\ eff} + \kappa_{\rm abs}},
\end{equation}
and the absorption optical depth,
\begin{equation}
    \tau(\nu, r) = \kappa_{\rm abs} \Sigma,
\end{equation}
where the $(\nu, r)$ dependence is omitted on the right-hand sides for simplicity.

We approximate the radiative transfer using an isothermal, geometrically thin slab model. We employ the analytical solution derived by \citet{2020ApJ...892..136S} {using the two-stream approximation \citep[e.g.,][]{1993Icar..106...20M, 2019ApJ...877L..18Z, 2019ApJ...877L..22L}} to calculate the emergent intensity at each radius:
\begin{equation} \label{eq:Inur}
I(\nu, r) = B(\nu, T)\left[ 1 - \exp{\left( -\frac{\tau}{1-\omega} \right)} + \omega \mathcal{F}(\tau, \omega)\right],
\end{equation}
where $B(\nu, T)$ is the Planck function and $\mathcal{F}(\tau, \omega)$ is defined as,
\begin{eqnarray}
&\mathcal{F}&(\tau, \omega) \\
&=& \frac{1}{ ( \sqrt{1-\omega} - 1 )\exp{\left( -\sqrt{\frac{3}{1-\omega}} \tau \right)} - ( \sqrt{1-\omega} + 1 ) }  \\
&\times& \left\{ 
\frac{ 1 - \exp\left[ -( \sqrt{3(1-\omega)} + 1 )\frac{\tau}{ 1- \omega } \right] }{ \sqrt{3(1-\omega) }+1 } \right.  \\
&+& \left. \frac{ \exp{\left[ - \frac{\tau}{1- \omega} \right]} - \exp{ \left[ -\sqrt{\frac{3}{1-\omega}} \tau \right] } }{  \sqrt{3(1-\omega)} - 1 } \right\}.
\end{eqnarray}
These expressions can be modified or replaced based on the user's specific requirements.

\subsection{Hankel Transform and Visibilities}

The radial intensity profiles at each wavelength, $I(\nu, r)$, are converted into one-dimensional visibility profiles using a discrete Hankel transform. The detailed derivation of this numerical approach is {based on} \citet{2015JOSAA..32..611B} and \citet{2020MNRAS.495.3209J}; here, we briefly summarize the implementation. We utilize a Fourier-Bessel series expansion to express the model visibilities,
\begin{equation} \label{eq:hankel_main}
\mathcal{V}_{\rm mod}( \nu, q_{c, k} ) = \frac{4 \pi R_{\rm out}^2 }{j_{0,N+1}^2} \sum_{i=1}^N  J_0\left( \frac{j_{0,k} j_{0,i}}{ j_{0,N+1} } \right) \frac{ A_{\rm eff} (\nu, r_{c, i}) I(\nu, r_{c, i}) }{ J_1^2(j_{0,i}) },
\end{equation}
where $R_{\rm out}$ is the outer radius of the disk, $J_0(x)$ and $J_1(x)$ are the zeroth- and first-order Bessel functions of the first kind, respectively, and $j_{0,i}$ is the $i$-th root of $J_0(x)$ (i.e., $J_0(j_{0,i}) = 0$). 
We define $R_{\rm out}$ such that the intensity beyond $r>R_{\rm out}$ is zero.
The terms $q_{c,k}$ and $r_{c,i}$ represent the collocation points in the spatial frequency and spatial domains:
\begin{eqnarray}
    q_{c, k} &=& \frac{j_{0,k}}{2\pi R_{\rm out}}, \\
    r_{c, i} &=& \frac{j_{0,i}}{j_{0,N+1}}R_{\rm out}.
\end{eqnarray}
{$A_{\rm eff} (\nu, r_{c, i})$ is the effective primary beam factor at each disk radius.
This has not been considered in the previous studies, but is discussed and introduced in Appendix \ref{app:pb}.
In practice, we adopt the primary beam implemented in the Common Astronomy Software Application version 6.7 \citep[CASA;][]{2022PASP..134k4501C} \footnote{See also \url{https://help.almascience.org/kb/articles/how-do-i-model-the-alma-primary-beam-and-how-can-i-use-that-model-to-obtain-the-sensitivity-pr}}.}

While the intensities are evaluated at the collocation points $r_{c, i}$, it is necessary to compute the visibilities at the specific $uv$-distances that match the observations. To achieve this, we employ
\begin{equation} \label{eq:vmod}
    \mathcal{V}_{\rm mod}( \nu, q_{l} ) = \sum_{i=1}^N H_i(q_l) I(\nu, r_{c, i}), 
\end{equation}
where 
\begin{equation}
    H_i(q_l) = \frac{4 \pi R_{\rm out}^2 A_{\rm eff} (\nu, r_{c, i}) }{j^2_{0, N+1} J^2_1 (j_{0,i}) } J_0 \left( 2 \pi q_l R_{\rm out} \frac{j_{0,i}}{j_{0, N+1}} \right).
\end{equation}
Here, $q_l$ corresponds to the $uv$-distance of the $l$-th data point at frequency $\nu$. This formulation assumes that the disk intensity vanishes for all radii $r \geq R_{\rm out}$.

\subsection{Visibility Comparison and Posterior Sampling} \label{sec:sampling}

The modeled visibilities ($\mathcal{V}_{\rm mod}( \nu, q_{l} )$, Eq. \ref{eq:vmod}) are compared directly with the observed visibilities, $\mathcal{V}_{\rm obs}(\nu, q_l)$. Note that the visibilities are purely real since we assume an axisymmetric disk.
We define the log-likelihood function as:
\begin{equation}
\log P = - \frac{1}{2} \sum_{m, l} \left( \frac{ \mathcal{V}_{{\rm obs}}(\nu_m, q_l) - f_m\mathcal{V}_{{\rm mod}}(\nu_m, q_l) }{ \sigma_{\mathcal{V}}(\nu_m, q_l) } \right)^2.
\end{equation}
The summation is performed over all $uv$ points across all observed frequencies $\nu_m$. The factor $f_m$ represents the flux scaling uncertainty, which we treat as a free parameter sampled from a log-normal prior. This factor is assumed to be constant within a single observational epoch where a common absolute flux calibration applies. If multiple independent datasets are combined, a different $f_m$ is assigned to each set to account for independent calibration uncertainties.
{$\sigma_{\mathcal{V}}(\nu_m, q_l)$ is the uncertainty associated with the visibility point.}

To retrieve the physical parameters, we utilize the \texttt{numpyro} probabilistic programming library \citep{2019arXiv191211554P, bingham2019pyro}.
We use the No-U-Turn Sampler \citep[NUTS; ][]{Hoffman2014-hj}, a variant of the Hamiltonian Monte Carlo (HMC) algorithm.
The specific hyper-parameters for the MCMC procedure, such as the number of warmup and sampling steps, are adjusted depending on the complexity of the problem.

\section{Injection Recovery Test Using Mock Data} \label{sec:injrec}

\subsection{Mock Data}
In this section, we perform a injection recovery test using mock datasets to evaluate the performance of our retrieval framework.
We first construct a mock disk model.
This model mimics the characteristic features of the HD 169142 disk, which we analyze in Section \ref{sec:hd169}, although it is not intended to be a high-fidelity physical representation.
First, the dust surface density is defined as a sum of four Gaussian rings,
\begin{equation}
    \Sigma_d(r) = \sum_{i=1}^4 \Sigma_{0,i} \exp\left( -\frac{(r-r_i)^2}{2 w_i^2} \right),
\end{equation}
where the peak surface densities are $\Sigma_{0,i} = (10, 1, 1.5, 0.1)~\text{g cm}^{-2}$, the ring center radii are $r_i = (26.2, 57.8, 64.7, 76.6)~\text{au}$, and the widths are $w_i = (2, 1, 1, 2)~\text{au}$. The radial positions $r_i$ correspond to the ring locations in HD 169142 identified by \citet{2019ApJ...881..159M} and \citet{2019AJ....158...15P}. The temperature profile is prescribed as
\begin{equation} \label{eq:Tinput}
 T(r) = 77 \left(\frac{r}{10~\text{au}}\right)^{-0.5}~\text{K},
\end{equation}
following \citet{2019ApJ...881..159M}. For the maximum grain size, we assume a step function,
\begin{equation}
     a_{\rm max}(r) = \left\{
\begin{array}{ll}
3.0~\text{cm} & (r < 46~\text{au})\\
0.1~\text{cm}  & (r \geq 46~\text{au})
\end{array}
\right.
\end{equation}
We adopt a power-law index of $q=3.5$ for the grain size distribution and use the default DSHARP dust opacity model \citep{2018ApJ...869L..45B}.

Using this physical structure and the radiative transfer model described in Section \ref{sec:rt}, we generated the radial intensity profiles at ALMA Bands 3, 6, and 9, centered at 91.5, 231, and 671 GHz, respectively.
To mimic the realistic data format, we simulated nine channels around each central frequency with a width of 0.1 GHz. 
The intensity and corresponding optical depth profiles are presented in Appendix \ref{app:tau} for reference.
Then, the two dimensional intensity distributions were calculated for a geometrically thin, axisymmetric disk with a position angle of $6.0^\circ$ and an inclination of $6.0^\circ$ using Equation \ref{eq:Inur}.
The distance to the source was set to the same as HD 169142 \citep[114.9 pc;][]{2023A&A...674A...1G}.

The resulting image cubes were processed using the CASA task \texttt{simobserve} to generate interferometric visibilities. We adopted ALMA antenna configurations designed to achieve a spatial resolution of ${\sim} 50~\text{mas}$: C43-10 and C43-7 for Band 3, C43-8 and C43-5 for Band 6, and C43-6 and C43-3 for Band 9. Integration times were set to 10,000 s for the long-baseline (LB) configurations and 2,000 s for the short-baseline (SB) configurations. Thermal noise was added with the default setting of \texttt{simobserve}, where the precipitable water vaport is 0.5 mm and the ground ambient temperature is 270 K.
We created two distinct datasets: an "SB+LB" set by concatenating both configurations, and an "SB-only" set using only the short-baseline data.

\begin{figure*}
\centering
\includegraphics[width=1.0\linewidth]{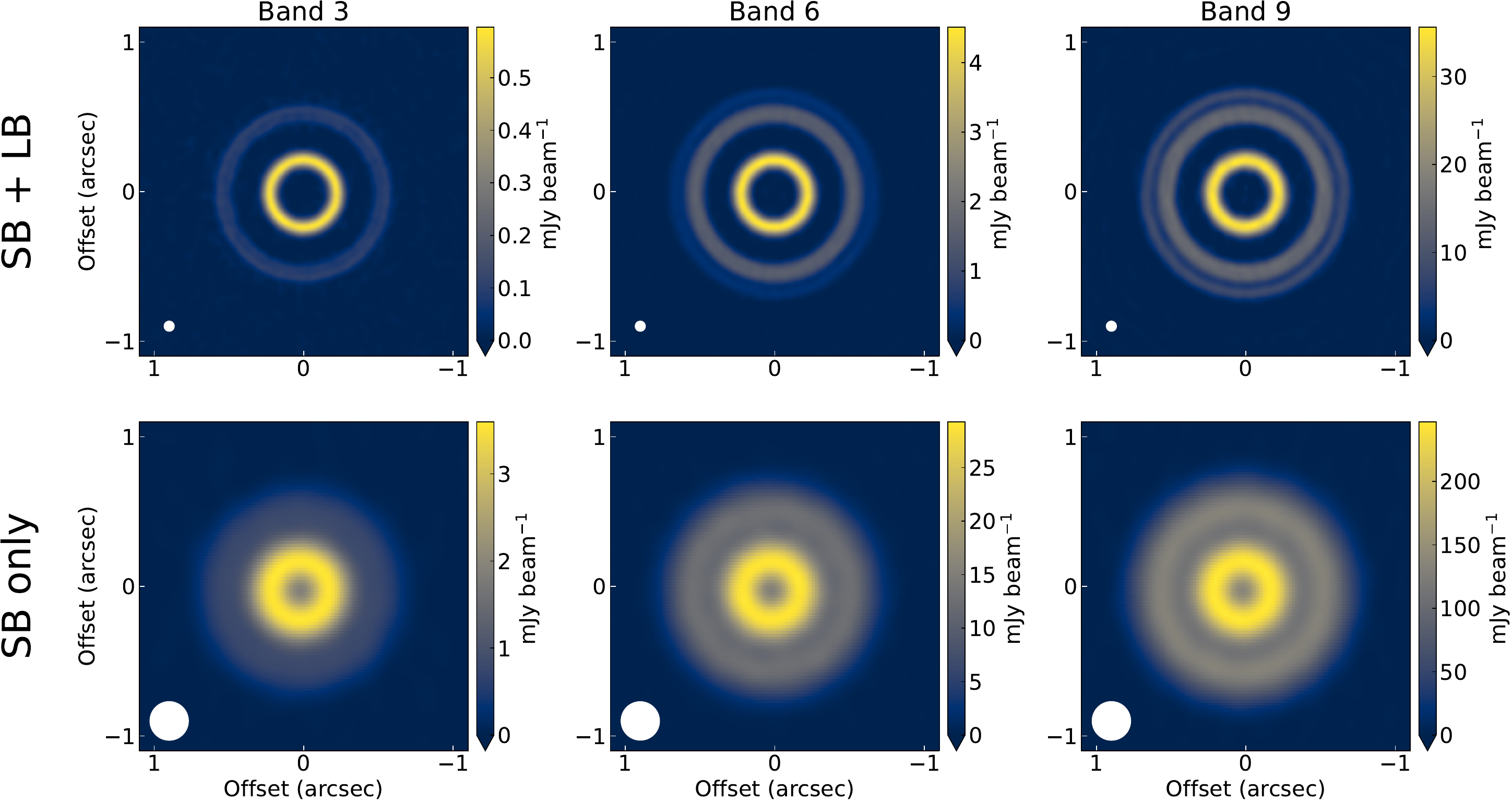}
\caption{CLEAN images of the mock datasets across three ALMA bands. The upper and lower panels show the SB+LB and SB-only datasets, respectively. The white ellipses in the bottom-left corners indicate the synthesized beam sizes. Note that all images within each row are smoothed to a common beam (${\sim}0\farcs06$ for SB+LB and ${\sim}0\farcs25$ for SB-only). Note that the images are not deprojected.}
\label{fig:mockdata}
\end{figure*}

Figure \ref{fig:mockdata} presents the CLEAN images generated from these visibility cubes. For comparison, the images in each set were smoothed to a common beam that is circular when the images are deprojected ($0\farcs06$ for SB+LB and $0\farcs25$ for SB-only).
The rms noise levels are $4.4\ {\rm \mu Jy\ beam^{-1}}$, $7.0\ {\rm \mu Jy\ beam^{-1}}$, and $110\ {\rm \mu Jy\ beam^{-1}}$ for the Band 3, 6, and 9 images from the SB+LB data, respectively.
The noise levels of the SB only data are ${\sim}$3-5 times larger than the SB+LB data for each band.
Even in the SB+LB data, the rings are not perfectly resolved due to the beam size ($\sim 6~\text{au}$ at 114.9 pc). The SB-only data, hampered by lower resolution, totally fails to distinguish the closely spaced rings at $r \sim 0\farcs 5$, showing only two broad features.

\subsection{Conventional Method} \label{sec:conv}

As a baseline for comparison, we first applied a conventional image-based analysis to the mock data. We azimuthally averaged the CLEAN images using the \texttt{GoFish} package \citep{2019JOSS....4.1632T} to produce radial intensity profiles. The uncertainties included both the image noise and a flux scaling factor $f$, modeled as a log-normal distribution $f \sim \exp(\mathcal{N}(0, \sigma_f))$, where $\mathcal{N}(0, \sigma_f)$ represents the normal distribution with the mean of zero and standard deviation of $\sigma_f$.
We adopted calibration uncertainties of $\sigma_f = 0.025, 0.05,$ and $0.1$ for Bands 3, 6, and 9, respectively.

At each radial bin, we fitted the SED model described in Section \ref{sec:rt} and sampled the posterior distribution using the NUTS algorithm in \texttt{numpyro}. We utilized the same DSHARP opacity model as used in the mock data generation. The free parameters were $\log_{10} \Sigma_d$, $\log_{10} T$, and $\log_{10} a_{\rm max}$, with $q=3.5$ fixed. We employed uniform priors: $\log_{10} \Sigma_d \in [-7, 3]~\text{g cm}^{-2}$, $\log_{10} a_{\rm max} \in [-3, 2]~\text{cm}$, and a temperature range $[T_{\rm min}, T_{\rm max}]$ defined as five times lower and higher than the input profile (Equation \ref{eq:Tinput}).
The MCMC chains consisted of 2,000 warmup steps and 1,000 sampling steps across 32 chains.

\begin{figure*}
\centering
\includegraphics[width=1.0\linewidth]{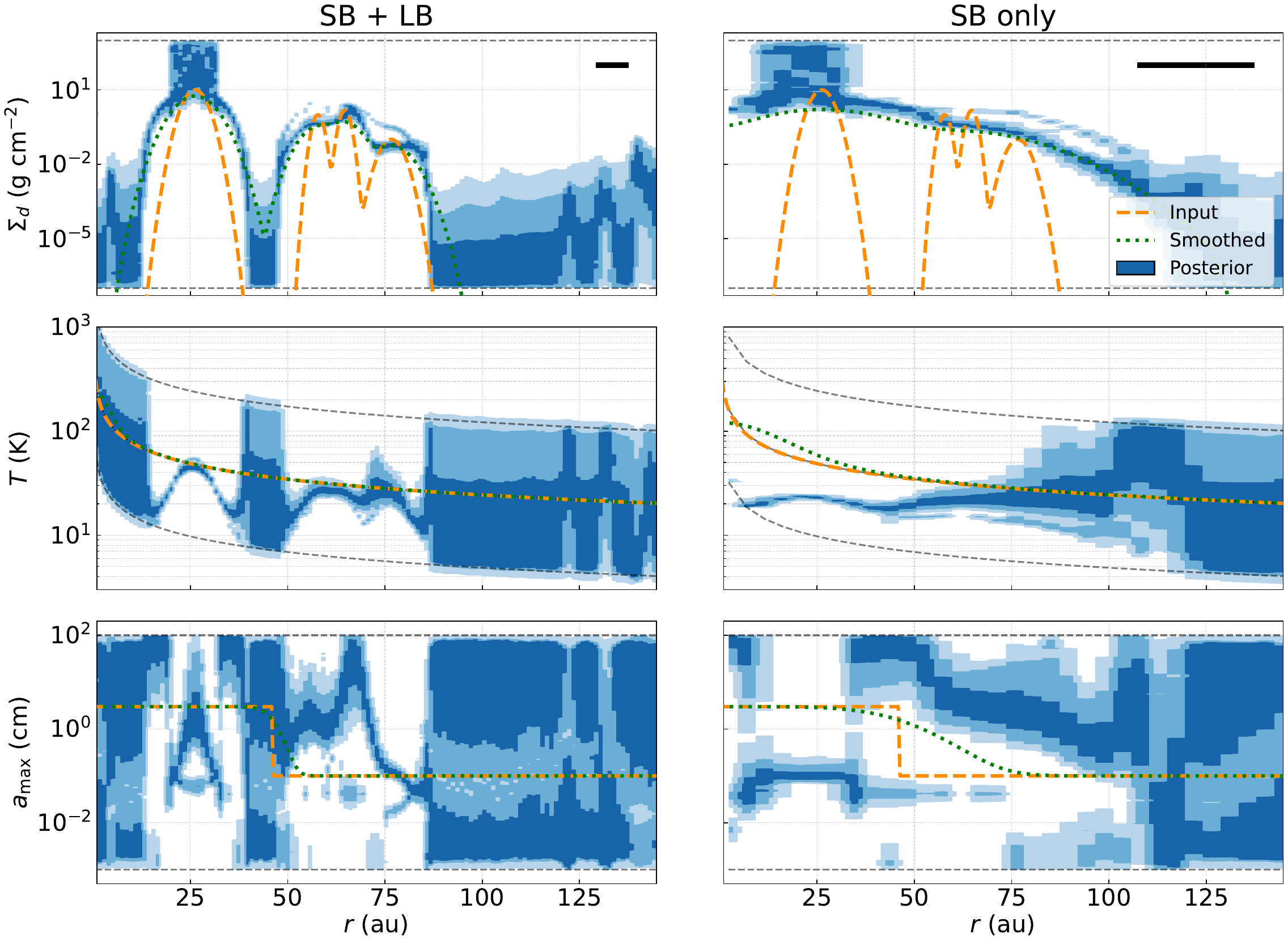}
\caption{Marginal posterior distributions for the conventional image-based fitting. The orange dashed lines and green dotted lines represent the true input profiles and the profiles smoothed with the image beam sizes, respectively. The left and right columns show results for the SB+LB and SB-only datasets, respectively. Shaded regions indicate the $68.3\%$, $95.4\%$, and $99.7\%$ (1, 2, and 3$\sigma$) highest density intervals. The black bars at the top indicate the FWHM of the synthesized beams.}
\label{fig:mock_results_imagebased}
\end{figure*}

Figure \ref{fig:mock_results_imagebased} shows the resulting marginal posterior distributions. Comparison with the input profiles reveals several systematic biases. First, the spatial resolution of the CLEAN beam limits the ability to recover tightly packed ring structures. Second, beam smearing causes an overestimation of the surface density across most radii, although {those} are similar to the input profiles smoothed with the imaging beam sizes.
Third, the inferred temperatures are biased toward lower values due to beam dilution of the emission peaks. Finally, the retrieved $a_{\rm max}$ values are almost always inaccurate, showing both over- and underestimations.
Furthermore, these biases also depend on the spatial resolution; in the SB-only case, the effect is more severe than the SB+LB case.
These results are basically consistent with a similar exercise by \citet{2025A&A...695A.147V}.

To quantitatively estimate the uncertainty, we calculate the total dust mass for $r > 46~\text{au}$, where the surface density appears better constrained, {from 10,000 random samples from each radius of the posterior.
We present the posterior distribution in Appendix \ref{app:mass}.
While the true input mass is $100~M_\oplus$, the image-based fitting yields $200_{-21}^{+25}~M_\oplus$ and $290_{-38}^{+57}~M_\oplus$ for the SB+LB and SB-only datasets, respectively. This demonstrates that conventional methods can lead to a severe overestimation of the total dust budget in disks.
}


\subsection{ \frappe\ Fitting }

\subsubsection{Visibility Reduction} \label{sec:datared}

Using the same mock datasets, we evaluate the performance of our proposed method.
In \frappe, we directly fit the one-dimensional visibility profiles.
While it is possible to project each individual $(u, v)$ point to its corresponding deprojected $uv$-distance at each frequency, this needs expensive Hankel transform calcuration for large datasets during posterior sampling. Therefore, in advance, we compress the data by calculating visibility values at a representative set of frequencies and $uv$-distances.

First, the $uv$-space in the wavelength unit is binned with a separation of 
\begin{equation}
    \Delta q_{\rm uv} = \frac{1}{\Theta},
\end{equation}
where we adopt $\Theta = 40'' (\simeq1.9 \times 10^{-4}\ {\rm rad})$ to prevent aliasing effects \citep[e.g.,][]{2017isra.book.....T}. 
Within each cell, the visibility points at $(\nu, q_{\rm uv})$ are fitted using a first-order Taylor expansion:
\begin{equation}
    \mathcal{\bar{V}}(\nu, q_{\rm uv}) = \mathcal{\bar{V}}_0 + \frac{\nu - \nu_0}{\nu_0} c_1 + \frac{ q_{\rm uv} - q_{\rm uv, 0} }{ q_{\rm uv, 0}} c_2,
\end{equation}
where $\nu_0$ is the representative frequency and $q_{\rm uv, 0}$ is the center of the $uv$-cell. The parameters $\mathcal{\bar{V}}_0, c_1,$ and $c_2$ are free parameters in the local fit, where $\mathcal{\bar{V}}_0$ represents the representative visibility value at the cell center. This procedure is implemented using \texttt{scipy.optimize.curve\_fit} \citep{2020NatMe..17..261V} to obtain $\mathcal{\bar{V}}_0$ and its associated uncertainty across all $uv$-cells.
The fitting is performed for cells with non-zero visibility points.
Those reduction of visibility data are conducted for each band.

\subsubsection{Sampling} \label{sec:sampling_mock}

\begin{figure}
\centering
\includegraphics[width=0.9\linewidth]{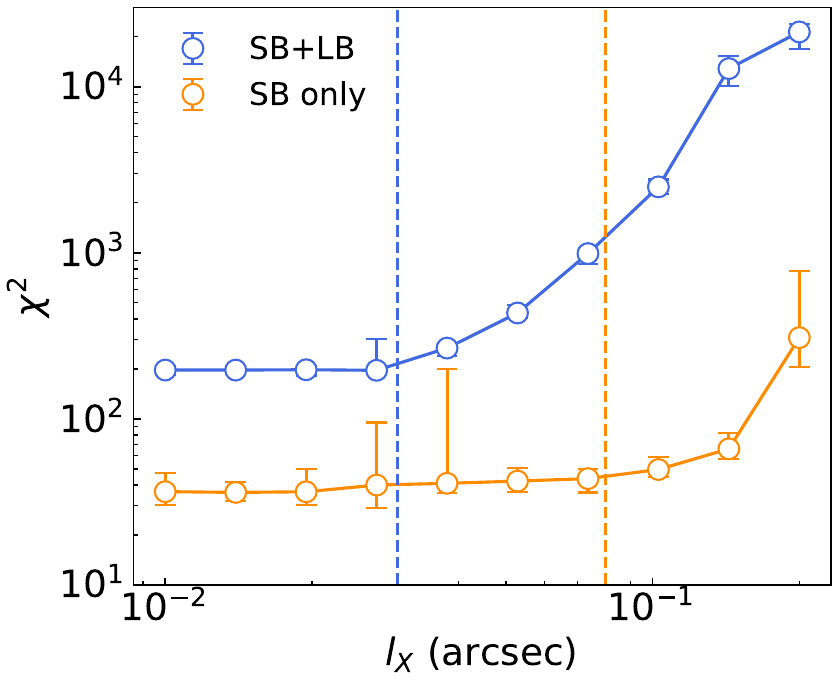}
\caption{ $\chi^2$ values as a function of the GP length scale $l_X$ from the cross-validation analysis. For the SB+LB data, $\chi^2$ remains nearly constant for length scales $l_X < 0\farcs03$ (indicated by the blue dashed line), suggesting that models within this regime are equally consistent with the data. }
\label{fig:cv_mock}
\end{figure}

We apply \frappe\ to the prepared one-dimensional visibilities, fixing $q=3.5$ for direct comparison with the conventional method but keeping other parameters free.
{In Appendix \ref{app:qfree}, we also demonstrate an additional fitting with $q$ being a free parameter.}
We adopt 130 radial grid points extending to $R_{\rm out} = 1\farcs3$.
As discussed in Section \ref{sec:gp}, the GP length scale $l_X$ is an important hyperparameter that determines the smoothness of the retrieved profiles.
To determine $l_X$ in a data-driven manner, we perform a ten-fold cross-validation analysis.

The data for each band are randomly partitioned into ten groups. We perform maximum a posteriori (MAP) estimation using Stochastic Variational Inference (SVI) on nine groups for a fixed $l_X$ and calculate the resulting $\chi^2$ against the remaining test group. This process is repeated for all groups to determine the median and $16-84$th percentiles of $\chi^2$ for various values of $l_X$ ($0\farcs01$--$0\farcs2$). As shown in Figure \ref{fig:cv_mock}, $\chi^2$ is nearly minimized and constant for $l_X \lesssim 0\farcs03$ in the SB+LB case, while it increases significantly for larger $l_X$. Consequently, we select $l_X = 0\farcs03$ for the SB+LB data and $l_X = 0\farcs08$ for the SB-only data.
We discuss this selection in Section \ref{sec:lx}.

\begin{figure*}
\centering
\includegraphics[width=1.0\linewidth]{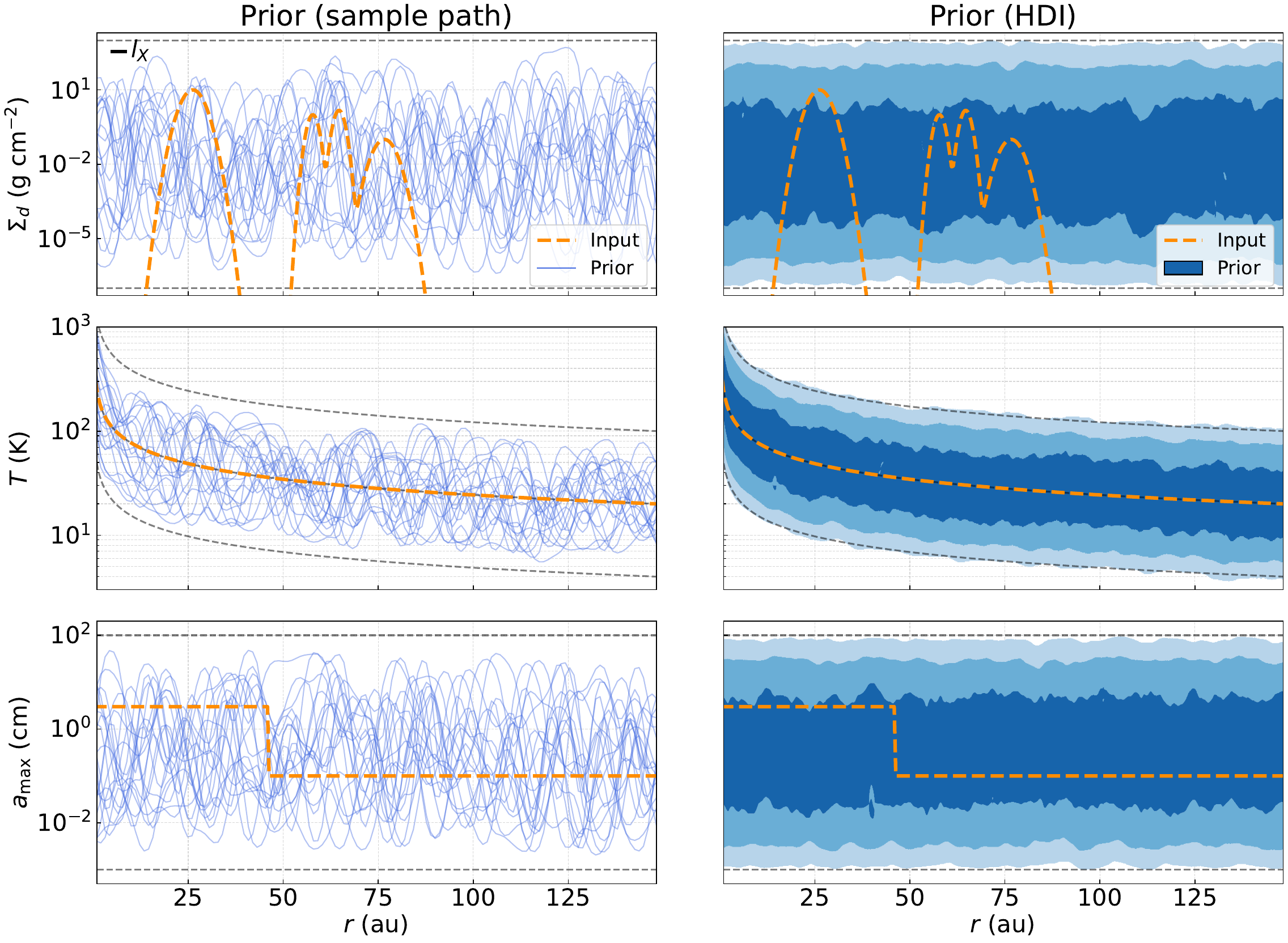}
\caption{ (Left) Selected 20 prior sample paths of the physical parameters used for the \frappe\ retrieval. The ensemble of paths demonstrates that the prior distribution sufficiently encompasses the true input profiles (orange dashed lines).
{The black bar in the top left panel corresponds to the length scale $l_X = 0\farcs03$.}
(Right) The same prior distribution visualized using Highest Density Intervals (HDI). The three shaded regions indicate the $68.3\%$, $95.4\%$, and $99.7\%$ ($1\sigma$, $2\sigma$, and $3\sigma$) HDI at each radius, respectively. }
\label{fig:prior_physical}
\end{figure*}
For the fixed hyperparameter, the prior distributions are visualized in Figure \ref{fig:prior_physical}.
The left panel shows individual sample paths drawn from the Gaussian Process prior, illustrating the wide range of radial variations permitted by the framework.
{The black bar at the top left panel indicates the adopted length scale ($l_X = 0\farcs03$).}
The right panel represents the same distributions using Highest Density Intervals (HDI). For the sake of clarity, we adopt the HDI visualization for the remainder of this work.

We perform posterior sampling using MCMC/NUTS with 32 chains.
Each chain consists of 5,000 warmup steps and 1,000 sampling steps.
The entire computation takes ${\sim}1$ hour on a workstation equipped with an AMD EPYC {7343} processor.
{We note that the computation time significantly depends on the number of radial points.}

\begin{figure*}
\centering
\includegraphics[width=1.0\linewidth]{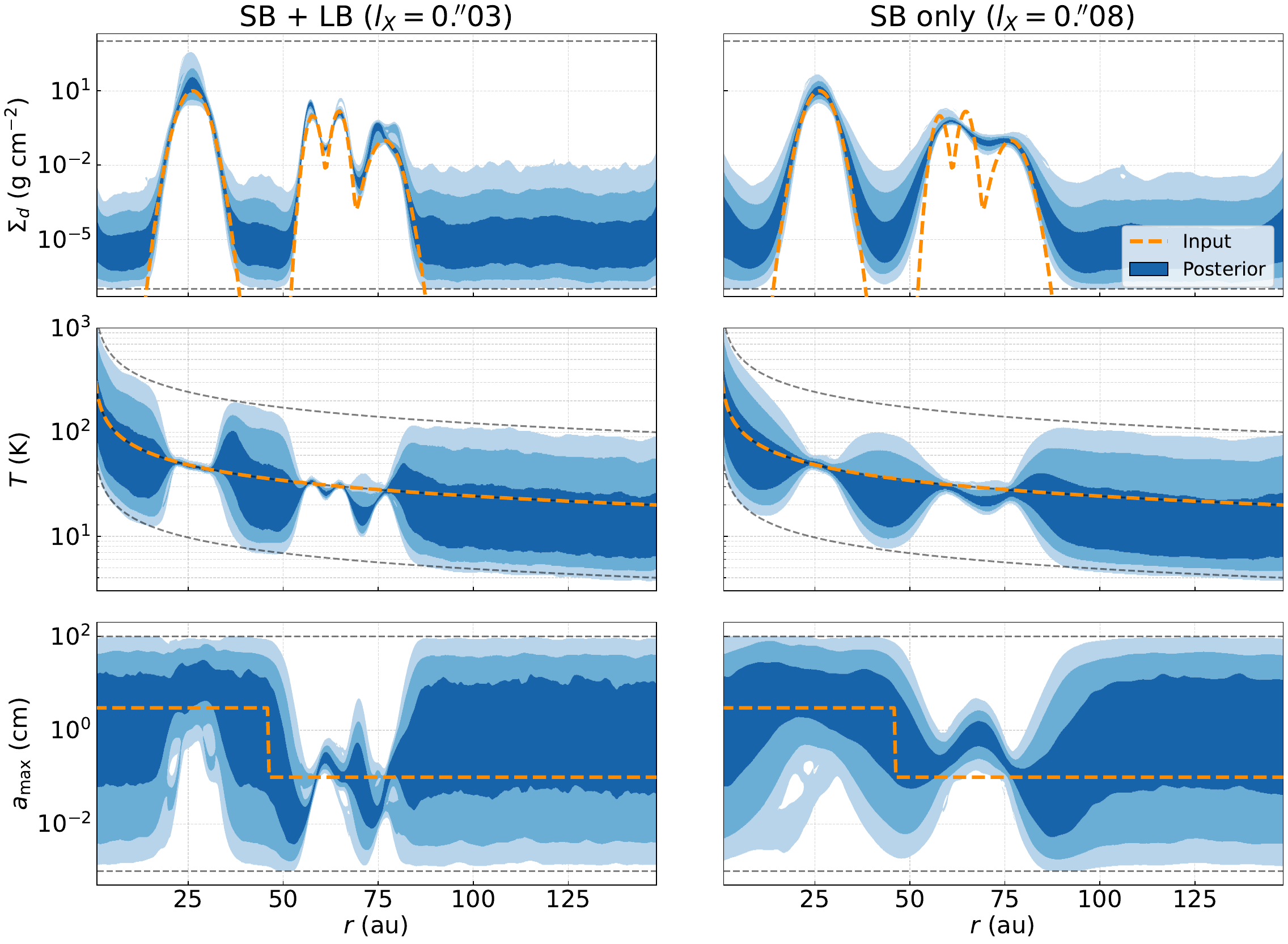}
\caption{Marginal posterior distributions obtained with \frappe. Shaded regions indicate the $68.3\%$, $95.4\%$, and $99.7\%$ (1, 2, and 3$\sigma$) highest density intervals. Compared to the conventional method (Figure \ref{fig:mock_results_imagebased}), the results are significantly less biased and accurately recover the input parameters in both configurations. }
\label{fig:mock_results_frappe}
\end{figure*}

Figure \ref{fig:mock_results_frappe} displays the posterior distributions from the \frappe\ fitting. In contrast to the conventional image-based analysis, {the results are much closer to the input profiles, suggesting that} the method exhibits significantly reduced biases.
Notably, in the SB+LB case, the tightly packed outer rings, which were unresolved in the CLEAN images, are successfully recovered in the surface density profile.
{There is deviations of the posterior median from the input surface density and temperature profiles at $r{\sim}70$ au, however, they are reasonably covered by the uncertainty ranges.}
{It is notable that the surface density posterior in the regions where the input value is almost zero ($r < 20, \sim 50, > 100$ au) gives the upper limits.
Although the median values are higher than the lower boundary, this is mathematically natural and should not be considered as constraints, given that the lower sides of the posterior are just determined by the prior (Figure \ref{fig:prior_physical}).
}

For the SB-only case, the adopted $l_X$ is insufficient to express the actual radial variation of the surface density profile, resulting that the individual rings remain unresolved and underestimation of the uncertainties.
Still, compared to the conventional method, the resulting profiles are much more closer to the input ones, which is a preferable feature.
Futhermore, in both cases, the temperature is much better constrained, and the maximum dust size is also better behaved, compared to the conventional method.

{In Appendix \ref{app:mass}, we present the posterior distribution of the dust mass at $r>46$ au for a comparison with the conventional method (Section \ref{sec:sampling_mock}), suggesting that the total dust mass is also reasnably recovered.}

\subsubsection{{A case with different configurations among bands}} \label{app:diffconf}

So far, we performed analysis for datasets with the same antenna configurations among bands.
However, \frappe\ is not limited to this ideal case and treat different configurations at the same time.
To demonstrate this, we perform the same injection recovery test in Section \ref{sec:injrec} but change the combination of the configurations to 1) SB+LB for Band 6 and 9, and SB-only for Band 3, and 2) SB+LB for Band 6, and SB-only for Band 3 and 9.

\begin{figure*}
\centering
\includegraphics[width=1.0\linewidth]{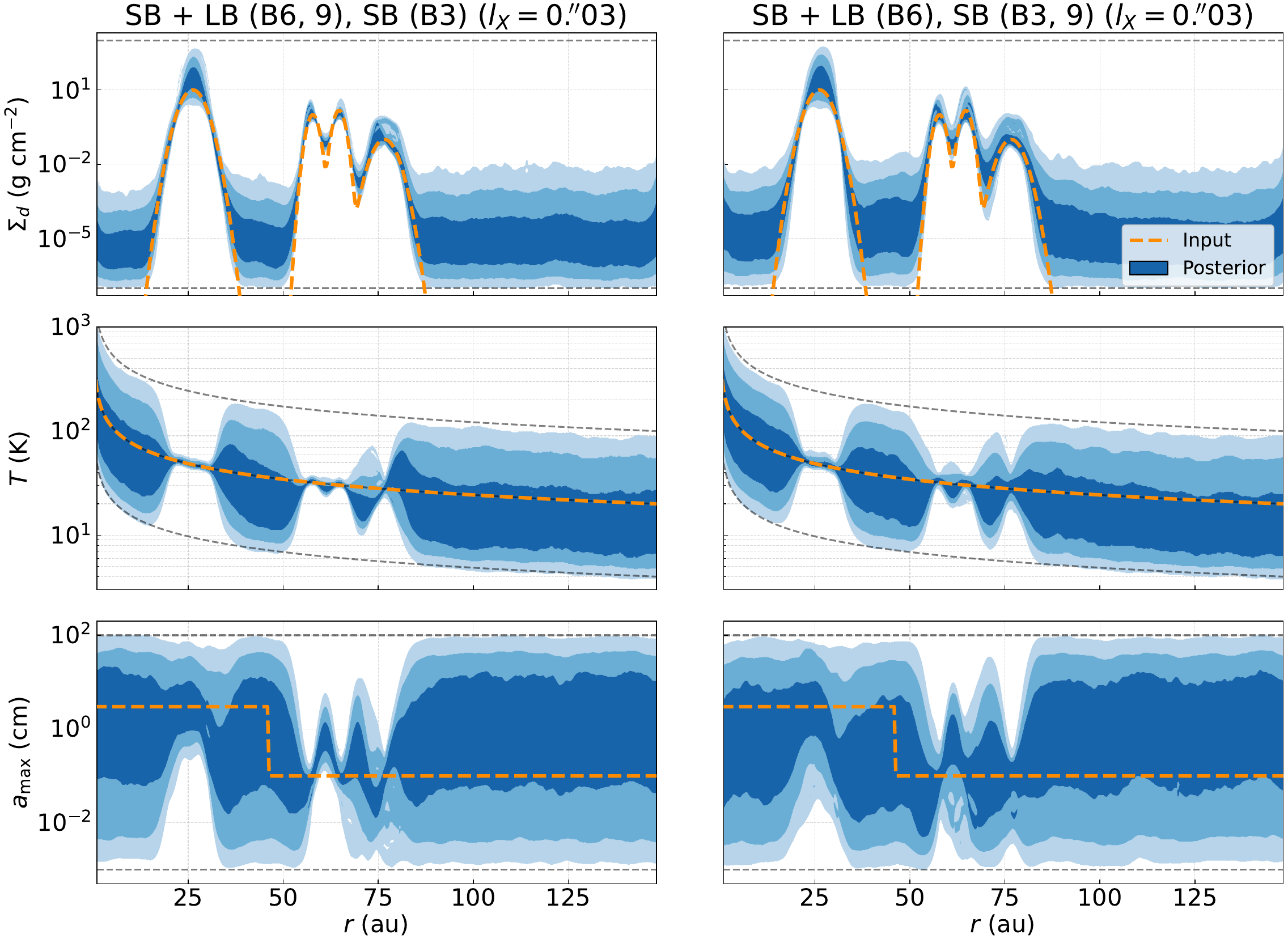}
\caption{Marginal posterior distributions obtained with \frappe\ for the case with different configurations among bands. Shaded regions indicate the $68.3\%$, $95.4\%$, and $99.7\%$ (1, 2, and 3$\sigma$) highest density intervals. }
\label{fig:mock_results_frappe_diffconfig}
\end{figure*}

In Figure \ref{fig:mock_results_frappe_diffconfig}, the results are presented.
In both cases, the input profiles are well recovered.
One difference is the uncertainty range; as information of the long baselines decreases, resultant uncertainty increases.

\subsubsection{On the Selection of the Length Scale Parameter} \label{sec:lx}

\begin{figure*}
\centering
\includegraphics[width=1.0\linewidth]{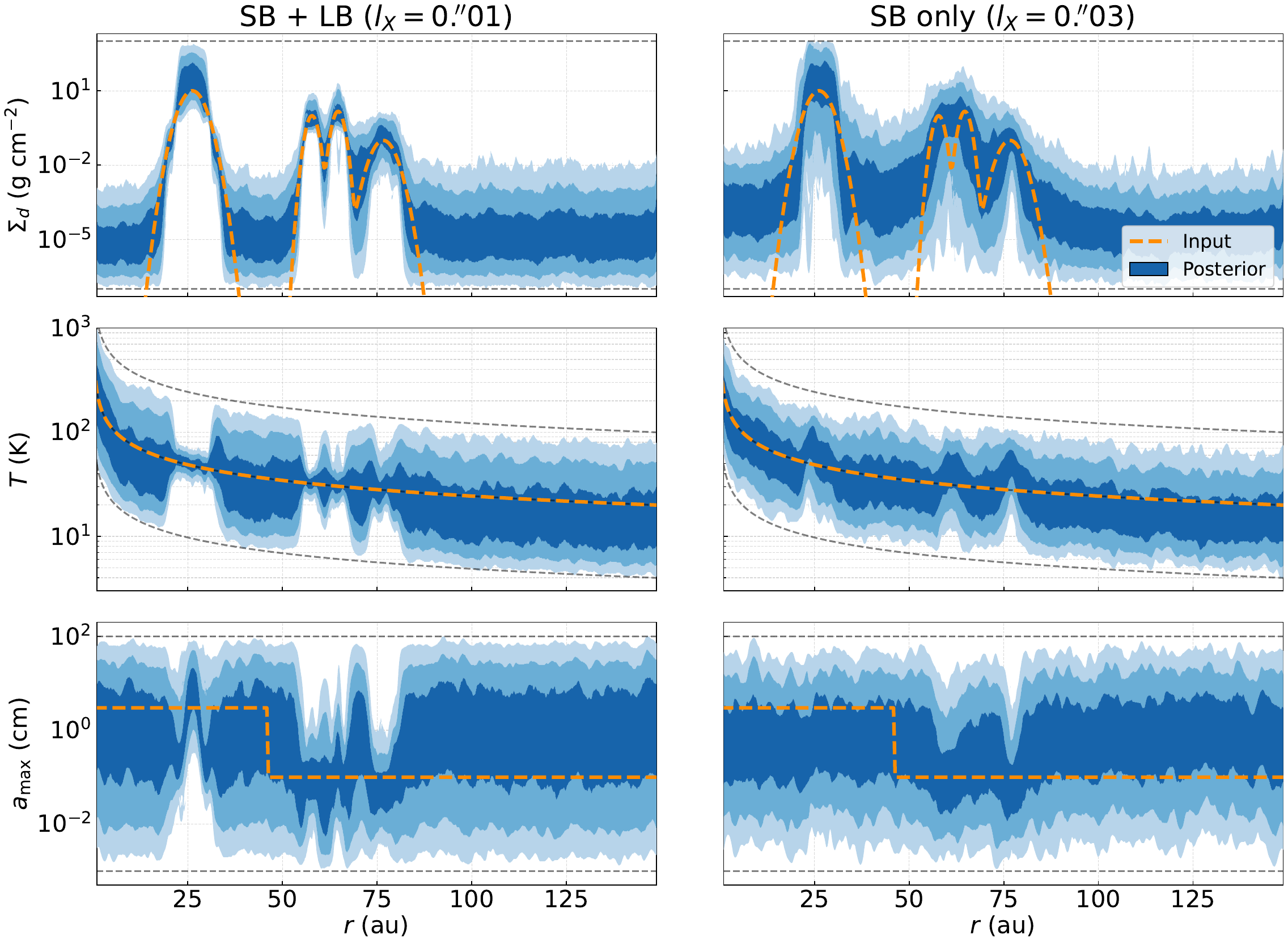}
\caption{Marginal posterior distributions for additional sampling with alternative length scales ($l_X = 0\farcs01$). }
\label{fig:different_lX}
\end{figure*}

To investigate the sensitivity of our results to the choice of $l_X$, we perform additional sampling with $l_X = 0\farcs01$ (Figure \ref{fig:different_lX}). 
For the SB+LB data, the results remain consistent with the input model, although the uncertainties are larger compared to the $l_X = 0\farcs03$ case.
This increased uncertainty arises from the higher flexibility allowed by the smaller length scale, which permits small-scale oscillations.
Crucially, however, the lack of strong bias remains.
In the SB-only case, the higher flexibility allows that now the uncertainty range covers the input profiles.
{We plot additional examples with different $l_X$ in Appendix \ref{app:diff_lX}.}

{
These examples show that increasing $l_X$ affects the retrieved posteriors since variations on scales smaller than $l_X$ are effectively smoothed.
However, the results can only be significantly deviated from reality if the scale length is larger than the scale of the physical variation, which is happening in the SB only case of Figure \ref{fig:mock_results_frappe}. In other cases, the results reasonably cover the truth within the uncertainty range, although the uncertainty itself depends on the length scale (i.e., smaller length scales make larger uncertainties).}

{This behavior could be leveraged by setting $l_X$ based on physical motivations. For instance, one idea is that setting $l_X$ to the local gas scale height.
Still, it would require sufficiently deep high-resolution observations for meaningful results.}
{Therefore, in most cases, determining $l_X$ via cross-validation is sufficient, provided it is acknowledged that the model is insensitive to variations below this scale.
We note that, while it is possible to treat $l_X$ as a free parameter, which is a common practice in GP regression, the constant $\chi^2$ values at small $l_X$ (as seen in the cross-validation; Figure \ref{fig:cv_mock}) often make it difficult to derive meaningful constraints on $l_X$ directly from the data.
}

\section{Application to ALMA observations of the HD 169142 disk} \label{sec:hd169}

In this section, we apply the developed framework to actual ALMA observations of the protoplanetary disk around HD 169142.
HD 169142 is a Herbig Ae star located at a distance of 114.9 pc from the Earth \citep{2023A&A...674A...1G} with a well-known protoplanetary disk.
The dust component of the disk has been extensively studied across various wavelengths \citep{2010PASJ...62..347F, 2012ApJ...752..143H, 2013ApJ...766L...2Q, 2014ApJ...791L..36O, 2015PASJ...67...83M, 2017A&A...600A..72F, 2019ApJ...881..159M, 2019AJ....158...15P, 2025AJ....170..278L}, consistently revealing that the disk consists of two major rings at ${\sim}26$ au and ${\sim}60$ au from the central star.
The highest spatial resolution observations to date further revealed that the outer ring is composed of at least three narrow (${\sim}$5 au) rings \citep{2019AJ....158...15P}.

\subsection{Observations and Data Reduction} \label{sec:obs}

We analyze multi-band continuum data from ALMA Band 3 ($\lambda \sim 3.1$ mm), Band 6 ($\lambda \sim 1.3$ mm), and Band 9 ($\lambda \sim 0.45$ mm). The observational details are summarized in Appendix \ref{app:obsdet}.
In the following, we describe the calibration and data reduction procedures.

\subsubsection{Band 3}

The Band 3 continuum observations ($\lambda = 3.1$ mm) were conducted under program \#2018.1.01716.S (PI: E. Mac\'{i}as) and originally published by \citet{2024A&A...684A.134R}. We retrieved the pipeline-calibrated measurement sets from the NRAO archive and performed additional self-calibration using the \texttt{exoALMA} pipeline \citep{2025ApJ...984L...7L}.

The self-calibration process involved six rounds of phase-only calibration for the short-baseline data, with solution intervals ranging from the execution block (EB) length down to 10 s. After concatenating the short-baseline data with the long-baseline configurations, we performed six additional rounds of phase-only self-calibration and one final round of amplitude and phase calibration using the EB length as the interval. Although our analysis is primarily performed in the visibility domain, we generated a CLEAN image (robust=0.0) for visualization, yielding a synthesized beam size of $41 \times 38$ mas (${\rm PA} = 83^\circ$). To ensure a consistent multi-wavelength comparison, the image was smoothed to a resolution that becomes a $45$ mas circular beam when deprojected to a face-on geometry. The RMS noise level of the smoothed image is $4.4~\mu\text{Jy beam}^{-1}$.

\subsubsection{Band 6}
For Band 6, we utilized archival data from four different projects. These datasets were selected based on the criterion that they cover the CO $J=2-1$ line, the detailed analysis of which will be presented in a forthcoming paper (Yoshida et al. in prep.). 
Following the standard pipeline calibration for each dataset, we applied the \texttt{exoALMA} self-calibration pipeline to the concatenated continuum data. This process included four rounds of phase-only self-calibration for both the short- and long-baseline data, with intervals decreasing to 60 s, followed by one round of amplitude and phase self-calibration. A CLEAN image was generated with robust$=0.0$, resulting in a synthesized beam of $42 \times 26$ mas (${\rm PA} = 73^\circ$). Similar to the Band 3 data, the image was smoothed to a deprojected face-on resolution of $45$ mas. The resulting RMS noise level is $11.3~\mu\text{Jy beam}^{-1}$.

\subsubsection{Band 9}

For Band 9, we incorporated data from projects \#2017.1.00727.S (PI: J. Szulágyi) and \#2024.1.01254.S (PI: T. C. Yoshida).
{The former and later program corresponds to the long and short baselines, and the both originally consist of two EBs. However, we found that the second EB of the short baseline program was affected by significant phase decoherence, resulting in a large (a factor of two) descrepancy in the total flux. Therefore, we decided to flag the second EB and only use the first EB.
}
After pipeline calibration, the data were self-calibrated using the \texttt{exoALMA} pipeline.
We performed seven rounds of phase-only self-calibration for the short-baseline data, with intervals ranging from the EB length down to 10 s. This was followed by {one round} of amplitude and phase self-calibration using a EB-length interval, respectively, to specifically calibrate the flux offsets between EBs.
This calibrated short-baseline dataset was then concatenated with the long-baseline data, and the entire self-calibration procedure was repeated {with an additional second round of amplitude and phase self-calibration using a scan-length interval.}
{The final CLEAN image (robust=0.5) has a synthesized beam of $35 \times 32$ mas (${\rm PA} = 86^\circ$).
After smoothing to the $45$ mas beam, the RMS noise level is estimated to be $0.20~\text{mJy beam}^{-1}$.}

\subsection{Data-driven Estimation of Absolute Flux Uncertainty}

Since the uncertainty in absolute flux calibration is a dominant factor in the total flux accuracy, a careful assessment is essential.
While the ALMA Technical Handbook \citep{cortes_2025_14933753} provides general stability guidelines for calibrators, actual uncertainties can vary significantly \citep{2020AJ....160..270F, 2025ApJ...980...50Y}.
Therefore, we estimate the flux uncertainty in a data-driven manner for each band.

First, we measure the flux ratios of each EB relative to a reference EB within the same band just after their phase-only self-calibration, utilizing azimuthally averaged visibility profiles as implemented in the \texttt{exoALMA} calibration script \citep{2018ApJ...869L..41A, 2025ApJ...984L...7L}.
The short-baseline EB with the highest signal-to-noise ratio is selected as the reference, consistent with our self-calibration procedure.
We then group these ratios for EBs observed within a 30-day window, representing the typical duration of an observing project, and calculate the mean for each group. 

The final mean flux scaling factor for a given band, $\bar{f}$, is obtained by averaging the mean values across all groups. The associated uncertainty is estimated as:
\begin{equation}
   \sigma_{\bar{f}} = \frac{\max( \sigma_g, \sigma_n )}{\sqrt{N_b}},
\end{equation}
where $\sigma_g$ is the standard deviation of the group averages, $\sigma_n$ is the nominal accuracy reported in the ALMA Technical Handbook ($\sigma_n = 0.025, 0.05,$ and $0.2$ for Bands 3, 6, and 9, respectively), and $N_b$ is the number of independent groups in the band. 

\begin{deluxetable}{cccc}
\tablecaption{Absolute flux uncertainties derived from the observational data. \label{tab:fs}}
\tablewidth{0pt}
\setlength{\tabcolsep}{15pt} 
\tablehead{
\colhead{Band} & \colhead{$\bar{f}$} & \colhead{$\sigma_{\bar{f}}$} & \colhead{$\sigma_{n}$}
}
\startdata
3 & 1.0 & 0.018 & 0.025 \\
6 & 1.1 & 0.063 & 0.050 \\
9 & 0.98 & 0.14 & 0.20 \\
\enddata
\end{deluxetable}

The resulting mean values and standard deviations are summarized in Table \ref{tab:fs}. We note that $\bar{f}$ serves as a nuisance parameter, as its absolute value depends on the choice of the reference EB. We confirmed that these results are not very sensitive to the grouping duration.
These empirically derived values are adopted for the subsequent retrieval analysis.

\subsection{Constraining Disk Geometry}

Since \frappe\ operates on visibility data reduced to one-dimensional radial profiles, the disk geometry must be accurately constrained beforehand. We employ the \texttt{protomidpy} package \citep{2024MNRAS.532.1361A} to fit axisymmetric intensity models to the two-dimensional visibilities. This allows us to constrain the disk center coordinates, inclination $i$, and position angle PA independently for each band.

After identifying the geometric parameters, we align the disk center with the phase center using the CASA task \texttt{fixvis}. For the final visibility deprojection and azimuthal averaging, we adopt the inclination and position angle derived from the Band 6 data ($i = 6.3^\circ$ and $\text{PA} = 6.3^\circ$; coincidentally the same numbers), which provided the highest signal-to-noise ratio among the available datasets.

\subsection{Overview of the Images and Radial Profiles}

\begin{figure*}
\centering
\includegraphics[width=1.0\linewidth]{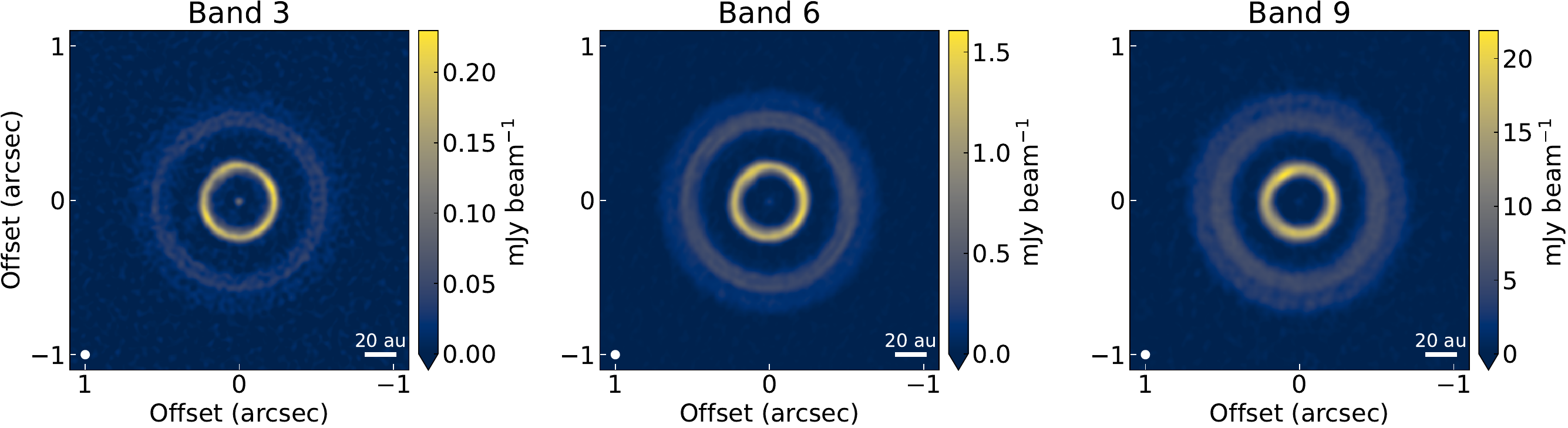}
\caption{CLEAN images of the HD 169142 disk across three ALMA bands. The white ellipses at the bottom left indicate the synthesized beam size (45 mas). {The white bars represent a scale of 20 au.} Note that the images are not deprojected. }
\label{fig:actualdata}
\end{figure*}

\begin{figure*}
\centering
\includegraphics[width=1.0\linewidth]{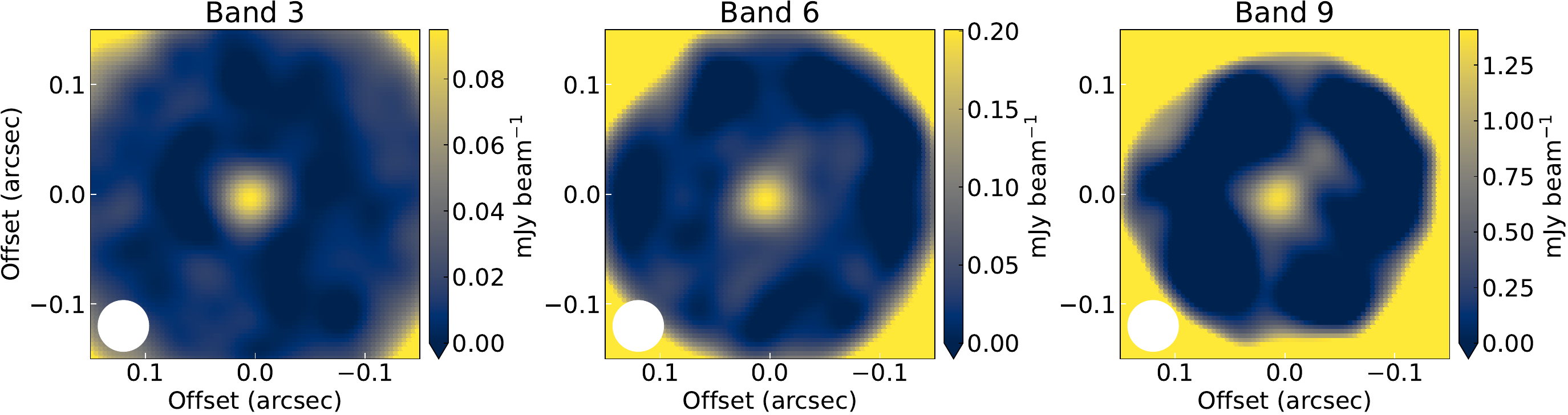}
\caption{Zoom-in views of the innermost regions of the HD 169142 disk corresponding to the images in Figure \ref{fig:actualdata}. }
\label{fig:actualdata_zoomin}
\end{figure*}

Figure \ref{fig:actualdata} displays the resulting CLEAN images of the processed datasets. To facilitate a consistent multi-wavelength comparison, all images were smoothed to a common effective resolution of 45 mas after deprojecting to a face-on geometry. Figure \ref{fig:actualdata_zoomin} provides zoom-in views of the innermost regions, confirming that the central disk component is detected across all observed bands.

\begin{figure}
\centering
\includegraphics[width=0.9\linewidth]{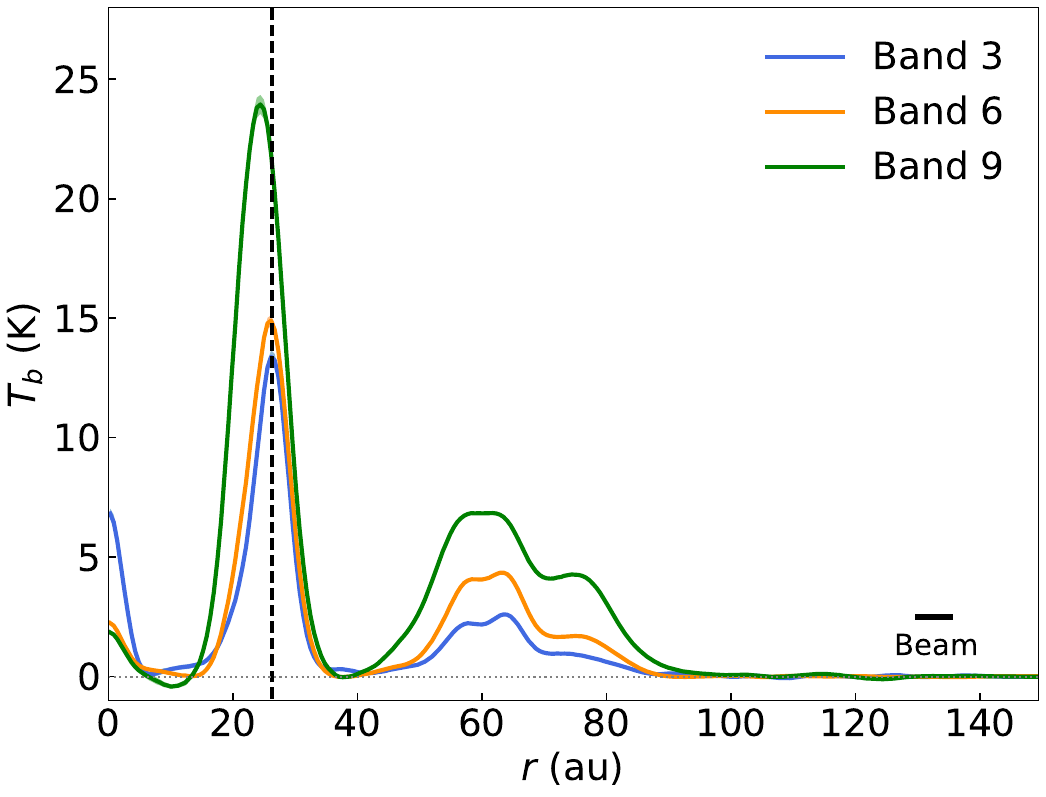}
\caption{Radial brightness temperature profiles of the HD 169142 disk derived from the CLEAN images. The vertical dashed line indicates the peak location of the inner ring in Band 6 ($\sim 26$ au).}
\label{fig:radprofhd169}
\end{figure}

While the continuum images exhibit several asymmetric features that have been discussed in previous studies \citep{2019AJ....158...15P, 2025AJ....170..278L}, this work focuses on the azimuthally averaged radial properties. Figure \ref{fig:radprofhd169} shows the radial brightness temperature profiles in the Rayleigh–Jeans approximation extracted from the CLEAN images. We find that the peak of the inner ring at $\sim 26$ au in Band 9 is located at a slightly smaller radius compared to its counterparts in Bands 3 and 6.
Furthermore, the inner ring in Band 9 appears substantially broader, and the outer disk emission is smoother than in the other bands. These morphological differences are potentially indicative of differential dust trapping and grain growth as a function of radius \citep[e.g.,][]{2023ApJ...957...11D} and/or an exposed edge of the ring from the central star with a high optical depth in Band 9.

In the innermost region ($r < 10$ au), a compact source is detected in all bands. While the brightness temperatures in Bands 6 and 9 are comparable, the brightness temperature in Band 3 is significantly higher. This relative enhancement at longer wavelengths suggests that an emission mechanism other than thermal dust emission, such as free-free emission from ionized gas, contributes to the Band 3 flux in the central region.

\subsection{Retrieving Dust Properties with \frappe}

\begin{figure*}
\centering
\includegraphics[width=0.8\linewidth]{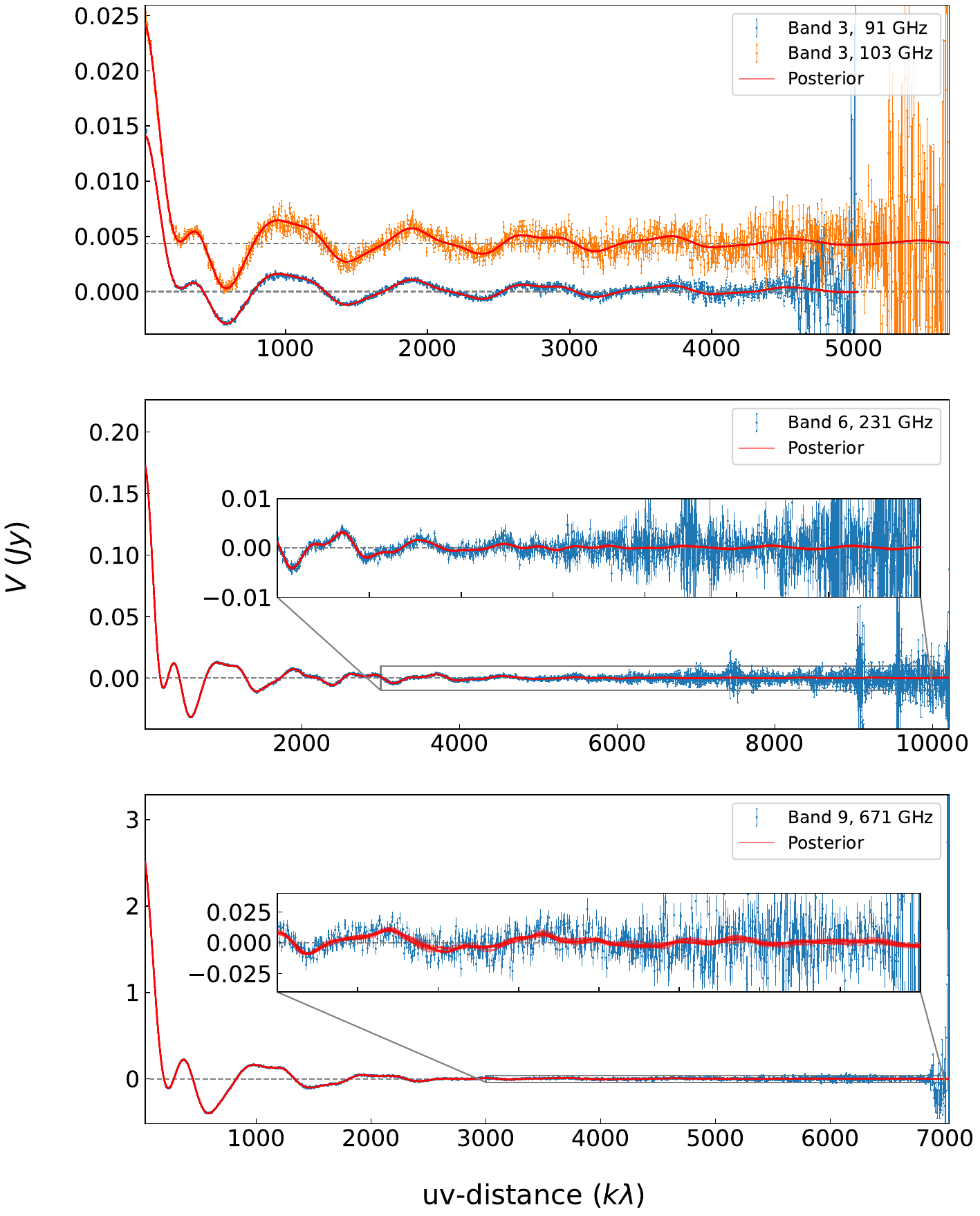}
\caption{Visibility profiles of the HD 169142 disk across three ALMA bands. The data points include $1\sigma$ error bars. The solid red lines represent ten randomly selected posterior samples from the \frappe\ fitting using the Ricci opacity model, demonstrating the excellent agreement between the model and the interferometric data.}
\label{fig:visplot}
\end{figure*}

The retrieval procedure for the observational data follows the methodology validated through the mock tests (Section \ref{sec:injrec}). We generated one-dimensional visibility profiles. For Band 3, however, we adopted two representative frequencies corresponding to the lower and upper sidebands. This accounts for the significant fractional bandwidth of Band 3, where a single-frequency approximation may not sufficiently capture the spectral slope of the emission. The resulting visibility profiles are shown in Figure \ref{fig:visplot}.

In this retrieval, we expand the parameter space by treating the power-law index of the dust grain size distribution, $q$, as a free parameter, in addition to the dust surface density ($\log_{10} \Sigma_d$), dust temperature ($\log_{10} T$), and maximum grain size ($\log_{10} a_{\rm max}$).
{We also test this problem setting using the mock datasets in Appendix \ref{app:qfree}.}
We adopt the same prior boundaries as used in the mock tests, with the index $q$ constrained between 1 and 5. Based on a dedicated cross-validation analysis for this dataset, we selected a Gaussian Process length scale of $l_X = 0\farcs03$.

To assess the systematic uncertainties arising from dust property assumptions, we performed the retrieval using two distinct opacity models: the DSHARP model \citep{2018ApJ...869L..45B} and the model presented by \citet{2010A&A...512A..15R} (assuming zero porosity). Both opacity tables were generated using the \texttt{dsharp\_opac} module \citep{2018ApJ...869L..45B}.

The central region of HD 169142 harbors a compact point source that may include contributions from free-free emission from ionized gas, as suggested by previous studies \citep{2017ApJ...838...97M, 2024A&A...684A.134R}. To account for this without making assumptions about its physical origin, we incorporate an additional point-source component into the visibility model:
\begin{equation}
    F_{\rm pt}(\nu) = F_0 \left( \frac{\nu}{100~\text{GHz}} \right)^{\beta},
\end{equation}
where $F_0$ and $\beta$ are the flux at $\nu = 100$ GHz and spectral slope.
As priors, we set broad Gaussian distributions for these two parameters, $\log_{10} (F_0/ {\rm mJy}) \sim \mathcal{N}(-1.0, 1.0)$ and $\beta \sim \mathcal{N}(0.0, 4.0)$.
This power-law term represents the non-thermal-dust contribution at the disk center, effectively allowing the framework to isolate the thermal dust emission from the total flux. 
We utilized the flux scaling uncertainties derived in Section \ref{sec:obs}. The posterior distribution was sampled using {10,000} warmup steps and 1,000 sampling steps across 32 chains.
{
After sampling, for both opacity models, we identified that one chain was trapped in a region where the log-probability is ${\sim}1,000$ lower than that of the other chains.
To ensure a proper evaluation of the posterior, this chain was excluded from the subsequent analysis as this behavior is purely a numerical artifact.
}

\subsection{Results}

In Figure \ref{fig:visplot}, the ten randomly selected visibility profiles from the posterior distribution {with the Ricci opacity model} are plotted, demonstrating that they are in excellent agreement with the interferometric data across all bands.

\begin{figure*}
\centering
\includegraphics[width=1.0\linewidth]{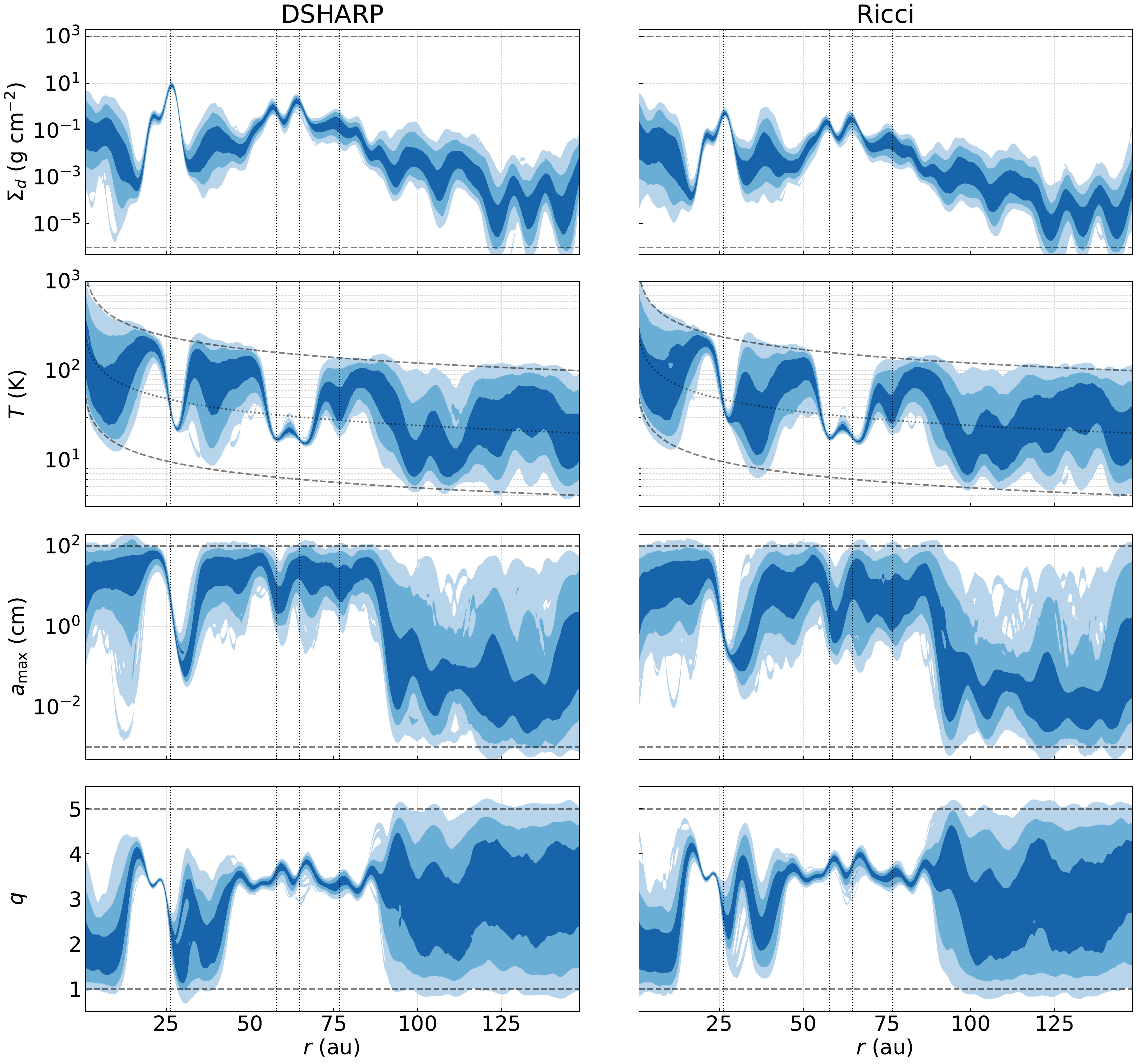}
\caption{Marginal posterior distributions for the physical properties of the HD 169142 disk. Shaded regions indicate the $68.3\%$, $95.4\%$, and $99.7\%$ (1, 2, and 3$\sigma$) highest density intervals. Results are compared between the DSHARP (left) and Ricci (right) opacity models. Black dashed lines indicate the prior boundaries. Vertical lines show the ring locations previously specified. Black dotted lines in the $T$ panels show the temperatures expected from a passive disk model (Eq. \ref{eq:Tinput}).}
\label{fig:hd169_results_frappe}
\end{figure*}

The retrieved physical quantities are presented in Figure \ref{fig:hd169_results_frappe}. Our framework successfully reconstructs the radial profiles for both opacity models. The dust surface density profiles reveal a complex structure consisting of an inner ring at $\sim 26$~au, three outer rings around $\sim 60$~au (see also the intensity profiles in Figure \ref{fig:radprofhd169}), and a broad background extending up to $\sim 120$~au. Notably, the surface density remains non-negligible ($10^{-3}$--$10^{-2}~\text{g cm}^{-2}$) even within the major gaps at $\sim 10$~au and $\sim 40$~au, as well as in the regions beyond the primary rings.
{These posteriors are clearly distinct from those in the injection-recovery test, where a zero-input surface density yields a posterior consistent with the prior plus an upper limit (Section \ref{sec:sampling_mock}).}
Furthermore, the results indicate a significant contribution from dust grains in the innermost disk ($r < 10$~au). While the two opacity models yield structurally similar profiles, the absolute values of the surface density are generally lower in the Ricci model.
\begin{figure*}
\centering
\includegraphics[width=1.0\linewidth]{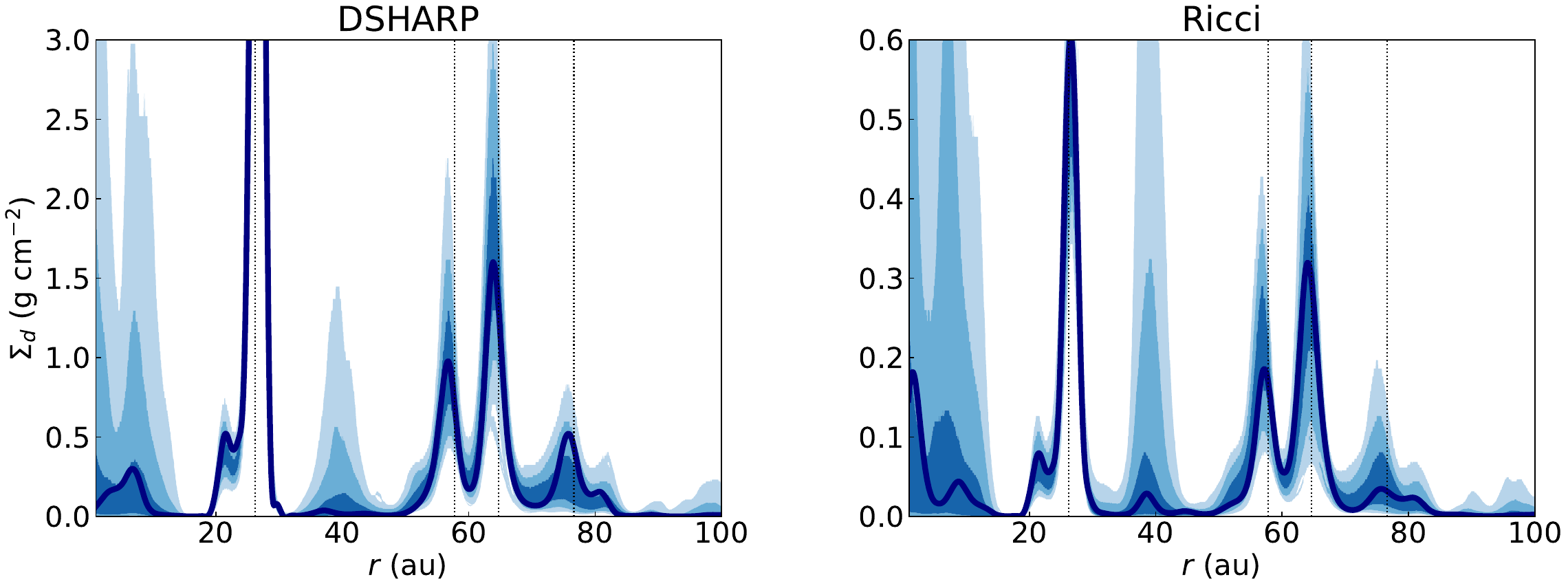}
\caption{Marginal posterior distributions for the dust surface densities shown on a linear scale. Results are shown for the DSHARP (left) and Ricci (right) opacity models. Solid lines indicate the maximum a posteriori (MAP) samples.}
\label{fig:sigma_lin}
\end{figure*}
For reference, Figure \ref{fig:sigma_lin} provides a linear-scale representation of the surface density profiles, which more clearly highlights the dominance of the multiple ring structures.
{The total {dust} mass is estimated to be $330^{+40}_{-30}\ {M_\oplus}$ and $49^{+7}_{-6}\ {M_\oplus}$ for the DSHARP and Ricci opacities, respectively, with Gaussian-like marginal posterior distributions.}
This descrepancy between two models is primarily because of the higher opacity of the Ricci model.

The temperature profiles are well constrained primarily within the ring regions. 
A steep radial temperature gradient is suggested just inward of the 26 au ring.
At the peaks of both the inner and outer rings, the retrieved temperatures are consistent with, or slightly lower than, the values expected from a standard passive disk model (dotted line in Figure \ref{fig:hd169_results_frappe}; Equation \ref{eq:Tinput}).

The maximum grain size at the inner ring is estimated to be $\sim 1$~cm. In contrast, the outer rings and the innermost disk exhibit a lower limit of approximately $\sim 1$~mm. Beyond the outer rings, the grain size appears to decrease toward $\sim 0.1$~mm, although the uncertainties in this region are considerable. The power-law index $q$ of the grain size distribution is largely consistent with the interstellar medium (ISM) value of 3.5 across most radii. While the DSHARP model allows for values as low as $q \sim 2$ at certain radii, this trend is not observed when the Ricci model is adopted.

\begin{figure*}
\centering
\includegraphics[width=1.0\linewidth]{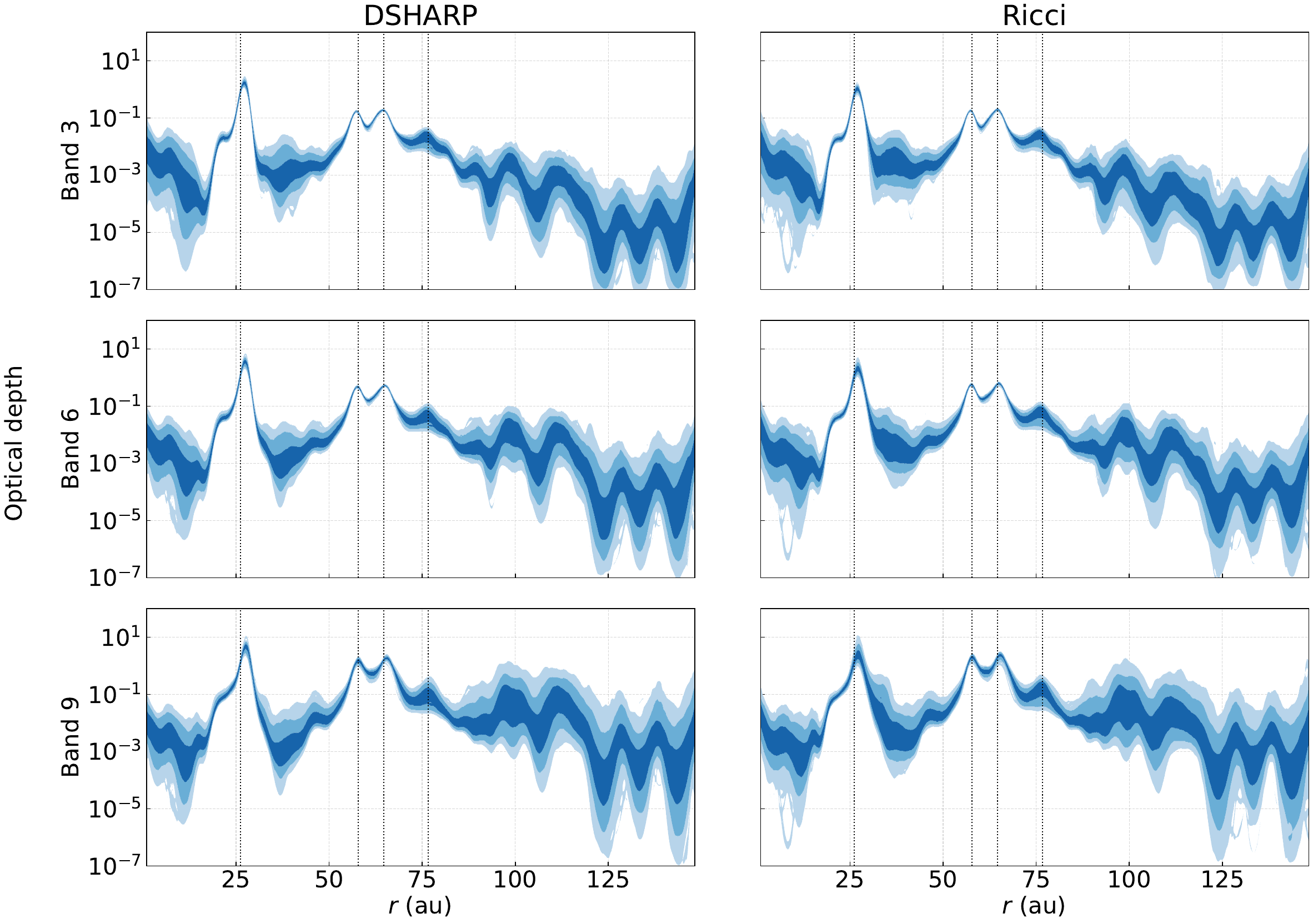}
\caption{Marginal posterior distributions for the optical depths at each band. Shaded regions indicate the $68.3\%$, $95.4\%$, and $99.7\%$ (1, 2, and 3$\sigma$) highest density intervals. Results are compared between the DSHARP (left) and Ricci (right) opacity models. Vertical lines show the ring locations previously specified.  }
\label{fig:tau_hd169142}
\end{figure*}
In Figure \ref{fig:tau_hd169142}, we show the corresponding optical depth distributions at each band.
The peak optical depths at the inner ring is ${\sim}2$, therefore, the intensity {still} has a sensitivity to the optical depth.
The outer rings are essentially optically thin in the all bands.
The both opacity models suggest well-consistent results because the optical depths can be determined from data without assuming an opacity.

\begin{figure}
\centering
\includegraphics[width=1.0\linewidth]{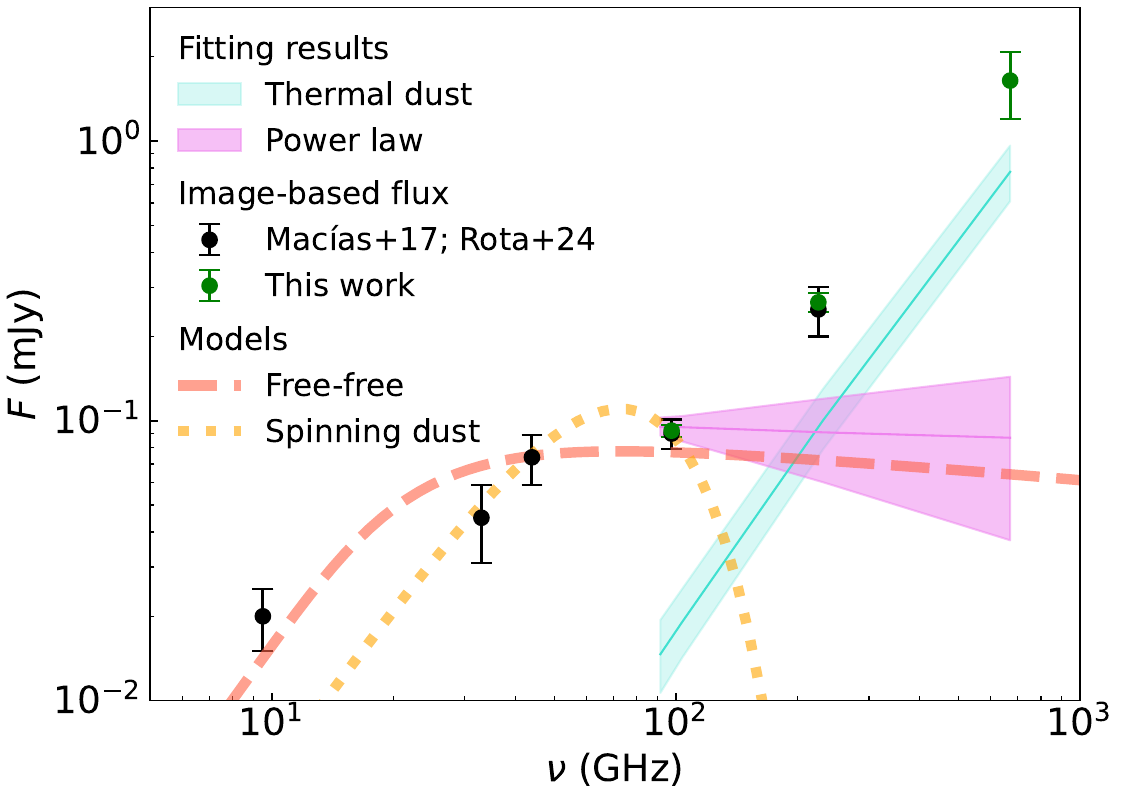}
\caption{{SED} of the innermost region ($r < 45$~mas) and comparison with theoretical models for the Ricci opacity case. Blue and magenta lines represent the retrieved {SED} for the thermal dust and additional power-law components, respectively. Black and green points show image-based flux estimates from \citet{2017ApJ...838...97M, 2024A&A...684A.134R} and this work, respectively. The free-free emission and spinning dust models are indicated by the red dashed and orange dotted lines.}
\label{fig:ps}
\end{figure}

Figure \ref{fig:ps} displays the spectra for the central region ($r < 45$~mas), distinguishing between the retrieved non-thermal-dust point source and the thermal dust emission from the innermost disk, based on the Ricci opacity results. We also overplot previous image-based flux measurements \citep{2017ApJ...838...97M, 2024A&A...684A.134R} and our own aperture photometry. The data strongly suggest that the Band 3 flux originates primarily from the additional non-thermal-dust component rather than thermal dust. This additional component exhibits a substantially flatter spectral slope compared to the thermal dust emission. These spectral characteristics remain consistent regardless of the chosen opacity model.

\begin{figure}
\centering
\includegraphics[width=0.9\linewidth]{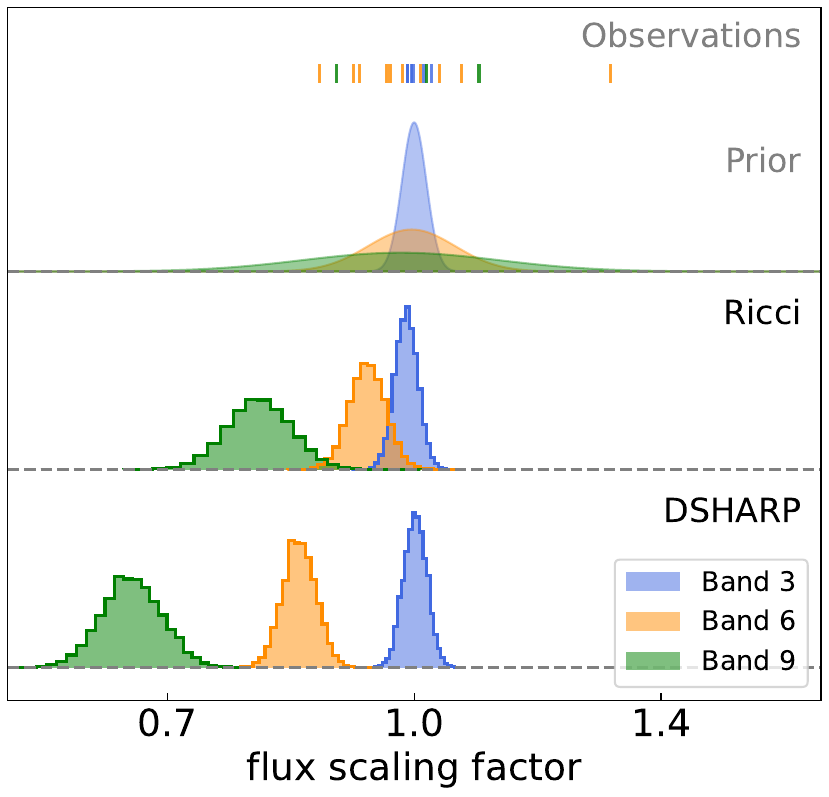}
\caption{Prior and posterior distributions of the flux scaling parameters for each dataset.}
\label{fig:fs}
\end{figure}

Finally, the posterior distributions for the flux scaling parameters are shown in Figure \ref{fig:fs}. While the priors are centered on the observed values, the DSHARP model favors lower scaling factors, whereas the Ricci model remains closer to unity. Although both models provide reasonable fits to the data after adjusting for flux calibration, the closer alignment of the Ricci model with the nominal calibration suggests it may be more consistent with the observations.

\section{Discussion} \label{sec:disc}

\subsection{Comparison of \frappe\ to Other Approaches}\label{sec:comp}

In Section \ref{sec:injrec}, we demonstrated that \frappe\ provides significantly less-biased estimates of dust properties compared to the conventional radius-by-radius fitting approach. This advantage stems primarily from the high radial flexibility of our framework, whereas the conventional approach is fundamentally limited by the spatial resolution of the CLEAN beam. In this sense, \frappe\ achieves ``super-resolution'', thereby maximizing the scientific potential of ALMA datasets.

It is instructive to compare our approach with other existing methods. \texttt{frankenstein} \citep{2020MNRAS.495.3209J} is a pioneering and widely used tool for reconstructing radial intensity profiles using GPs.
While \frappe\ also utilizes a GP-based framework, a key difference exists. 
\texttt{frankenstein} regularizes the solution by assuming that the power of visibilities at non-measured uv-distances is essentially zero, and derives the radial intensity profile.
On the other hand, \frappe\ regularizes the solution by assuming that the physical quantities, such as dust surface densities, are smooth, and directly derives the profiles of the physical quantities.
To do so, \frappe\ is built upon a physical SED model that enables the simultaneous fitting of multiple frequencies.
While our method is sensitive to the accuracy of the assumed physical model, this is a necessary trade-off that allows us to leverage physical constraints to better regularize the problem.

This physically motivated modeling is {one of} the cornerstones of our super-resolution capability, {although the visibility-based analysis itself can already achieve a super-CLEAN beam resolution \citep{2020MNRAS.495.3209J}.}
A conceptual analogy can be found in stellar observations: while a single telescope cannot resolve the surface of a distant star, its radius can be accurately estimated from its SED by applying appropriate stellar models. Similarly, even when rings and gaps in a protoplanetary disk are not spatially resolved, their underlying physical properties can be inferred by combining multi-wavelength interferometric data with radiative transfer models.

Our approach shares some similarities with forward-modeling techniques that assume specific functional forms for the intensity profiles \citep[e.g.,][]{2019ApJ...881..159M}. However, \frappe\ only assumes that the radial variations are relatively smooth and can be described by a GP. This makes our framework significantly more flexible and less prone to the biases inherent in parametric models. Furthermore, the implementation of \frappe\ in \texttt{JAX} allows for auto-differentiation and accelerated sampling, enabling a robust Bayesian posterior estimation rather than a simple point estimate.
{This essentially addresses the overfitting issue that creates artifacts, making the method more reliable.}

\subsection{Limitations of \frappe\ }

As discussed in Section \ref{sec:comp}, the primary limitation of \frappe\ lies in its dependence on model assumptions. In this study, we utilized the radiative transfer model described in Section \ref{sec:rt}. Exploring alternative models, such as those including vertical structure or more complex dust settling, would be beneficial for assessing how results vary under different physical conditions.
It is also important that \frappe\ modeling assumes the disk is essentially axisymmetric.

Additionally, as discussed in Section \ref{sec:lx}, it is notable that the results depend on choise of the {scale length} that essentially determines the flexibility of the model.

Regarding dust opacities, we compared two distinct models in Section \ref{sec:hd169}. Such comparisons are essential to avoid underestimating the total uncertainty in the retrieved parameters. Moreover, our framework can potentially serve as a tool to evaluate the validity of various dust opacity models themselves \citep[e.g.,][]{2025A&A...702A..56Z}.

The computational cost of \frappe\ is relatively modest when the GP hyperparameters are fixed (${\sim}40$ minutes for the demonstration in Section \ref{sec:injrec}). However, the execution time increases significantly if these hyperparameters are treated as free parameters or if the number of radial sampling points is substantially increased.

\subsection{ Physical Implications for the HD 169142 Disk }

\subsubsection{Global Disk Structure}
Previous studies \citep[e.g.,][]{2017A&A...600A..72F, 2019ApJ...881..159M, 2019AJ....158...15P} have established the presence of two primary dust rings in HD 169142, with the outer ring resolved into three tightly packed sub-rings.
Our reconstructed profiles are broadly consistent with these findings while providing several new insights.

The inner ring at $r \sim 26$~au is not perfectly Gaussian-like but exhibits an ``inner shoulder'' in the dust surface density (Figures \ref{fig:hd169_results_frappe} and \ref{fig:sigma_lin}), a feature previously pointed out by \citet{2019AJ....158...15P}. Similar asymmetric ring profiles have been observed in other disks \citep[e.g.,][]{2021ApJ...916L...2B, 2024ApJ...974L..25D} and may be explained by leaky dust traps, where small grains drift inward from the main trap \citep{2024A&A...686A.135P}. The apparent inward shift of the ring peak in Band 9 (Figure \ref{fig:radprofhd169}) likely reflects this size-dependent distribution. Notably, \citet{2025AJ....170..278L} found the ring peak at $\sim 21$~au in infrared scattered light compared to $\sim 27$~au in Band 6, attributing this to a leaky trap. Our Band 9 peak at $r = 24$~au fits well within this sequence, tracing grains smaller than those at Band 6 but larger than those seen in the infrared.

In the outer disk, our results suggest that the relative surface densities of the three sub-rings depend significantly on the choice of dust opacity model. Hydrodynamic simulations by \citet{2019AJ....158...15P} suggest that the relative surface density of these rings may evolve over time, assuming they were formed by an embedded mini-Neptune-sized planet. Future comparisons between our retrieved surface densities and hydrodynamic models will be crucial for constraining these formation mechanisms.

The posterior distributions (Figure \ref{fig:hd169_results_frappe}) indicate that the dust surface densities within the gaps at $\sim 20$~au and $\sim 40$~au are non-negligible. This supports the leaky trap scenario, where small grains cross the traps. While these grains are expected to be significantly smaller than those in the rings, current sensitivity limits preclude a precise measurement of grain sizes within the gaps.

Beyond the rings ($80 < r < 120$~au), we detect a diffuse outer disk with non-negligible surface density, roughly coincident with the $\rm C^{18}O$ emission extent \citep{2022MNRAS.517.5942G}. The median dust size profile shows a sharp transition at $r \sim 80$~au toward smaller grains ($\sim 0.1$--$1$~mm). This extended, small-grain disk is consistent with the analysis of the TW Hya disk \citep{2022MNRAS.515L..23I} and the predictions of \citet{2019MNRAS.486.4829R}, suggesting that previous measurements of continuum disk radii were primarily sensitivity-limited. 

The total dust mass of ${\sim}280$ and ${\sim}45\ M_\oplus$ for the DSHARP and Ricci opacities, respectively, are also broadely consistent with the results by \citet{2019ApJ...881..159M}, $160^{+250}_{-90}\ M_\oplus$, who performed a parametric model fitting.

The temperature profiles reveal a sharp drop inward of the inner ring at 26~au, with a similar, albeit less pronounced, trend at the inner edge of the outer rings. This likely reflects efficient heating at the illuminated inner edges of the rings \citep[e.g.,][]{2013A&A...559A..46B}, consistent with infrared observations suggesting these rings are directly irradiated by the central star \citep{2023MNRAS.522L..51H, 2025AJ....170..278L}.
In the inner side of the exposed cavity wall, the temperature exceeds ${\sim}100$ K, which is consistent with the brightness temperature of the CO line \citep{2022A&A...663A..23L}.
Futhermore, such high temperature may lead sublimation of water ice. Indeed, \citet{2023A&A...678A.146B} detected molecules that should be enhanced at the water snowline, methanol and SO, at this radius.
The correspondency of the water snowline and exposed inner cavity wall has been also found in the HD 100546 disk \citep{2026ApJ...996L..17R}.

Finally, we find that the power-law index of the grain size distribution is remarkably consistent with $q \sim 3.5$, the ISM value \citep{1977ApJ...217..425M}, across the disk. In well-structured disks where grain growth is expected, a value of $q \sim 3.5$ often points to a steady-state collisional cascade, where the size distribution is maintained by the fragmentation of larger bodies \citep{1969JGR....74.2531D, 1996Icar..123..450T, 2011A&A...525A..11B}.

\subsubsection{ Inner most disk }
We retrieved the spectrum of the innermost disk at $r < 10$ au simultaneously with the outer disk (Figure \ref{fig:ps}), where the non-thermal-dust component exhibits a very shallow or even negative spectral index. At wavelengths longer than Band 3, bright emission has been detected in this same region \citep{2017ApJ...838...97M, 2024A&A...684A.134R}.
To explain such long-wavelength emission near a disk center, free-free emission is a well-considered machanism \citep[e.g.,][]{2024A&A...684A.134R}.
Another possibility would be emission from spinning dust grains \citep{2018NewAR..80....1D}.
These spectra often degenerate at frequencies $\nu < 100$ GHz. Indeed, \citet{2018ApJ...862..116H} proposed that a spinning dust spectrum fits the spatially integrated SED of the HD 169142 disk, while \citet{2024A&A...684A.134R} suggested free-free emission based on a tight correlation between the emission and the stellar accretion rate across a large sample including HD 169142.

The degeneracy, however, can be broken by observing the flux at higher frequencies; spinning dust emission exhibits a sharp drop-off above its peak rotation frequency, while free-free emission can continue with a relatively flat spectral slope.

To distinguish between these models, we compare the spectral models with the retrieved spectrum. We include a data point at 671 GHz ($F = 0.087 ^{+0.06}_{-0.05}$ mJy) from our posterior, in addition to the measurements by \citet{2017ApJ...838...97M} and \citet{2024A&A...684A.134R}. We then search for models that reproduce this combined spectrum. For the free-free emission, we adopt the model described by \citet{1967ApJ...147..471M}, \citet{2003ApJ...599.1196K}, and \citet{2024ApJ...972..163L}:
\begin{equation}
    F_{\rm ff} = B(\nu, T)( 1 - e^{-\tau_{\rm ff}} ) \pi r_{\rm ff}^2
\end{equation}
where $r_{\rm ff}$ is the radius of the free-free emission and the optical depth is given by
\begin{equation}
\tau_{\rm ff} = 8.235 \times 10^{-2} \left( \frac{T_{\rm ff}}{\rm K} \right)^{-1.35} \left(\frac{\nu}{\rm GHz}\right)^{-2.1} \left( \frac{\rm EM}{\rm pc\ cm^{-6}} \right),
\end{equation}
with $T_{\rm ff}$ and $\rm EM$ representing the temperature and emission measure, respectively. Due to the strong degeneracy between parameters, we fixed $r_{\rm ff} = 1.0$ au. For the spinning dust emission model, we employ the equation from \citet{2018ApJ...862..116H}:
\begin{equation}
    F_{\rm sd} = F_{\rm sd,0} \left(\frac{\nu}{\nu_{\rm pk}} \right)^2 \exp \left\{ 1 - \left(\frac{\nu}{\nu_{\rm pk}}\right)^2 \right \},
\end{equation}
where $F_{\rm sd,0}$ is the peak flux density and $\nu_{\rm pk}$ is the peak frequency.

We compared these models via $\chi^2$ analysis against the retrieved spectrum, sampling the posterior distribution under broad priors using MCMC. In Figure \ref{fig:ps}, we plot the resulting best-fit (median) models. Under the free-free emission hypothesis, a combination of $T_{\rm ff} \sim 10^{3}$ K and ${\rm EM} \sim 10^8\ {\rm pc\ cm^{-6}}$ well reproduces the spectrum. Conversely, while the best-fit spinning dust model ($F_{\rm sd, 0} \sim 0.1$ mJy and $\nu_{\rm pk} \sim 72$ GHz) is broadly consistent with the peak values reported by \citet{2018ApJ...862..116H} ($F_{\rm sd, 0} \sim 0.6$ mJy, $\nu_{\rm pk} = 60$ GHz), it fails to reproduce the flat spectral slope at frequencies higher than the peak. In contrast, the free-free emission model remains consistent with the data across the entire range. Overall, we conclude that free-free emission is the more likely mechanism for the innermost region of the HD 169142 disk.

\section{Summary} \label{sec:sum}

In this paper, we have presented \frappe, a new open-source framework designed to retrieve the physical properties of dust grains in protoplanetary disks directly from multi-wavelength interferometric continuum observations. We validated the method using mock datasets and demonstrated its scientific capabilities by applying it to ALMA observations of the HD 169142 disk. The key features and results of \frappe\ are summarized as follows:

\begin{itemize}
 \item Flexible physics-based modeling: The radial profiles of dust properties, including surface density, temperature, and grain size distribution, are modeled as sample paths of Gaussian processes. This approach enables a highly flexible representation of disk structures while ensuring the intensity profiles are calculated through radiative transfer.
 
 \item Direct visibility fitting: One-dimensional interferometric visibilities are derived from the intensity profiles via Hankel transforms. By comparing these models directly with observed visibilities across multiple frequencies, \frappe\ avoids the biases inherent in image-domain analysis. {This also enables simultaneous model fitting to observations with different angular resolutions.}
 
 \item Differentiable programming: The framework is implemented using \texttt{JAX}, enabling end-to-end automatic differentiation. This significantly accelerates the sampling of posterior distributions, making robust Bayesian inference feasible for complex, multi-parameter disk models.
 
 \item Super-resolution and bias reduction: Through injection-recovery tests, we demonstrated that \frappe\ provides significantly less biased estimates of dust properties than conventional radius-by-radius fitting. Notably, the method can resolve structures finer than the CLEAN beam, effectively achieving ``super-resolution.''
 
 \item Application to HD 169142: Applying \frappe\ to real ALMA data, we successfully retrieved the radial profiles of dust surface density, temperature, and grain size distribution index. Our results identified multiple rings and gaps, provided evidence for leaky dust traps, and distinguished between thermal dust and free-free emission in the innermost disk. Additionally, we found that the temperature profile has a steep gradient at the inner edge of the inner ring. This is likely due to irradiation from the central star. The temperature at the inner side exceeds ${\sim}100$ K and the inner edge may correspond to the water snowline.
\end{itemize}

As ALMA and future facilities continue to provide high-resolution, multi-band, and high-sensitivity datasets, such physics-based retrieval tools will become increasingly essential for deciphering the conditions of planet formation. \frappe\ is publicly available and being actively developped at \url{https://github.com/tomyoshida/frap}, with documentation at \url{https://frap.readthedocs.io/}. We welcome contributions from the community to further enhance the capabilities of this tool.
{The code version used for this paper is stored in a Zenodo repository \citep{zenodo_version}.}

\begin{acknowledgments}
{We would like to thank the anonymous referee for helpful comments and suggestions.}
This paper makes use of the following ALMA data: ADS/JAO.ALMA\#2018.1.01716.S, \#2013.1.00592.S, \#2015.1.00490.S, \#2015.1.01301.S, \#2016.1.00344.S, \#2017.1.00727.S, and \#2024.1.01254.S.
ALMA is a partnership of ESO (representing its member states), NSF (USA), and NINS (Japan), together with NRC (Canada), NSTC and ASIAA (Taiwan), and KASI (Republic of Korea), in cooperation with the Republic of Chile.
The Joint ALMA Observatory is operated by ESO, AUI/NRAO, and NAOJ.
T.C.Y. was supported by the ALMA Japan Research Grant of NAOJ ALMA Project, NAOJ-ALMA-385.
This work was supported by Grant-in-Aid for JSPS Fellows, JP23KJ1008 (T.C.Y.).
{T.C.Y. and G.R. acknowledge support from the European Union (ERC Starting Grant DiscEvol, project number 101039651) and from Fondazione Cariplo, grant No. 2022-1217. Views and opinions expressed are, however, those of the author(s) only and do not necessarily reflect those of the European Union or the European Research Council. Neither the European Union nor the granting authority can be held responsible for them.}
\end{acknowledgments}

%
\facilities{ALMA}

\software{JAX \citep{jax2018github}, numpyro \citep{bingham2019pyro, 2019arXiv191211554P}, astropy\citep{2013A&A...558A..33A, 2018AJ....156..123A, 2022ApJ...935..167A},  CASA\citep{2022PASP..134k4501C}
}

\appendix 

{\section{Visibilities of axisymmetric source with primary beam correction}\label{app:pb}} 
In previous studies such as {\tt frankenstein} \citep{2020MNRAS.495.3209J}, 
the antenna primary beam was not taken into account during modeling, 
leaving the retrieved intensity profiles affected by primary beam attenuation. 
This has not been a significant issue because the primary beam is sufficiently 
larger than the typical disk radius in most cases. 
Additionally, since these studies focused on the intensity profile at a single 
frequency, which is proportional to the primary beam, the resultant profiles 
could simply be understood as the product of the actual profile and the primary beam.

However, in {\tt FRAP}, we should explicitly take this effect into account because the retrieved quantities are now physical parameters, which is no longer proportional to the primary beam.
The measurement equation of interferometry with the primary beam correction is
\begin{equation}
\mathcal{V}(u, v) = \int_{-\infty}^{\infty} dl\int_{-\infty}^{\infty} dm\ A(l, m) I(l, m) e^{-2\pi i (ul + vm)} ,
\end{equation}
where $\mathcal{V}(u, v)$ is the visibility at spatial frequencies $(u, v)$, $A(l, m)$ is the primary beam at a sky coordinate $(l, m)$, and $I$ is the intensity distribution on the plane of the sky.
In our problem setting, we consider a geometrically thin, axisynmetric disk with a position angle $\theta_{\rm PA}$ and inclination angle $\theta_{\rm incl}$.
We also assume that the disk center is located at the phase center.
Here, we define spatial coordinates on the disk plane $(l', m')$ and coresponding coordinates in the fourier domain $(u', v')$.
Those coordinates are related to the original ones with
\begin{equation}
\begin{pmatrix}
l' \\
m'
\end{pmatrix}
=
\begin{pmatrix}
1/ \cos \theta_{\rm incl} & 0 \\
0 & 1
\end{pmatrix}
\begin{pmatrix}
\cos\theta_{\rm PA} & -\sin\theta_{\rm PA} \\
\sin\theta_{\rm PA} & \cos\theta_{\rm PA}
\end{pmatrix}
\begin{pmatrix}
l \\
m
\end{pmatrix},
\end{equation}
and
\begin{equation}
\begin{pmatrix}
u' \\
v'
\end{pmatrix}
=
\begin{pmatrix}
\cos \theta_{\rm incl} & 0 \\
0 & 1
\end{pmatrix}
\begin{pmatrix}
\cos\theta_{\rm PA} & -\sin\theta_{\rm PA} \\
\sin\theta_{\rm PA} & \cos\theta_{\rm PA}
\end{pmatrix}
\begin{pmatrix}
u \\
v
\end{pmatrix}.
\end{equation}
The deprojected equation can be expressed as
\begin{equation} \label{eq:V2}
\mathcal{V'}(u', v') = \cos\theta_{\rm incl} \int_{-\infty}^{\infty} dl' \int_{-\infty}^{\infty} dm' A'(l', m') I'(l', m') e^{-2\pi i (u'l' + v'm')} ,
\end{equation}
where $A'$ and $I'$ are the primary beam and intensity distribution in the deprojected coordinate.
We assume that those values are the same as that on the original coordinate \citep[e.g.,][]{2024MNRAS.532.1361A};
\begin{eqnarray}
A'(l', m') &=& A(l, m) \\
I'(l', m') &=& I(l, m)
\end{eqnarray}
We can convert Equation \ref{eq:V2} to the polar coordinate;
\begin{equation} \label{eq:Vp}
\mathcal{V''}(q, \varphi) = |\cos\theta_{\rm incl} | \int_{-\pi}^{\pi} d \phi \int_{0}^{\infty} rdr A''(r, \phi) I''(r) e^{-2\pi i q r \cos(\phi - \varphi)},
\end{equation}
where $q = \sqrt{u'^2 + v'^2}$, $\varphi = \arctan{(v'/u')}$, $r = \sqrt{l'^2 + m'^2}$, and $\phi = \arctan{(l'/m')}$.
$A''$ and $I''$ are the primary beam and intensity distribution on the polar coordinate.
Since we consider axisymmetric disk, $I''$ depends only on $r$.
However, the ``deprojected'' primary beam, which was initially circular, is now instead elliptical and has a dependency on $\phi$.

Here, we introduce Fourier series expansion of $A''$ in the $\phi$ direction;
\begin{equation} \label{eq:A''}
    A''(r, \phi) = \sum_{m=-\infty}^{\infty} A''_m(r) e^{im\phi},
\end{equation}
where $m$ indicates a mode, and $A''_m(r)$ is the fourier coefficient,
\begin{equation} \label{eq:pb}
A''_m(r) = \frac{1}{2\pi} \int_0^{2\pi}d\phi\ A''(r, \phi) e^{-im\phi} .
\end{equation}

By substituting Equation \ref{eq:A''} to \ref{eq:Vp}, we obtain
\begin{equation} \label{eq:Vp2}
\mathcal{V''}(q, \varphi) = |\cos\theta_{\rm incl}|  \int_{0}^{\infty} dr\ I''(r) r \sum_{m=-\infty}^{\infty}A''_m(r)  \int_{-\pi}^{\pi} d \phi\    e^{i \{ m\phi -2\pi q r \cos(\phi - \varphi) \}}.
\end{equation}
Using the $m$-th order Bessel function of the first kind, $J_m$, the $\phi-$integral can be further reduced to be
\begin{equation}
\int_{-\pi}^{\pi} d \phi\  e^{i \{ m\phi -2\pi q r \cos(\phi - \varphi) \}} = 2 \pi (-i)^m e^{im\varphi} J_m(2 \pi q r),
\end{equation}
therefore,
\begin{equation} \label{eq:V_bessel}
\mathcal{V''}(q, \varphi) = 2 \pi |\cos\theta_{\rm incl}| \int_{0}^{\infty}  dr\  I''(r) r \sum_{m=-\infty}^{\infty}A''_m(r)   (-i)^m e^{im\varphi} J_m(2 \pi q r).
\end{equation}

In {\tt FRAP}, we aim to fit the azimuthally averaged deprojected visibility profiles.
Therefore, the quantity we compare is
\begin{eqnarray}
\mathcal{V_{\rm mod}} &=& \langle \mathcal{V''}(q, \varphi) \rangle_\phi \\
&=& \frac{1}{2\pi} \int_{-\pi}^{\pi} d\varphi\  2 \pi |\cos\theta_{\rm incl}|  \int_{0}^{\infty}  dr\  I''(r) r\sum_{m=-\infty}^{\infty}A''_m(r)   (-i)^m e^{im\varphi} J_m(2 \pi q r) \\
&=& |\cos\theta_{\rm incl}| \int_{0}^{\infty} dr\ I''(r) r A''_0(r) J_0(2 \pi q r),
\end{eqnarray}
because $e^{im\varphi}$ vanishes for $m\neq0$ after integration by $\varphi$.
Here, $A''_0(r)$ is the averaged primary beam at each disk radius and can be calcurated by Equation \ref{eq:pb}.

These results are consistent with the conventional Hankel transform expression of visibilities \citep[e.g.,][]{2000fta..book.....B,  2020MNRAS.495.3209J, 2024MNRAS.532.1361A}, except for the primary beam correction.
The equation becomes identical to the previous one if we take $A''_0(r) = 1$.
By replacing the intensity profile in Equation (3) of \citet{2020MNRAS.495.3209J} by $A''_0(r)I''(r)$, we obtain Equation \ref{eq:hankel_main}. Note that we use $A_{\rm eff}(\nu, r)$ instead of $A''_0(r)$ in the main text for a consistent notation.

\section{Intensities and Optical Depths of the Mock Data} \label{app:tau}

\begin{figure}
\centering
\includegraphics[width=0.4\linewidth]{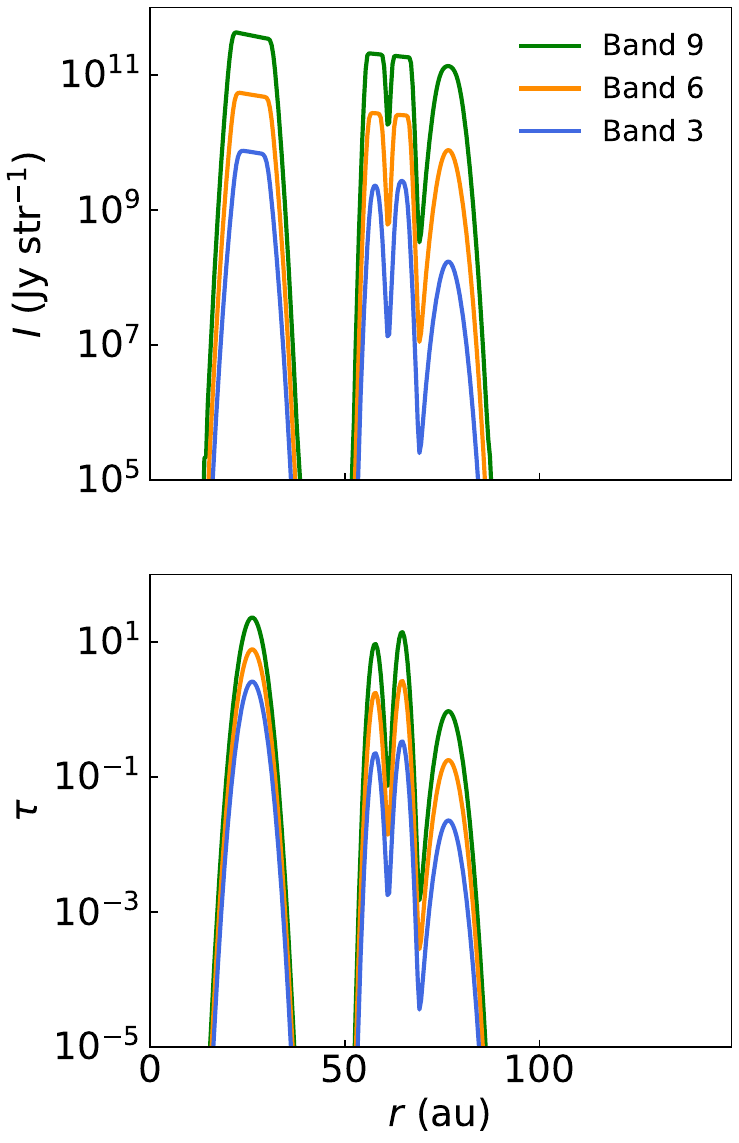}
\caption{ Input radial profiles of the intensities and optical depths for the mock data. }
\label{fig:tau}
\end{figure}

Figure \ref{fig:tau} presents the radial profiles of the intensities and optical depths for each band in the mock datasets analyzed in Section \ref{sec:injrec}. The inner ring is marginally optically thick even in Band 3. In contrast, the outer rings remain optically thin, particularly in Bands 3, but become moderately optically thick in other bands.

\section{{Estimated dust mass in the injection recovery test}} \label{app:mass}

\begin{figure}
\centering
\includegraphics[width=0.4\linewidth]{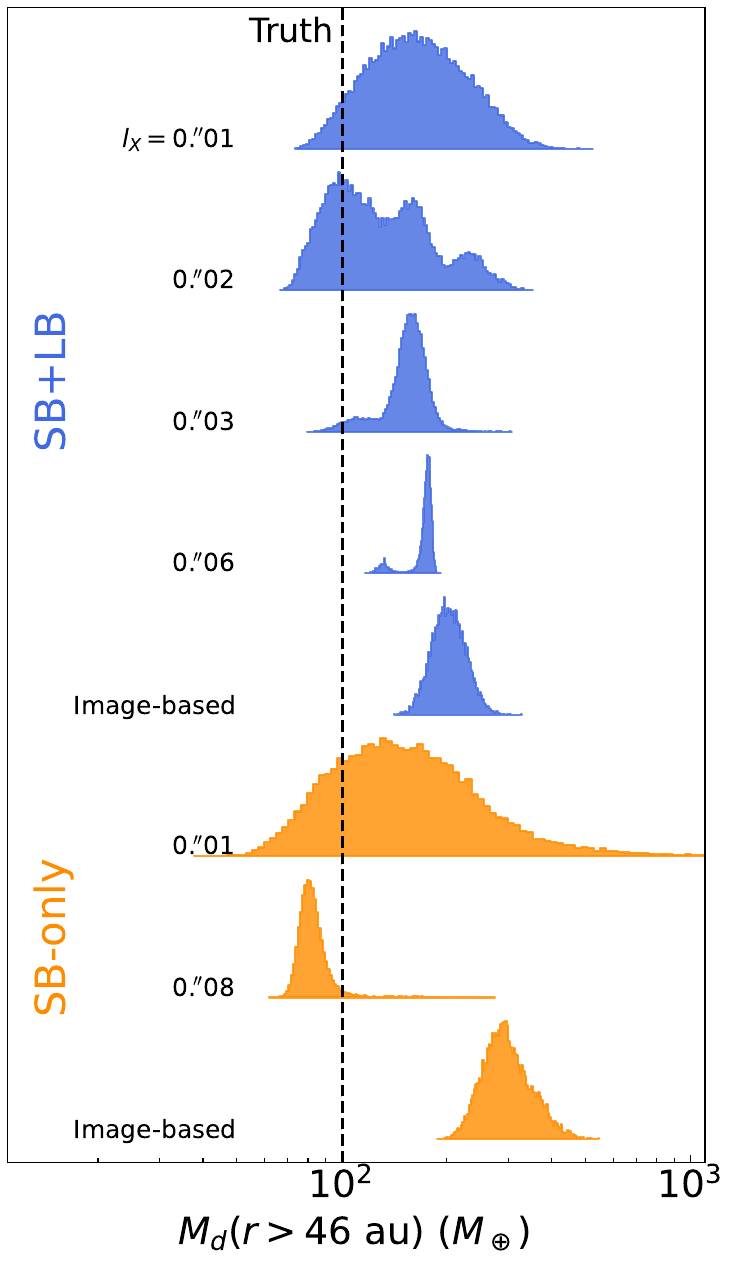}
\caption{ Posterior distributions of the dust mass at $r>46$ au for various $l_X$ and the image-based conventional method. The black vertical line indicates the input truth. }
\label{fig:mass}
\end{figure}
Figure \ref{fig:mass} presents the posterior distribution of the dust mass at $r>46$ au for various $l_X$ in the injection recovery tests (Section \ref{sec:sampling_mock}).
We also plot the results from the image-based conventional method (Section \ref{sec:conv}) for comparison.
The estimated masses are reasonably consistent with the input value if $l_X < 0\farcs03$, while the larger $l_X$ and conventional methods significantly overestmate it.
Note that the $l_X = 0\farcs03$ (SB+LB) case has a prominent peak of the distribution at a higher mass than the truth, however, this is primarily due to the marginalization effect of the Bayesian analysis and the true value is still covered with in the distribution.

\section{{Injection recovery test with \lowercase{$q$} as a free parameter}}
\label{app:qfree}

In Section \ref{sec:sampling_mock}, we demonstrate \frappe\ with the power law index of the grain size distribution, $q$, being a fixed paremeter for simpler interpretation of the results.
We also perform the same posterior sampling but make $q$ free.
Here, the prior range of $q$ is set to be $[1, 5]$. The other configurations are the same as the $l_X = 0\farcs03$ case in Section \ref{sec:sampling_mock}.
\begin{figure}
\centering
\includegraphics[width=0.5\linewidth]{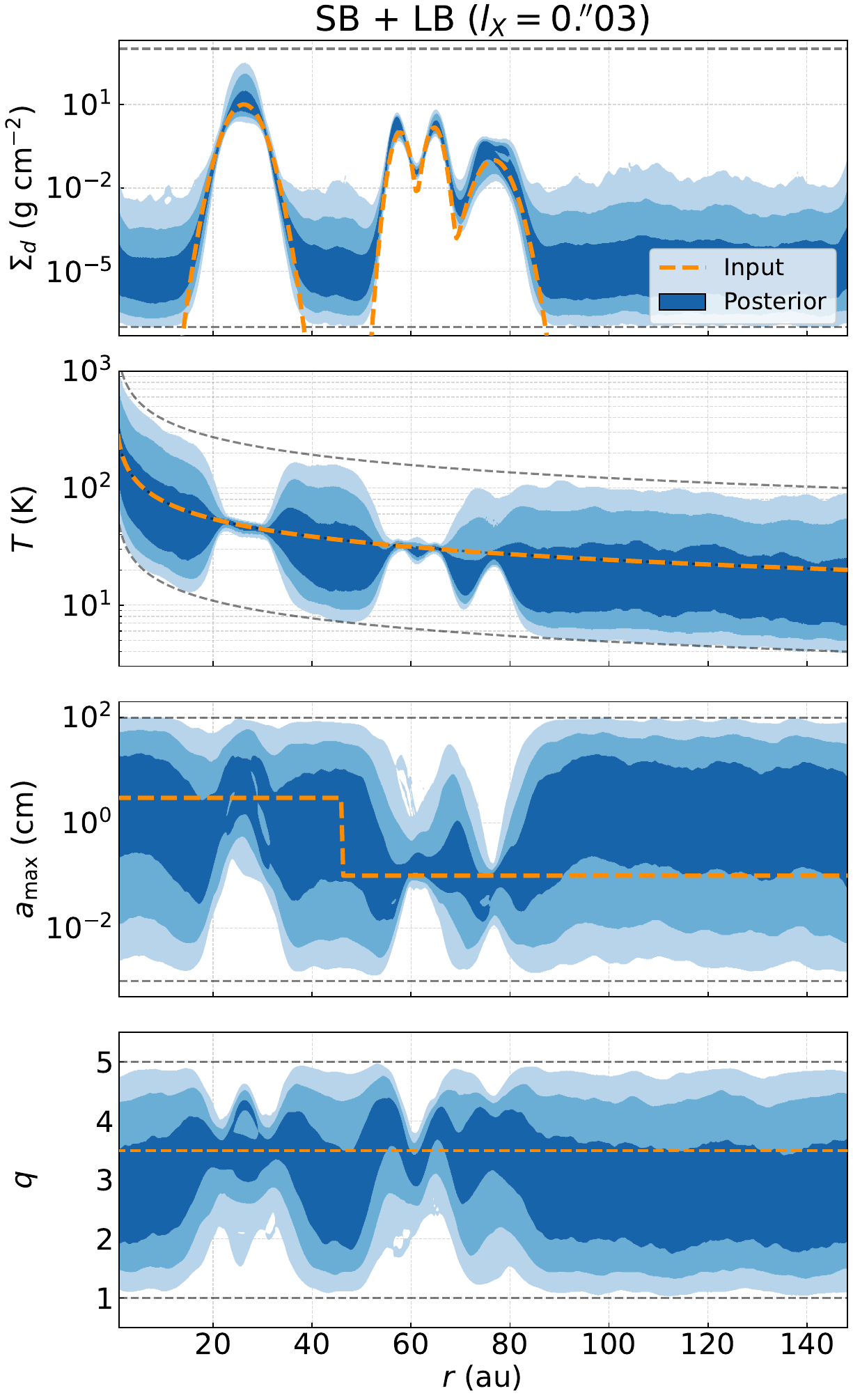}
\caption{ Same as the left panel of Figure \ref{fig:mock_results_frappe}, but with $q$ added as a free parameter. }
\label{fig:qfree}
\end{figure}
Figure \ref{fig:qfree} shows the posterior distribution.
The uncertainties become larger than the fixed $q$ case in Figure \ref{fig:mock_results_frappe}, however, the posteriors are still consistent with the input profiles.

\section{{Sensitivity on the length scale parameter}} \label{app:diff_lX}
\begin{figure}
\centering
\includegraphics[width=0.5\linewidth]{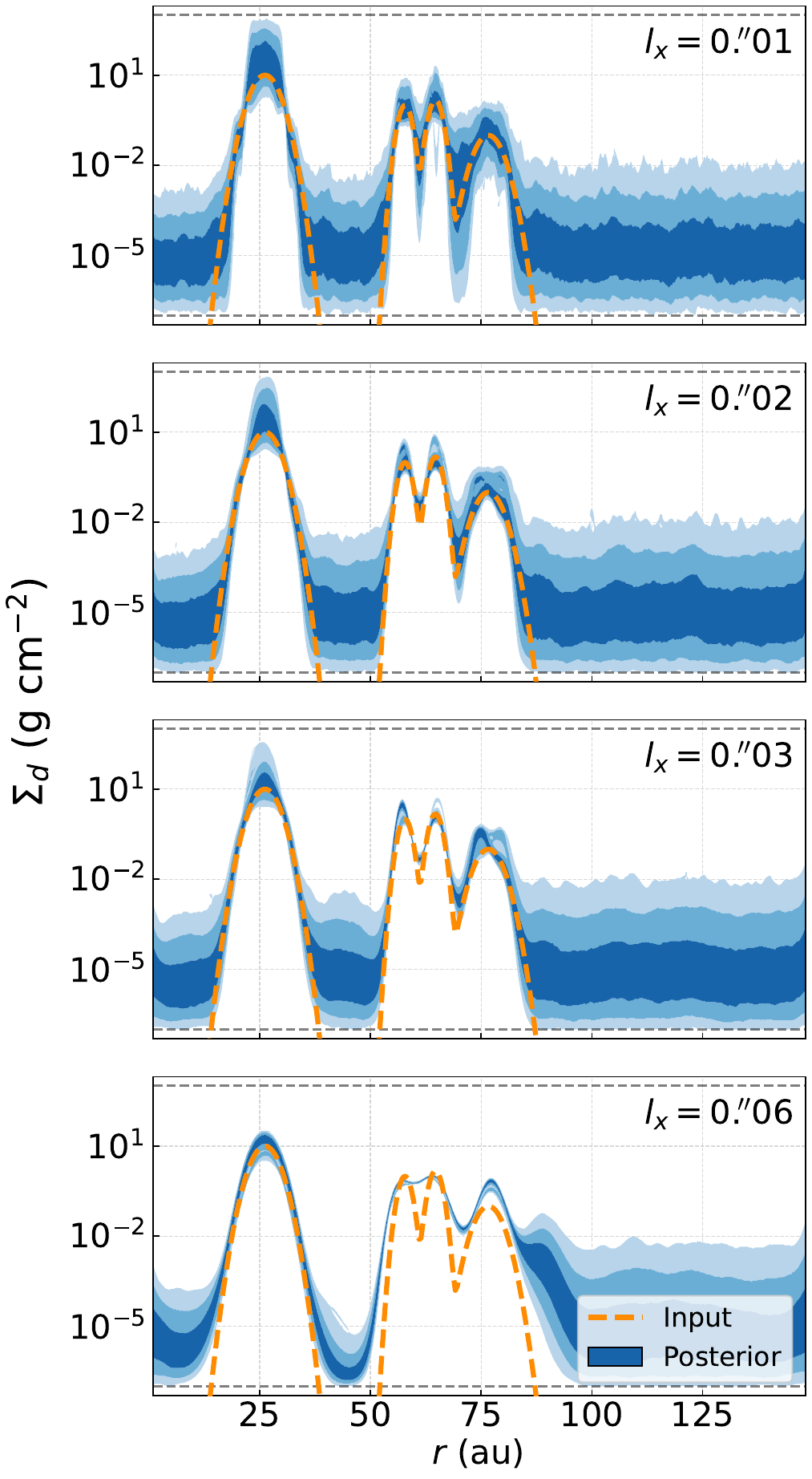}
\caption{ Comparison of the posterior distribution of $\Sigma_d$ for various $l_X$. }
\label{fig:sigma_g_diff_lX}
\end{figure}
In Section \ref{sec:lx}, we mentioned that a smaller length scale parameter allows higher flexibility, resulting in less biased but larger uncertainty.
To show this behaivior more clearly, we additionally ran posterior sampling with $l_X = 0\farcs02, 0\farcs06, $ and $0\farcs08$.
In Figure \ref{fig:sigma_g_diff_lX}, a comparison of the resultant dust surface density profiles, including the cases with $l_X=0\farcs01$ and $0\farcs03$, is presented.
This shows the sensitivity of the results on the length scale parameter.
As long as the prior can express the physical profiles (i.e., sufficiently small $l_X$), the results are not biased.

\section{Observational details} \label{app:obsdet}

\begin{deluxetable*}{cccccc}
\tablecaption{Summary of ALMA Observations. \label{tab:obs}}
\tablewidth{0pt}
\tablehead{
\colhead{\makecell{Project \\ code}} & \colhead{P.I.} & \colhead{Date} & \colhead{\makecell{On-source time \\ (min)}} & \colhead{\makecell{Baselines \\ (m)}} & \colhead{ \makecell{Frequencies\tablenotemark{a}\\ (GHz)}}
}
\startdata
\hline
\multicolumn{6}{c}{Band 3} \\
\hline
2018.1.01716.S & E. Mac\'{i}as & 2019 Jun 21 & 41.5 & 80 -- 14373 & 90.5, 92.4, 102.5, 104.5 \\
~ & ~ & 2019 Jun 25 & 41.4 & 82 -- 16196 & 90.5, 92.4, 102.5, 104.5 \\
~ & ~ & 2019 Jun 25 & 41.6 & 250 -- 16185 & 90.5, 92.4, 102.5, 104.5 \\
~ & ~ & 2019 Jun 25 & 41.4 & 81 -- 15306 & 90.5, 92.4, 102.5, 104.5 \\
~ & ~ & 2019 Sep 01 & 21.3 & 31 -- 3562 & 90.5, 92.4, 102.5, 104.5 \\
~ & ~ & 2019 Sep 03 & 21.3 & 38 -- 3617 & 90.5, 92.4, 102.5, 104.5 \\
\hline
\multicolumn{6}{c}{Band 6} \\
\hline
2013.1.00592.S & D. Fedele & 2015 Aug 30 & 4.7 & 13 -- 1447 & \makecell[c]{216.1, 219.5, 220.4, 230.5, 233.0} \\
2015.1.00490.S & M. Honda & 2016 Sep 14 & 3.6 & 14 -- 3195 & 219.5, 220.4, 230.5, 233.0 \\
~ & ~ & 2016 Sep 14 & 3.6 & 13 -- 2962 & 219.5, 220.4, 230.5, 233.0 \\
~ & ~ & 2016 Sep 14 & 3.6 & 15 -- 2648 & 219.5, 220.4, 230.5, 233.0 \\
2015.1.01301.S & J. Hashimoto & 2016 Sep 17 & 3.1 & 14 -- 2293 & 219.5, 220.4, 230.5, 232.5 \\
2016.1.00344.S & S. P\'{e}rez & 2016 Oct 04 & 4.5 & 18 -- 2675 & \makecell[c]{218.0, 219.5, 220.4, 230.5, 232.0} \\
~ & ~ & 2017 Jul 05 & 4.5 & 17 -- 2626 & \makecell[c]{218.0, 219.6, 220.4, 230.5, 232.0} \\
~ & ~ & 2017 Sep 18 & 19.1 & 41 -- 11877 & \makecell[c]{218.0, 219.5, 220.4, 230.5, 232.0} \\
~ & ~ & 2017 Sep 19 & 19.1 & 41 -- 11471 & \makecell[c]{218.0, 219.5, 220.4, 230.5, 232.0} \\
~ & ~ & 2017 Nov 09 & 19.2 & 101 -- 13207 & \makecell[c]{218.0, 219.5, 220.4, 230.5, 232.0} \\
\hline
\multicolumn{6}{c}{Band 9} \\
\hline
2017.1.00727.S & J. Szul\'{a}gyi & 2019 Aug 14 & 15.4 & 41 -- 3111 & \makecell[c]{660.9, 661.0, 664.0, 664.2, 677.0 \\ 677.2, 680.1, 680.2} \\
~ & ~ & 2019 Aug 14 & 13.4 & 39 -- 3013 & \makecell[c]{660.9, 661.0, 664.0, 664.2, 677.0 \\ 677.2, 680.1, 680.2} \\
2024.1.01254.S & T. Yoshida & 2024 Oct 11 & 6.5 & 14 -- 475 & \makecell[c]{660.0, 662.0, 664.0, 666.0, 676.0 \\ 678.0, 680.0, 682.0} \\
\enddata
\tablenotetext{a}{Mean frequency of spectral windows.}
\tablecomments{This table is generated by {\tt tabgenalma} (\url{https://github.com/tomyoshida/tabgenalma}).}
\end{deluxetable*}
Table \ref{tab:obs} summarizes the details of the observations for the HD 169142 disk.



\bibliography{reference_ads, miscellaneous}{}

@article{bingham2019pyro,
  author    = {Eli Bingham and
               Jonathan P. Chen and
               Martin Jankowiak and
               Fritz Obermeyer and
               Neeraj Pradhan and
               Theofanis Karaletsos and
               Rohit Singh and
               Paul A. Szerlip and
               Paul Horsfall and
               Noah D. Goodman},
  title     = {Pyro: Deep Universal Probabilistic Programming},
  journal   = {J. Mach. Learn. Res.},
  volume    = {20},
  pages     = {28:1--28:6},
  year      = {2019},
  url       = {http://jmlr.org/papers/v20/18-403.html}
}

@BOOK{2000fta..book.....B,
       author = {{Bracewell}, Ronald N.},
        title = "{The Fourier transform and its applications}",
         year = 2000,
       adsurl = {https://ui.adsabs.harvard.edu/abs/2000fta..book.....B}
}

@software{zenodo_version,
  author       = {{Yoshida}, Tomohiro C. and {Mac{\'\i}as}, Enrique and {Rosotti}, Giovanni and {Facchini}, Stefano and {Viscardi}, Elena M. and {Doi}, Kiyoaki },
  title = "{Flexible Radial Analysis of Protoplanetary disks: FRAP}",
  year         = 2026,
  publisher    = {Zenodo},
  version      = {v1.0.0},
  doi          = {10.5281/zenodo.20689546},
  url          = {https://doi.org/10.5281/zenodo.20689546}
}

@software{jax2018github,
  author = {James Bradbury and Roy Frostig and Peter Hawkins and Matthew James Johnson and Chris Leary and Dougal Maclaurin and George Necula and Adam Paszke and Jake Vander{P}las and Skye Wanderman-{M}ilne and Qiao Zhang},
  title = {{JAX}: composable transformations of {P}ython+{N}um{P}y programs},
  url = {http://github.com/jax-ml/jax},
  version = {0.3.13},
  year = {2018},
}

@ARTICLE{Hoffman2014-hj,
  title   = "The no-u-turn sampler: adaptively setting path lengths in
             Hamiltonian Monte Carlo",
  author  = "Hoffman, Matthew D and Gelman, Andrew",
  journal = "Journal of Machine Learning Research",
  year    =  2014
}

@manual{cortes_2025_14933753,
  title        = {ALMA Cycle 12 Technical Handbook},
  author       = {Cortes, Paulo and
                  Carpenter, John and
                  Kameno, Seiji and
                  Loomis, Ryan and
                  Vila Vilaro, Baltasar and
                  Immer, Katharina and
                  Vlahakis, Catherine and
                  Law, James and
                  Stoehr, Felix and
                  Saini, Kamaljeet and
                  Hales, Antonio and
                  Kneissl, Ruediger},
  month        = mar,
  year         = 2025,
  doi          = {10.5281/zenodo.14933753},
  url          = {https://doi.org/10.5281/zenodo.14933753},
}

@BOOK{2017isra.book.....T,
       author = {{Thompson}, A. Richard and {Moran}, James M. and {Swenson}, Jr., George W.},
        title = "{Interferometry and Synthesis in Radio Astronomy, 3rd Edition}",
         year = 2017,
          doi = {10.1007/978-3-319-44431-4},
       adsurl = {https://ui.adsabs.harvard.edu/abs/2017isra.book.....T}
}

@ARTICLE{2023A&A...678A.146B,
       author = {{Booth}, Alice S. and {Law}, Charles J. and {Temmink}, Milou and {Leemker}, Margot and {Mac{\'\i}as}, Enrique},
        title = "{Tracing snowlines and C/O ratio in a planet-hosting disk. ALMA molecular line observations towards the HD 169142 disk}",
      journal = {\aap},
         year = 2023,
        month = oct,
       volume = {678},
          eid = {A146},
        pages = {A146},
          doi = {10.1051/0004-6361/202346974},
archivePrefix = {arXiv},
       eprint = {2308.07910},
 primaryClass = {astro-ph.EP},
       adsurl = {https://ui.adsabs.harvard.edu/abs/2023A&A...678A.146B}
}

@ARTICLE{2024A&A...688A.204L,
       author = {{Li}, Dafa and {Liu}, Yao and {Wang}, Hongchi and {Fang}, Min and {Wang}, Lei},
        title = "{Uncertainties of the dust grain size in protoplanetary disks retrieved from millimeter continuum observations}",
      journal = {\aap},
         year = 2024,
        month = aug,
       volume = {688},
          eid = {A204},
        pages = {A204},
          doi = {10.1051/0004-6361/202449253},
archivePrefix = {arXiv},
       eprint = {2410.04079},
 primaryClass = {astro-ph.EP},
       adsurl = {https://ui.adsabs.harvard.edu/abs/2024A&A...688A.204L}
}

@ARTICLE{2003ApJ...599.1196K,
       author = {{Keto}, Eric},
        title = "{The Formation of Massive Stars by Accretion through Trapped Hypercompact H II Regions}",
      journal = {\apj},
         year = 2003,
        month = dec,
       volume = {599},
       number = {2},
        pages = {1196-1206},
          doi = {10.1086/379545},
archivePrefix = {arXiv},
       eprint = {astro-ph/0309131},
 primaryClass = {astro-ph},
       adsurl = {https://ui.adsabs.harvard.edu/abs/2003ApJ...599.1196K}
}

@ARTICLE{1967ApJ...147..471M,
       author = {{Mezger}, P.~G. and {Henderson}, A.~P.},
        title = "{Galactic H II Regions. I. Observations of Their Continuum Radiation at the Frequency 5 GHz}",
      journal = {\apj},
         year = 1967,
        month = feb,
       volume = {147},
        pages = {471},
          doi = {10.1086/149030},
       adsurl = {https://ui.adsabs.harvard.edu/abs/1967ApJ...147..471M}
}

@ARTICLE{2018ApJ...862..116H,
       author = {{Hoang}, Thiem and {Lan}, Nguyen-Quynh and {Vinh}, Nguyen-Anh and {Kim}, Yun-Jeong},
        title = "{Spinning Dust Emission from Circumstellar Disks and Its Role In Excess Microwave Emission}",
      journal = {\apj},
         year = 2018,
        month = aug,
       volume = {862},
       number = {2},
          eid = {116},
        pages = {116},
          doi = {10.3847/1538-4357/aaccf0},
archivePrefix = {arXiv},
       eprint = {1803.11028},
 primaryClass = {astro-ph.GA},
       adsurl = {https://ui.adsabs.harvard.edu/abs/2018ApJ...862..116H}
}

@ARTICLE{1969JGR....74.2531D,
       author = {{Dohnanyi}, J.~S.},
        title = "{Collisional Model of Asteroids and Their Debris}",
      journal = {\jgr},
         year = 1969,
        month = may,
       volume = {74},
        pages = {2531-2554},
          doi = {10.1029/JB074i010p02531},
       adsurl = {https://ui.adsabs.harvard.edu/abs/1969JGR....74.2531D}
}

@ARTICLE{1996Icar..123..450T,
       author = {{Tanaka}, Hidekazu and {Inaba}, Satoshi and {Nakazawa}, Kiyoshi},
        title = "{Steady-State Size Distribution for the Self-Similar Collision Cascade}",
      journal = {\icarus},
         year = 1996,
        month = oct,
       volume = {123},
       number = {2},
        pages = {450-455},
          doi = {10.1006/icar.1996.0170},
       adsurl = {https://ui.adsabs.harvard.edu/abs/1996Icar..123..450T}
}

@ARTICLE{1977ApJ...217..425M,
       author = {{Mathis}, J.~S. and {Rumpl}, W. and {Nordsieck}, K.~H.},
        title = "{The size distribution of interstellar grains.}",
      journal = {\apj},
         year = 1977,
        month = oct,
       volume = {217},
        pages = {425-433},
          doi = {10.1086/155591},
       adsurl = {https://ui.adsabs.harvard.edu/abs/1977ApJ...217..425M}
}

@ARTICLE{2019MNRAS.486.4829R,
       author = {{Rosotti}, Giovanni P. and {Tazzari}, Marco and {Booth}, Richard A. and {Testi}, Leonardo and {Lodato}, Giuseppe and {Clarke}, Cathie},
        title = "{The time evolution of dusty protoplanetary disc radii: observed and physical radii differ}",
      journal = {\mnras},
         year = 2019,
        month = jul,
       volume = {486},
       number = {4},
        pages = {4829-4844},
          doi = {10.1093/mnras/stz1190},
archivePrefix = {arXiv},
       eprint = {1905.00019},
 primaryClass = {astro-ph.EP},
       adsurl = {https://ui.adsabs.harvard.edu/abs/2019MNRAS.486.4829R}
}

@ARTICLE{2024A&A...686A.135P,
       author = {{Pinilla}, Paola and {Benisty}, Myriam and {Waters}, Rens and {Bae}, Jaehan and {Facchini}, Stefano},
        title = "{Survival of the long-lived inner disk of PDS70}",
      journal = {\aap},
         year = 2024,
        month = jun,
       volume = {686},
          eid = {A135},
        pages = {A135},
          doi = {10.1051/0004-6361/202348707},
archivePrefix = {arXiv},
       eprint = {2403.07057},
 primaryClass = {astro-ph.EP},
       adsurl = {https://ui.adsabs.harvard.edu/abs/2024A&A...686A.135P}
}

@ARTICLE{2018NewAR..80....1D,
       author = {{Dickinson}, Clive and {Ali-Ha{\"\i}moud}, Y. and {Barr}, A. and {Battistelli}, E.~S. and {Bell}, A. and {Bernstein}, L. and {Casassus}, S. and {Cleary}, K. and {Draine}, B.~T. and {G{\'e}nova-Santos}, R. and {Harper}, S.~E. and {Hensley}, B. and {Hill-Valler}, J. and {Hoang}, Thiem and {Israel}, F.~P. and {Jew}, L. and {Lazarian}, A. and {Leahy}, J.~P. and {Leech}, J. and {L{\'o}pez-Caraballo}, C.~H. and {McDonald}, I. and {Murphy}, E.~J. and {Onaka}, T. and {Paladini}, R. and {Peel}, M.~W. and {Perrott}, Y. and {Poidevin}, F. and {Readhead}, A.~C.~S. and {Rubi{\~n}o-Mart{\'\i}n}, J.-A. and {Taylor}, A.~C. and {Tibbs}, C.~T. and {Todorovi{\'c}}, M. and {Vidal}, Matias},
        title = "{The State-of-Play of Anomalous Microwave Emission (AME) research}",
      journal = {\nar},
         year = 2018,
        month = feb,
       volume = {80},
        pages = {1-28},
          doi = {10.1016/j.newar.2018.02.001},
archivePrefix = {arXiv},
       eprint = {1802.08073},
 primaryClass = {astro-ph.GA},
       adsurl = {https://ui.adsabs.harvard.edu/abs/2018NewAR..80....1D}
}

@ARTICLE{2024ApJ...972..163L,
       author = {{Liu}, Hauyu Baobab and {Casassus}, Simon and {Dong}, Ruobing and {Doi}, Kiyoaki and {Hashimoto}, Jun and {Muto}, Takayuki},
        title = "{First JVLA Radio Observation on PDS 70}",
      journal = {\apj},
         year = 2024,
        month = sep,
       volume = {972},
       number = {2},
          eid = {163},
        pages = {163},
          doi = {10.3847/1538-4357/ad5dab},
archivePrefix = {arXiv},
       eprint = {2406.19843},
 primaryClass = {astro-ph.EP},
       adsurl = {https://ui.adsabs.harvard.edu/abs/2024ApJ...972..163L}
}

@ARTICLE{2024MNRAS.532.1361A,
       author = {{Aizawa}, Masataka and {Muto}, Takayuki and {Momose}, Munetake},
        title = "{Revealing asymmetry on mid-plane of protoplanetary disc through modelling of axisymmetric emission: methodology}",
      journal = {\mnras},
         year = 2024,
        month = aug,
       volume = {532},
       number = {2},
        pages = {1361-1390},
          doi = {10.1093/mnras/stae1549},
archivePrefix = {arXiv},
       eprint = {2406.14006},
 primaryClass = {astro-ph.EP},
       adsurl = {https://ui.adsabs.harvard.edu/abs/2024MNRAS.532.1361A}
}

@ARTICLE{2025AJ....170..278L,
       author = {{Lucas}, Miles and {Bottom}, Michael and {Dong}, Ruobing and {Benisty}, Myriam and {Flock}, Mario and {Vincent}, Maria and {Williams}, Jonathan and {Ahn}, Kyohoon and {Currie}, Thayne and {Deo}, Vincent and {Guyon}, Olivier and {Kudo}, Tomoyuki and {Lilley}, Lucinda and {Lozi}, Julien and {Millar-Blanchaer}, Maxwell and {Norris}, Barnaby and {P{\'e}rez}, Sebasti{\'a}n and {Safonov}, Boris and {Tuthill}, Peter and {Uyama}, Taichi and {Vievard}, S{\'e}bastien and {Zhang}, Manxuan},
        title = "{Dynamical Analysis of the HD 169142 Planet-forming Disk: Twelve Years of High-contrast Polarimetry}",
      journal = {\aj},
         year = 2025,
        month = nov,
       volume = {170},
       number = {5},
          eid = {278},
        pages = {278},
          doi = {10.3847/1538-3881/ae093c},
archivePrefix = {arXiv},
       eprint = {2509.15323},
 primaryClass = {astro-ph.EP},
       adsurl = {https://ui.adsabs.harvard.edu/abs/2025AJ....170..278L}
}

@ARTICLE{2018MNRAS.474.5105B,
       author = {{Bertrang}, G.~H.-M. and {Avenhaus}, H. and {Casassus}, S. and {Montesinos}, M. and {Kirchschlager}, F. and {Perez}, S. and {Cieza}, L. and {Wolf}, S.},
        title = "{HD 169142 in the eyes of ZIMPOL/SPHERE}",
      journal = {\mnras},
         year = 2018,
        month = mar,
       volume = {474},
       number = {4},
        pages = {5105-5113},
          doi = {10.1093/mnras/stx3052},
archivePrefix = {arXiv},
       eprint = {1711.09040},
 primaryClass = {astro-ph.EP},
       adsurl = {https://ui.adsabs.harvard.edu/abs/2018MNRAS.474.5105B}
}

@ARTICLE{2018MNRAS.473.1774L,
       author = {{Ligi}, R. and {Vigan}, A. and {Gratton}, R. and {de Boer}, J. and {Benisty}, M. and {Boccaletti}, A. and {Quanz}, S.~P. and {Meyer}, M. and {Ginski}, C. and {Sissa}, E. and {Gry}, C. and {Henning}, T. and {Beuzit}, J.-L. and {Biller}, B. and {Bonnefoy}, M. and {Chauvin}, G. and {Cheetham}, A.~C. and {Cudel}, M. and {Delorme}, P. and {Desidera}, S. and {Feldt}, M. and {Galicher}, R. and {Girard}, J. and {Janson}, M. and {Kasper}, M. and {Kopytova}, T. and {Lagrange}, A.-M. and {Langlois}, M. and {Lecoroller}, H. and {Maire}, A.-L. and {M{\'e}nard}, F. and {Mesa}, D. and {Peretti}, S. and {Perrot}, C. and {Pinilla}, P. and {Pohl}, A. and {Rouan}, D. and {Stolker}, T. and {Samland}, M. and {Wahhaj}, Z. and {Wildi}, F. and {Zurlo}, A. and {Buey}, T. and {Fantinel}, D. and {Fusco}, T. and {Jaquet}, M. and {Moulin}, T. and {Ramos}, J. and {Suarez}, M. and {Weber}, L.},
        title = "{Investigation of the inner structures around HD 169142 with VLT/SPHERE}",
      journal = {\mnras},
         year = 2018,
        month = jan,
       volume = {473},
       number = {2},
        pages = {1774-1783},
          doi = {10.1093/mnras/stx2318},
archivePrefix = {arXiv},
       eprint = {1709.01734},
 primaryClass = {astro-ph.EP},
       adsurl = {https://ui.adsabs.harvard.edu/abs/2018MNRAS.473.1774L}
}

@ARTICLE{2015PASJ...67...83M,
       author = {{Momose}, Munetake and {Morita}, Ayaka and {Fukagawa}, Misato and {Muto}, Takayuki and {Takeuchi}, Taku and {Hashimoto}, Jun and {Honda}, Mitsuhiko and {Kudo}, Tomoyuki and {Okamoto}, Yoshiko K. and {Kanagawa}, Kazuhiro D. and {Tanaka}, Hidekazu and {Grady}, Carol A. and {Sitko}, Michael L. and {Akiyama}, Eiji and {Currie}, Thayne and {Follette}, Katherine B. and {Mayama}, Satoshi and {Kusakabe}, Nobuhiko and {Abe}, Lyu and {Brandner}, Wolfgang and {Brandt}, Timothy D. and {Carson}, Joseph C. and {Egner}, Sebastian and {Feldt}, Markus and {Goto}, Miwa and {Guyon}, Olivier and {Hayano}, Yutaka and {Hayashi}, Masahiko and {Hayashi}, Saeko S. and {Henning}, Thomas and {Hodapp}, Klaus W. and {Ishii}, Miki and {Iye}, Masanori and {Janson}, Markus and {Kandori}, Ryo and {Knapp}, Gillian R. and {Kuzuhara}, Masayuki and {Kwon}, Jungmi and {Matsuo}, Taro and {McElwain}, Michael W. and {Miyama}, Shoken and {Morino}, Jun-Ichi and {Moro-Martin}, Amaya and {Nishimura}, Tetsuo and {Pyo}, Tae-Soo and {Serabyn}, Eugene and {Suenaga}, Takuya and {Suto}, Hiroshi and {Suzuki}, Ryuji and {Takahashi}, Yasuhiro H. and {Takami}, Michihiro and {Takato}, Naruhisa and {Terada}, Hiroshi and {Thalmann}, Christian and {Tomono}, Daigo and {Turner}, Edwin L. and {Watanabe}, Makoto and {Wisniewski}, John and {Yamada}, Toru and {Takami}, Hideki and {Usuda}, Tomonori and {Tamura}, Motohide},
        title = "{Detailed structure of the outer disk around HD 169142 with polarized light in H-band}",
      journal = {\pasj},
         year = 2015,
        month = oct,
       volume = {67},
       number = {5},
          eid = {83},
        pages = {83},
          doi = {10.1093/pasj/psv051},
archivePrefix = {arXiv},
       eprint = {1505.04937},
 primaryClass = {astro-ph.SR},
       adsurl = {https://ui.adsabs.harvard.edu/abs/2015PASJ...67...83M}
}

@ARTICLE{2017A&A...600A..72F,
       author = {{Fedele}, D. and {Carney}, M. and {Hogerheijde}, M.~R. and {Walsh}, C. and {Miotello}, A. and {Klaassen}, P. and {Bruderer}, S. and {Henning}, Th. and {van Dishoeck}, E.~F.},
        title = "{ALMA unveils rings and gaps in the protoplanetary system <ASTROBJ>HD 169142</ASTROBJ>: signatures of two giant protoplanets}",
      journal = {\aap},
         year = 2017,
        month = apr,
       volume = {600},
          eid = {A72},
        pages = {A72},
          doi = {10.1051/0004-6361/201629860},
archivePrefix = {arXiv},
       eprint = {1702.02844},
 primaryClass = {astro-ph.SR},
       adsurl = {https://ui.adsabs.harvard.edu/abs/2017A&A...600A..72F}
}

@ARTICLE{2013ApJ...766L...2Q,
       author = {{Quanz}, Sascha P. and {Avenhaus}, Henning and {Buenzli}, Esther and {Garufi}, Antonio and {Schmid}, Hans Martin and {Wolf}, Sebastian},
        title = "{Gaps in the HD 169142 Protoplanetary Disk Revealed by Polarimetric Imaging: Signs of Ongoing Planet Formation?}",
      journal = {\apjl},
         year = 2013,
        month = mar,
       volume = {766},
       number = {1},
          eid = {L2},
        pages = {L2},
          doi = {10.1088/2041-8205/766/1/L2},
archivePrefix = {arXiv},
       eprint = {1302.3029},
 primaryClass = {astro-ph.GA},
       adsurl = {https://ui.adsabs.harvard.edu/abs/2013ApJ...766L...2Q}
}

@ARTICLE{2012ApJ...752..143H,
       author = {{Honda}, M. and {Maaskant}, Koen and {Okamoto}, Y.~K. and {Kataza}, H. and {Fukagawa}, M. and {Waters}, L.~B.~F.~M. and {Dominik}, C. and {Tielens}, A.~G.~G.~M. and {Mulders}, G.~D. and {Min}, M. and {Yamashita}, T. and {Fujiyoshi}, T. and {Miyata}, T. and {Sako}, S. and {Sakon}, I. and {Fujiwara}, H. and {Onaka}, T.},
        title = "{Mid-infrared Imaging of the Transitional Disk of HD 169142: Measuring the Size of the Gap}",
      journal = {\apj},
         year = 2012,
        month = jun,
       volume = {752},
       number = {2},
          eid = {143},
        pages = {143},
          doi = {10.1088/0004-637X/752/2/143},
archivePrefix = {arXiv},
       eprint = {1204.5364},
 primaryClass = {astro-ph.EP},
       adsurl = {https://ui.adsabs.harvard.edu/abs/2012ApJ...752..143H}
}

@ARTICLE{2010PASJ...62..347F,
       author = {{Fukagawa}, Misato and {Tamura}, Motohide and {Itoh}, Yoichi and {Oasa}, Yumiko and {Kudo}, Tomoyuki and {Hayashi}, Saeko S. and {Kato}, Eri and {Ootsubo}, Takafumi and {Itoh}, Yusuke and {Shibai}, Hiroshi and {Hayashi}, Masahiko},
        title = "{Subaru Near-Infrared Imaging of Herbig Ae Stars}",
      journal = {\pasj},
         year = 2010,
        month = apr,
       volume = {62},
        pages = {347},
          doi = {10.1093/pasj/62.2.347},
       adsurl = {https://ui.adsabs.harvard.edu/abs/2010PASJ...62..347F}
}

@ARTICLE{2023A&A...674A...1G,
       author = {{Gaia Collaboration} and {Vallenari}, A. and {Brown}, A.~G.~A. and {Prusti}, T. and {de Bruijne}, J.~H.~J. and {Arenou}, F. and {Babusiaux}, C. and {Biermann}, M. and {Creevey}, O.~L. and {Ducourant}, C. and {Evans}, D.~W. and {Eyer}, L. and {Guerra}, R. and {Hutton}, A. and {Jordi}, C. and {Klioner}, S.~A. and {Lammers}, U.~L. and {Lindegren}, L. and {Luri}, X. and {Mignard}, F. and {Panem}, C. and {Pourbaix}, D. and {Randich}, S. and {Sartoretti}, P. and {Soubiran}, C. and {Tanga}, P. and {Walton}, N.~A. and {Bailer-Jones}, C.~A.~L. and {Bastian}, U. and {Drimmel}, R. and {Jansen}, F. and {Katz}, D. and {Lattanzi}, M.~G. and {van Leeuwen}, F. and {Bakker}, J. and {Cacciari}, C. and {Casta{\~n}eda}, J. and {De Angeli}, F. and {Fabricius}, C. and {Fouesneau}, M. and {Fr{\'e}mat}, Y. and {Galluccio}, L. and {Guerrier}, A. and {Heiter}, U. and {Masana}, E. and {Messineo}, R. and {Mowlavi}, N. and {Nicolas}, C. and {Nienartowicz}, K. and {Pailler}, F. and {Panuzzo}, P. and {Riclet}, F. and {Roux}, W. and {Seabroke}, G.~M. and {Sordo}, R. and {Th{\'e}venin}, F. and {Gracia-Abril}, G. and {Portell}, J. and {Teyssier}, D. and {Altmann}, M. and {Andrae}, R. and {Audard}, M. and {Bellas-Velidis}, I. and {Benson}, K. and {Berthier}, J. and {Blomme}, R. and {Burgess}, P.~W. and {Busonero}, D. and {Busso}, G. and {C{\'a}novas}, H. and {Carry}, B. and {Cellino}, A. and {Cheek}, N. and {Clementini}, G. and {Damerdji}, Y. and {Davidson}, M. and {de Teodoro}, P. and {Nu{\~n}ez Campos}, M. and {Delchambre}, L. and {Dell'Oro}, A. and {Esquej}, P. and {Fern{\'a}ndez-Hern{\'a}ndez}, J. and {Fraile}, E. and {Garabato}, D. and {Garc{\'\i}a-Lario}, P. and {Gosset}, E. and {Haigron}, R. and {Halbwachs}, J.-L. and {Hambly}, N.~C. and {Harrison}, D.~L. and {Hern{\'a}ndez}, J. and {Hestroffer}, D. and {Hodgkin}, S.~T. and {Holl}, B. and {Jan{\ss}en}, K. and {Jevardat de Fombelle}, G. and {Jordan}, S. and {Krone-Martins}, A. and {Lanzafame}, A.~C. and {L{\"o}ffler}, W. and {Marchal}, O. and {Marrese}, P.~M. and {Moitinho}, A. and {Muinonen}, K. and {Osborne}, P. and {Pancino}, E. and {Pauwels}, T. and {Recio-Blanco}, A. and {Reyl{\'e}}, C. and {Riello}, M. and {Rimoldini}, L. and {Roegiers}, T. and {Rybizki}, J. and {Sarro}, L.~M. and {Siopis}, C. and {Smith}, M. and {Sozzetti}, A. and {Utrilla}, E. and {van Leeuwen}, M. and {Abbas}, U. and {{\'A}brah{\'a}m}, P. and {Abreu Aramburu}, A. and {Aerts}, C. and {Aguado}, J.~J. and {Ajaj}, M. and {Aldea-Montero}, F. and {Altavilla}, G. and {{\'A}lvarez}, M.~A. and {Alves}, J. and {Anders}, F. and {Anderson}, R.~I. and {Anglada Varela}, E. and {Antoja}, T. and {Baines}, D. and {Baker}, S.~G. and {Balaguer-N{\'u}{\~n}ez}, L. and {Balbinot}, E. and {Balog}, Z. and {Barache}, C. and {Barbato}, D. and {Barros}, M. and {Barstow}, M.~A. and {Bartolom{\'e}}, S. and {Bassilana}, J.-L. and {Bauchet}, N. and {Becciani}, U. and {Bellazzini}, M. and {Berihuete}, A. and {Bernet}, M. and {Bertone}, S. and {Bianchi}, L. and {Binnenfeld}, A. and {Blanco-Cuaresma}, S. and {Blazere}, A. and {Boch}, T. and {Bombrun}, A. and {Bossini}, D. and {Bouquillon}, S. and {Bragaglia}, A. and {Bramante}, L. and {Breedt}, E. and {Bressan}, A. and {Brouillet}, N. and {Brugaletta}, E. and {Bucciarelli}, B. and {Burlacu}, A. and {Butkevich}, A.~G. and {Buzzi}, R. and {Caffau}, E. and {Cancelliere}, R. and {Cantat-Gaudin}, T. and {Carballo}, R. and {Carlucci}, T. and {Carnerero}, M.~I. and {Carrasco}, J.~M. and {Casamiquela}, L. and {Castellani}, M. and {Castro-Ginard}, A. and {Chaoul}, L. and {Charlot}, P. and {Chemin}, L. and {Chiaramida}, V. and {Chiavassa}, A. and {Chornay}, N. and {Comoretto}, G. and {Contursi}, G. and {Cooper}, W.~J. and {Cornez}, T. and {Cowell}, S. and {Crifo}, F. and {Cropper}, M. and {Crosta}, M. and {Crowley}, C. and {Dafonte}, C. and {Dapergolas}, A. and {David}, M. and {David}, P. and {de Laverny}, P. and {De Luise}, F. and {De March}, R.},
        title = "{Gaia Data Release 3. Summary of the content and survey properties}",
      journal = {\aap},
         year = 2023,
        month = jun,
       volume = {674},
          eid = {A1},
        pages = {A1},
          doi = {10.1051/0004-6361/202243940},
archivePrefix = {arXiv},
       eprint = {2208.00211},
 primaryClass = {astro-ph.GA},
       adsurl = {https://ui.adsabs.harvard.edu/abs/2023A&A...674A...1G}
}

@ARTICLE{2025A&A...695A.147V,
       author = {{Viscardi}, Elena M. and {Mac{\'\i}as}, Enrique and {Zagaria}, Francesco and {Sierra}, Anibal and {Jiang}, Haochang and {Yoshida}, Tomohiro C. and {Curone}, Pietro},
        title = "{Dust characterization of protoplanetary disks: A guide to multi-wavelength analyses and accurate dust mass measurements}",
      journal = {\aap},
         year = 2025,
        month = mar,
       volume = {695},
          eid = {A147},
        pages = {A147},
          doi = {10.1051/0004-6361/202452935},
archivePrefix = {arXiv},
       eprint = {2501.13877},
 primaryClass = {astro-ph.EP},
       adsurl = {https://ui.adsabs.harvard.edu/abs/2025A&A...695A.147V}
}

@ARTICLE{2024ApJ...971..129C,
       author = {{Carvalho}, Adolfo S. and {P{\'e}rez}, Laura M. and {Sierra}, Anibal and {Mellado}, Maria Jesus and {Hillenbrand}, Lynne A. and {Andrews}, Sean and {Benisty}, Myriam and {Birnstiel}, Tilman and {Carpenter}, John M. and {Guzm{\'a}n}, Viviana V. and {Huang}, Jane and {Isella}, Andrea and {Kurtovic}, Nicolas and {Ricci}, Luca and {Wilner}, David J.},
        title = "{A Dust-trapping Ring in the Planet-hosting Disk of Elias 2-24}",
      journal = {\apj},
         year = 2024,
        month = aug,
       volume = {971},
       number = {2},
          eid = {129},
        pages = {129},
          doi = {10.3847/1538-4357/ad5a07},
archivePrefix = {arXiv},
       eprint = {2406.12819},
 primaryClass = {astro-ph.EP},
       adsurl = {https://ui.adsabs.harvard.edu/abs/2024ApJ...971..129C}
}

@ARTICLE{2024ApJ...974..306S,
       author = {{Sierra}, Anibal and {P{\'e}rez}, Laura M. and {Sotomayor}, Benjam{\'\i}n and {Benisty}, Myriam and {Chandler}, Claire J. and {Andrews}, Sean and {Carpenter}, John and {Henning}, Thomas and {Testi}, Leonardo and {Ricci}, Luca and {Wilner}, David},
        title = "{Constraints on the Physical Origin of Large Cavities in Transition Disks from Multiwavelength Dust Continuum Emission}",
      journal = {\apj},
         year = 2024,
        month = oct,
       volume = {974},
       number = {2},
          eid = {306},
        pages = {306},
          doi = {10.3847/1538-4357/ad7460},
archivePrefix = {arXiv},
       eprint = {2408.15407},
 primaryClass = {astro-ph.EP},
       adsurl = {https://ui.adsabs.harvard.edu/abs/2024ApJ...974..306S}
}

@ARTICLE{2024A&A...686A.298G,
       author = {{Guerra-Alvarado}, Osmar M. and {Carrasco-Gonz{\'a}lez}, Carlos and {Mac{\'\i}as}, Enrique and {van der Marel}, Nienke and {Houge}, Adrien and {Maud}, Luke T. and {Pinilla}, Paola and {Villenave}, Marion and {Asaki}, Yoshiharu and {Humphreys}, Elizabeth},
        title = "{Into the thick of it: ALMA 0.45 mm observations of HL Tau at a resolution of 2 au}",
      journal = {\aap},
         year = 2024,
        month = jun,
       volume = {686},
          eid = {A298},
        pages = {A298},
          doi = {10.1051/0004-6361/202349046},
archivePrefix = {arXiv},
       eprint = {2404.04164},
 primaryClass = {astro-ph.EP},
       adsurl = {https://ui.adsabs.harvard.edu/abs/2024A&A...686A.298G}
}

@ARTICLE{2023ApJ...953...96Z,
       author = {{Zhang}, Shangjia and {Zhu}, Zhaohuan and {Ueda}, Takahiro and {Kataoka}, Akimasa and {Sierra}, Anibal and {Carrasco-Gonz{\'a}lez}, Carlos and {Mac{\'\i}as}, Enrique},
        title = "{Porous Dust Particles in Protoplanetary Disks: Application to the HL Tau Disk}",
      journal = {\apj},
         year = 2023,
        month = aug,
       volume = {953},
       number = {1},
          eid = {96},
        pages = {96},
          doi = {10.3847/1538-4357/acdb4e},
archivePrefix = {arXiv},
       eprint = {2306.00158},
 primaryClass = {astro-ph.EP},
       adsurl = {https://ui.adsabs.harvard.edu/abs/2023ApJ...953...96Z}
}

@ARTICLE{2025MNRAS.538.2358S,
       author = {{Sierra}, Anibal and {Pinilla}, Paola and {P{\'e}rez}, Laura M. and {Benisty}, Myriam and {Agurto-Gangas}, Carolina and {Carrasco-Gonz{\'a}lez}, Carlos and {Curone}, Pietro and {Long}, Feng},
        title = "{High angular resolution evidence of dust traps from deep ALMA Band 3 observations of LkCa15}",
      journal = {\mnras},
         year = 2025,
        month = apr,
       volume = {538},
       number = {4},
        pages = {2358-2374},
          doi = {10.1093/mnras/staf393},
archivePrefix = {arXiv},
       eprint = {2503.03336},
 primaryClass = {astro-ph.EP},
       adsurl = {https://ui.adsabs.harvard.edu/abs/2025MNRAS.538.2358S}
}

@ARTICLE{2023ApJ...954..110O,
       author = {{Ohashi}, Satoshi and {Momose}, Munetake and {Kataoka}, Akimasa and {Higuchi}, Aya E. and {Tsukagoshi}, Takashi and {Ueda}, Takahiro and {Codella}, Claudio and {Podio}, Linda and {Hanawa}, Tomoyuki and {Sakai}, Nami and {Kobayashi}, Hiroshi and {Okuzumi}, Satoshi and {Tanaka}, Hidekazu},
        title = "{Dust Enrichment and Grain Growth in a Smooth Disk around the DG Tau Protostar Revealed by ALMA Triple Bands Frequency Observations}",
      journal = {\apj},
         year = 2023,
        month = sep,
       volume = {954},
       number = {2},
          eid = {110},
        pages = {110},
          doi = {10.3847/1538-4357/ace9b9},
archivePrefix = {arXiv},
       eprint = {2307.14526},
 primaryClass = {astro-ph.EP},
       adsurl = {https://ui.adsabs.harvard.edu/abs/2023ApJ...954..110O}
}

@ARTICLE{2025A&A...702A..56Z,
       author = {{Zagaria}, Francesco and {Facchini}, Stefano and {Curone}, Pietro and {Williams}, Jonathan P. and {Clarke}, Cathie J. and {Ribas}, {\'A}lvaro and {Tazzari}, Marco and {Mac{\'\i}as}, Enrique and {Booth}, Richard A. and {Rosotti}, Giovanni P. and {Testi}, Leonardo},
        title = "{Multi-frequency analysis of the ALMA and VLA high resolution continuum observations of the substructured disc around CI Tau: Preference for submillimetre-sized low-porosity amorphous carbon grains}",
      journal = {\aap},
         year = 2025,
        month = oct,
       volume = {702},
          eid = {A56},
        pages = {A56},
          doi = {10.1051/0004-6361/202452986},
archivePrefix = {arXiv},
       eprint = {2507.08797},
 primaryClass = {astro-ph.EP},
       adsurl = {https://ui.adsabs.harvard.edu/abs/2025A&A...702A..56Z}
}

@ARTICLE{2018AJ....156..123A,
       author = {{Astropy Collaboration} and {Price-Whelan}, A.~M. and {Sip{\H{o}}cz}, B.~M. and {G{\"u}nther}, H.~M. and {Lim}, P.~L. and {Crawford}, S.~M. and {Conseil}, S. and {Shupe}, D.~L. and {Craig}, M.~W. and {Dencheva}, N. and {Ginsburg}, A. and {VanderPlas}, J.~T. and {Bradley}, L.~D. and {P{\'e}rez-Su{\'a}rez}, D. and {de Val-Borro}, M. and {Aldcroft}, T.~L. and {Cruz}, K.~L. and {Robitaille}, T.~P. and {Tollerud}, E.~J. and {Ardelean}, C. and {Babej}, T. and {Bach}, Y.~P. and {Bachetti}, M. and {Bakanov}, A.~V. and {Bamford}, S.~P. and {Barentsen}, G. and {Barmby}, P. and {Baumbach}, A. and {Berry}, K.~L. and {Biscani}, F. and {Boquien}, M. and {Bostroem}, K.~A. and {Bouma}, L.~G. and {Brammer}, G.~B. and {Bray}, E.~M. and {Breytenbach}, H. and {Buddelmeijer}, H. and {Burke}, D.~J. and {Calderone}, G. and {Cano Rodr{\'\i}guez}, J.~L. and {Cara}, M. and {Cardoso}, J.~V.~M. and {Cheedella}, S. and {Copin}, Y. and {Corrales}, L. and {Crichton}, D. and {D'Avella}, D. and {Deil}, C. and {Depagne}, {\'E}. and {Dietrich}, J.~P. and {Donath}, A. and {Droettboom}, M. and {Earl}, N. and {Erben}, T. and {Fabbro}, S. and {Ferreira}, L.~A. and {Finethy}, T. and {Fox}, R.~T. and {Garrison}, L.~H. and {Gibbons}, S.~L.~J. and {Goldstein}, D.~A. and {Gommers}, R. and {Greco}, J.~P. and {Greenfield}, P. and {Groener}, A.~M. and {Grollier}, F. and {Hagen}, A. and {Hirst}, P. and {Homeier}, D. and {Horton}, A.~J. and {Hosseinzadeh}, G. and {Hu}, L. and {Hunkeler}, J.~S. and {Ivezi{\'c}}, {\v{Z}}. and {Jain}, A. and {Jenness}, T. and {Kanarek}, G. and {Kendrew}, S. and {Kern}, N.~S. and {Kerzendorf}, W.~E. and {Khvalko}, A. and {King}, J. and {Kirkby}, D. and {Kulkarni}, A.~M. and {Kumar}, A. and {Lee}, A. and {Lenz}, D. and {Littlefair}, S.~P. and {Ma}, Z. and {Macleod}, D.~M. and {Mastropietro}, M. and {McCully}, C. and {Montagnac}, S. and {Morris}, B.~M. and {Mueller}, M. and {Mumford}, S.~J. and {Muna}, D. and {Murphy}, N.~A. and {Nelson}, S. and {Nguyen}, G.~H. and {Ninan}, J.~P. and {N{\"o}the}, M. and {Ogaz}, S. and {Oh}, S. and {Parejko}, J.~K. and {Parley}, N. and {Pascual}, S. and {Patil}, R. and {Patil}, A.~A. and {Plunkett}, A.~L. and {Prochaska}, J.~X. and {Rastogi}, T. and {Reddy Janga}, V. and {Sabater}, J. and {Sakurikar}, P. and {Seifert}, M. and {Sherbert}, L.~E. and {Sherwood-Taylor}, H. and {Shih}, A.~Y. and {Sick}, J. and {Silbiger}, M.~T. and {Singanamalla}, S. and {Singer}, L.~P. and {Sladen}, P.~H. and {Sooley}, K.~A. and {Sornarajah}, S. and {Streicher}, O. and {Teuben}, P. and {Thomas}, S.~W. and {Tremblay}, G.~R. and {Turner}, J.~E.~H. and {Terr{\'o}n}, V. and {van Kerkwijk}, M.~H. and {de la Vega}, A. and {Watkins}, L.~L. and {Weaver}, B.~A. and {Whitmore}, J.~B. and {Woillez}, J. and {Zabalza}, V. and {Astropy Contributors}},
        title = "{The Astropy Project: Building an Open-science Project and Status of the v2.0 Core Package}",
      journal = {\aj},
         year = 2018,
        month = sep,
       volume = {156},
       number = {3},
          eid = {123},
        pages = {123},
          doi = {10.3847/1538-3881/aabc4f},
archivePrefix = {arXiv},
       eprint = {1801.02634},
 primaryClass = {astro-ph.IM},
       adsurl = {https://ui.adsabs.harvard.edu/abs/2018AJ....156..123A}
}

@ARTICLE{2022ApJ...935..167A,
       author = {{Astropy Collaboration} and {Price-Whelan}, Adrian M. and {Lim}, Pey Lian and {Earl}, Nicholas and {Starkman}, Nathaniel and {Bradley}, Larry and {Shupe}, David L. and {Patil}, Aarya A. and {Corrales}, Lia and {Brasseur}, C.~E. and {N{\"o}the}, Maximilian and {Donath}, Axel and {Tollerud}, Erik and {Morris}, Brett M. and {Ginsburg}, Adam and {Vaher}, Eero and {Weaver}, Benjamin A. and {Tocknell}, James and {Jamieson}, William and {van Kerkwijk}, Marten H. and {Robitaille}, Thomas P. and {Merry}, Bruce and {Bachetti}, Matteo and {G{\"u}nther}, H. Moritz and {Aldcroft}, Thomas L. and {Alvarado-Montes}, Jaime A. and {Archibald}, Anne M. and {B{\'o}di}, Attila and {Bapat}, Shreyas and {Barentsen}, Geert and {Baz{\'a}n}, Juanjo and {Biswas}, Manish and {Boquien}, M{\'e}d{\'e}ric and {Burke}, D.~J. and {Cara}, Daria and {Cara}, Mihai and {Conroy}, Kyle E. and {Conseil}, Simon and {Craig}, Matthew W. and {Cross}, Robert M. and {Cruz}, Kelle L. and {D'Eugenio}, Francesco and {Dencheva}, Nadia and {Devillepoix}, Hadrien A.~R. and {Dietrich}, J{\"o}rg P. and {Eigenbrot}, Arthur Davis and {Erben}, Thomas and {Ferreira}, Leonardo and {Foreman-Mackey}, Daniel and {Fox}, Ryan and {Freij}, Nabil and {Garg}, Suyog and {Geda}, Robel and {Glattly}, Lauren and {Gondhalekar}, Yash and {Gordon}, Karl D. and {Grant}, David and {Greenfield}, Perry and {Groener}, Austen M. and {Guest}, Steve and {Gurovich}, Sebastian and {Handberg}, Rasmus and {Hart}, Akeem and {Hatfield-Dodds}, Zac and {Homeier}, Derek and {Hosseinzadeh}, Griffin and {Jenness}, Tim and {Jones}, Craig K. and {Joseph}, Prajwel and {Kalmbach}, J. Bryce and {Karamehmetoglu}, Emir and {Ka{\l}uszy{\'n}ski}, Miko{\l}aj and {Kelley}, Michael S.~P. and {Kern}, Nicholas and {Kerzendorf}, Wolfgang E. and {Koch}, Eric W. and {Kulumani}, Shankar and {Lee}, Antony and {Ly}, Chun and {Ma}, Zhiyuan and {MacBride}, Conor and {Maljaars}, Jakob M. and {Muna}, Demitri and {Murphy}, N.~A. and {Norman}, Henrik and {O'Steen}, Richard and {Oman}, Kyle A. and {Pacifici}, Camilla and {Pascual}, Sergio and {Pascual-Granado}, J. and {Patil}, Rohit R. and {Perren}, Gabriel I. and {Pickering}, Timothy E. and {Rastogi}, Tanuj and {Roulston}, Benjamin R. and {Ryan}, Daniel F. and {Rykoff}, Eli S. and {Sabater}, Jose and {Sakurikar}, Parikshit and {Salgado}, Jes{\'u}s and {Sanghi}, Aniket and {Saunders}, Nicholas and {Savchenko}, Volodymyr and {Schwardt}, Ludwig and {Seifert-Eckert}, Michael and {Shih}, Albert Y. and {Jain}, Anany Shrey and {Shukla}, Gyanendra and {Sick}, Jonathan and {Simpson}, Chris and {Singanamalla}, Sudheesh and {Singer}, Leo P. and {Singhal}, Jaladh and {Sinha}, Manodeep and {Sip{\H{o}}cz}, Brigitta M. and {Spitler}, Lee R. and {Stansby}, David and {Streicher}, Ole and {{\v{S}}umak}, Jani and {Swinbank}, John D. and {Taranu}, Dan S. and {Tewary}, Nikita and {Tremblay}, Grant R. and {de Val-Borro}, Miguel and {Van Kooten}, Samuel J. and {Vasovi{\'c}}, Zlatan and {Verma}, Shresth and {de Miranda Cardoso}, Jos{\'e} Vin{\'\i}cius and {Williams}, Peter K.~G. and {Wilson}, Tom J. and {Winkel}, Benjamin and {Wood-Vasey}, W.~M. and {Xue}, Rui and {Yoachim}, Peter and {Zhang}, Chen and {Zonca}, Andrea and {Astropy Project Contributors}},
        title = "{The Astropy Project: Sustaining and Growing a Community-oriented Open-source Project and the Latest Major Release (v5.0) of the Core Package}",
      journal = {\apj},
         year = 2022,
        month = aug,
       volume = {935},
       number = {2},
          eid = {167},
        pages = {167},
          doi = {10.3847/1538-4357/ac7c74},
archivePrefix = {arXiv},
       eprint = {2206.14220},
 primaryClass = {astro-ph.IM},
       adsurl = {https://ui.adsabs.harvard.edu/abs/2022ApJ...935..167A}
}

@ARTICLE{2022PASP..134k4501C,
       author = {{CASA Team} and {Bean}, Ben and {Bhatnagar}, Sanjay and {Castro}, Sandra and {Donovan Meyer}, Jennifer and {Emonts}, Bjorn and {Garcia}, Enrique and {Garwood}, Robert and {Golap}, Kumar and {Gonzalez Villalba}, Justo and {Harris}, Pamela and {Hayashi}, Yohei and {Hoskins}, Josh and {Hsieh}, Mingyu and {Jagannathan}, Preshanth and {Kawasaki}, Wataru and {Keimpema}, Aard and {Kettenis}, Mark and {Lopez}, Jorge and {Marvil}, Joshua and {Masters}, Joseph and {McNichols}, Andrew and {Mehringer}, David and {Miel}, Renaud and {Moellenbrock}, George and {Montesino}, Federico and {Nakazato}, Takeshi and {Ott}, Juergen and {Petry}, Dirk and {Pokorny}, Martin and {Raba}, Ryan and {Rau}, Urvashi and {Schiebel}, Darrell and {Schweighart}, Neal and {Sekhar}, Srikrishna and {Shimada}, Kazuhiko and {Small}, Des and {Steeb}, Jan-Willem and {Sugimoto}, Kanako and {Suoranta}, Ville and {Tsutsumi}, Takahiro and {van Bemmel}, Ilse M. and {Verkouter}, Marjolein and {Wells}, Akeem and {Xiong}, Wei and {Szomoru}, Arpad and {Griffith}, Morgan and {Glendenning}, Brian and {Kern}, Jeff},
        title = "{CASA, the Common Astronomy Software Applications for Radio Astronomy}",
      journal = {\pasp},
         year = 2022,
        month = nov,
       volume = {134},
       number = {1041},
          eid = {114501},
        pages = {114501},
          doi = {10.1088/1538-3873/ac9642},
archivePrefix = {arXiv},
       eprint = {2210.02276},
 primaryClass = {astro-ph.IM},
       adsurl = {https://ui.adsabs.harvard.edu/abs/2022PASP..134k4501C}
}

@ARTICLE{2013A&A...558A..33A,
       author = {{Astropy Collaboration} and {Robitaille}, Thomas P. and {Tollerud}, Erik J. and {Greenfield}, Perry and {Droettboom}, Michael and {Bray}, Erik and {Aldcroft}, Tom and {Davis}, Matt and {Ginsburg}, Adam and {Price-Whelan}, Adrian M. and {Kerzendorf}, Wolfgang E. and {Conley}, Alexander and {Crighton}, Neil and {Barbary}, Kyle and {Muna}, Demitri and {Ferguson}, Henry and {Grollier}, Fr{\'e}d{\'e}ric and {Parikh}, Madhura M. and {Nair}, Prasanth H. and {Unther}, Hans M. and {Deil}, Christoph and {Woillez}, Julien and {Conseil}, Simon and {Kramer}, Roban and {Turner}, James E.~H. and {Singer}, Leo and {Fox}, Ryan and {Weaver}, Benjamin A. and {Zabalza}, Victor and {Edwards}, Zachary I. and {Azalee Bostroem}, K. and {Burke}, D.~J. and {Casey}, Andrew R. and {Crawford}, Steven M. and {Dencheva}, Nadia and {Ely}, Justin and {Jenness}, Tim and {Labrie}, Kathleen and {Lim}, Pey Lian and {Pierfederici}, Francesco and {Pontzen}, Andrew and {Ptak}, Andy and {Refsdal}, Brian and {Servillat}, Mathieu and {Streicher}, Ole},
        title = "{Astropy: A community Python package for astronomy}",
      journal = {\aap},
         year = 2013,
        month = oct,
       volume = {558},
          eid = {A33},
        pages = {A33},
          doi = {10.1051/0004-6361/201322068},
archivePrefix = {arXiv},
       eprint = {1307.6212},
 primaryClass = {astro-ph.IM},
       adsurl = {https://ui.adsabs.harvard.edu/abs/2013A&A...558A..33A}
}

@ARTICLE{2018ApJ...869L..41A,
       author = {{Andrews}, Sean M. and {Huang}, Jane and {P{\'e}rez}, Laura M. and {Isella}, Andrea and {Dullemond}, Cornelis P. and {Kurtovic}, Nicol{\'a}s T. and {Guzm{\'a}n}, Viviana V. and {Carpenter}, John M. and {Wilner}, David J. and {Zhang}, Shangjia and {Zhu}, Zhaohuan and {Birnstiel}, Tilman and {Bai}, Xue-Ning and {Benisty}, Myriam and {Hughes}, A. Meredith and {{\"O}berg}, Karin I. and {Ricci}, Luca},
        title = "{The Disk Substructures at High Angular Resolution Project (DSHARP). I. Motivation, Sample, Calibration, and Overview}",
      journal = {\apjl},
         year = 2018,
        month = dec,
       volume = {869},
       number = {2},
          eid = {L41},
        pages = {L41},
          doi = {10.3847/2041-8213/aaf741},
archivePrefix = {arXiv},
       eprint = {1812.04040},
 primaryClass = {astro-ph.SR},
       adsurl = {https://ui.adsabs.harvard.edu/abs/2018ApJ...869L..41A}
}

@ARTICLE{2022A&A...663A..23L,
       author = {{Leemker}, M. and {Booth}, A.~S. and {van Dishoeck}, E.~F. and {P{\'e}rez-S{\'a}nchez}, A.~F. and {Szul{\'a}gyi}, J. and {Bosman}, A.~D. and {Bruderer}, S. and {Facchini}, S. and {Hogerheijde}, M.~R. and {Paneque-Carre{\~n}o}, T. and {Sturm}, J.~A.},
        title = "{Gas temperature structure across transition disk cavities}",
      journal = {\aap},
         year = 2022,
        month = jul,
       volume = {663},
          eid = {A23},
        pages = {A23},
          doi = {10.1051/0004-6361/202243229},
archivePrefix = {arXiv},
       eprint = {2204.03666},
 primaryClass = {astro-ph.EP},
       adsurl = {https://ui.adsabs.harvard.edu/abs/2022A&A...663A..23L}
}

@ARTICLE{2021A&A...648A..33M,
       author = {{Mac{\'\i}as}, E. and {Guerra-Alvarado}, O. and {Carrasco-Gonz{\'a}lez}, C. and {Ribas}, {\'A}. and {Espaillat}, C.~C. and {Huang}, J. and {Andrews}, S.~M.},
        title = "{Characterizing the dust content of disk substructures in TW Hydrae}",
      journal = {\aap},
         year = 2021,
        month = apr,
       volume = {648},
          eid = {A33},
        pages = {A33},
          doi = {10.1051/0004-6361/202039812},
archivePrefix = {arXiv},
       eprint = {2102.04648},
 primaryClass = {astro-ph.EP},
       adsurl = {https://ui.adsabs.harvard.edu/abs/2021A&A...648A..33M}
}

@ARTICLE{2021ApJS..257...14S,
       author = {{Sierra}, Anibal and {P{\'e}rez}, Laura M. and {Zhang}, Ke and {Law}, Charles J. and {Guzm{\'a}n}, Viviana V. and {Qi}, Chunhua and {Bosman}, Arthur D. and {{\"O}berg}, Karin I. and {Andrews}, Sean M. and {Long}, Feng and {Teague}, Richard and {Booth}, Alice S. and {Walsh}, Catherine and {Wilner}, David J. and {M{\'e}nard}, Fran{\c{c}}ois and {Cataldi}, Gianni and {Czekala}, Ian and {Bae}, Jaehan and {Huang}, Jane and {Bergner}, Jennifer B. and {Ilee}, John D. and {Benisty}, Myriam and {Le Gal}, Romane and {Loomis}, Ryan A. and {Tsukagoshi}, Takashi and {Liu}, Yao and {Yamato}, Yoshihide and {Aikawa}, Yuri},
        title = "{Molecules with ALMA at Planet-forming Scales (MAPS). XIV. Revealing Disk Substructures in Multiwavelength Continuum Emission}",
      journal = {\apjs},
         year = 2021,
        month = nov,
       volume = {257},
       number = {1},
          eid = {14},
        pages = {14},
          doi = {10.3847/1538-4365/ac1431},
archivePrefix = {arXiv},
       eprint = {2109.06433},
 primaryClass = {astro-ph.EP},
       adsurl = {https://ui.adsabs.harvard.edu/abs/2021ApJS..257...14S}
}

@ARTICLE{2021ApJ...916L...2B,
       author = {{Benisty}, Myriam and {Bae}, Jaehan and {Facchini}, Stefano and {Keppler}, Miriam and {Teague}, Richard and {Isella}, Andrea and {Kurtovic}, Nicolas T. and {P{\'e}rez}, Laura M. and {Sierra}, Anibal and {Andrews}, Sean M. and {Carpenter}, John and {Czekala}, Ian and {Dominik}, Carsten and {Henning}, Thomas and {Menard}, Francois and {Pinilla}, Paola and {Zurlo}, Alice},
        title = "{A Circumplanetary Disk around PDS70c}",
      journal = {\apjl},
         year = 2021,
        month = jul,
       volume = {916},
       number = {1},
          eid = {L2},
        pages = {L2},
          doi = {10.3847/2041-8213/ac0f83},
archivePrefix = {arXiv},
       eprint = {2108.07123},
 primaryClass = {astro-ph.EP},
       adsurl = {https://ui.adsabs.harvard.edu/abs/2021ApJ...916L...2B}
}

@ARTICLE{2019ApJ...877L..18Z,
       author = {{Zhu}, Zhaohuan and {Zhang}, Shangjia and {Jiang}, Yan-Fei and {Kataoka}, Akimasa and {Birnstiel}, Tilman and {Dullemond}, Cornelis P. and {Andrews}, Sean M. and {Huang}, Jane and {P{\'e}rez}, Laura M. and {Carpenter}, John M. and {Bai}, Xue-Ning and {Wilner}, David J. and {Ricci}, Luca},
        title = "{One Solution to the Mass Budget Problem for Planet Formation: Optically Thick Disks with Dust Scattering}",
      journal = {\apjl},
         year = 2019,
        month = jun,
       volume = {877},
       number = {2},
          eid = {L18},
        pages = {L18},
          doi = {10.3847/2041-8213/ab1f8c},
archivePrefix = {arXiv},
       eprint = {1904.02127},
 primaryClass = {astro-ph.EP},
       adsurl = {https://ui.adsabs.harvard.edu/abs/2019ApJ...877L..18Z}
}

@ARTICLE{2020ApJ...892..136S,
       author = {{Sierra}, Anibal and {Lizano}, Susana},
        title = "{Effects of Scattering, Temperature Gradients, and Settling on the Derived Dust Properties of Observed Protoplanetary Disks}",
      journal = {\apj},
         year = 2020,
        month = apr,
       volume = {892},
       number = {2},
          eid = {136},
        pages = {136},
          doi = {10.3847/1538-4357/ab7d32},
archivePrefix = {arXiv},
       eprint = {2003.02982},
 primaryClass = {astro-ph.EP},
       adsurl = {https://ui.adsabs.harvard.edu/abs/2020ApJ...892..136S}
}

@ARTICLE{1993Icar..106...20M,
       author = {{Miyake}, Kotaro and {Nakagawa}, Yoshitsugu},
        title = "{Effects of Particle Size Distribution on Opacity Curves of Protoplanetary Disks around T Tauri Stars}",
      journal = {\icarus},
         year = 1993,
        month = nov,
       volume = {106},
       number = {1},
        pages = {20-41},
          doi = {10.1006/icar.1993.1156},
       adsurl = {https://ui.adsabs.harvard.edu/abs/1993Icar..106...20M}
}

@ARTICLE{2022ApJ...930...56U,
       author = {{Ueda}, Takahiro and {Kataoka}, Akimasa and {Tsukagoshi}, Takashi},
        title = "{Massive Compact Dust Disk with a Gap around CW Tau Revealed by ALMA Multiband Observations}",
      journal = {\apj},
         year = 2022,
        month = may,
       volume = {930},
       number = {1},
          eid = {56},
        pages = {56},
          doi = {10.3847/1538-4357/ac634d},
archivePrefix = {arXiv},
       eprint = {2203.16236},
 primaryClass = {astro-ph.EP},
       adsurl = {https://ui.adsabs.harvard.edu/abs/2022ApJ...930...56U}
}

@ARTICLE{2018ApJ...869L..45B,
       author = {{Birnstiel}, Tilman and {Dullemond}, Cornelis P. and {Zhu}, Zhaohuan and {Andrews}, Sean M. and {Bai}, Xue-Ning and {Wilner}, David J. and {Carpenter}, John M. and {Huang}, Jane and {Isella}, Andrea and {Benisty}, Myriam and {P{\'e}rez}, Laura M. and {Zhang}, Shangjia},
        title = "{The Disk Substructures at High Angular Resolution Project (DSHARP). V. Interpreting ALMA Maps of Protoplanetary Disks in Terms of a Dust Model}",
      journal = {\apjl},
         year = 2018,
        month = dec,
       volume = {869},
       number = {2},
          eid = {L45},
        pages = {L45},
          doi = {10.3847/2041-8213/aaf743},
archivePrefix = {arXiv},
       eprint = {1812.04043},
 primaryClass = {astro-ph.SR},
       adsurl = {https://ui.adsabs.harvard.edu/abs/2018ApJ...869L..45B}
}

@ARTICLE{2013A&A...559A..46B,
       author = {{Bruderer}, Simon},
        title = "{Survival of molecular gas in cavities of transition disks. I. CO}",
      journal = {\aap},
         year = 2013,
        month = nov,
       volume = {559},
          eid = {A46},
        pages = {A46},
          doi = {10.1051/0004-6361/201321171},
archivePrefix = {arXiv},
       eprint = {1308.2966},
 primaryClass = {astro-ph.SR},
       adsurl = {https://ui.adsabs.harvard.edu/abs/2013A&A...559A..46B}
}

@ARTICLE{2022MNRAS.515L..23I,
       author = {{Ilee}, John D. and {Walsh}, Catherine and {Jennings}, Jeff and {Booth}, Richard A. and {Rosotti}, Giovanni P. and {Teague}, Richard and {Tsukagoshi}, Takashi and {Nomura}, Hideko},
        title = "{Unveiling the outer dust disc of TW Hya with deep ALMA observations}",
      journal = {\mnras},
         year = 2022,
        month = sep,
       volume = {515},
       number = {1},
        pages = {L23-L28},
          doi = {10.1093/mnrasl/slac048},
archivePrefix = {arXiv},
       eprint = {2205.01396},
 primaryClass = {astro-ph.EP},
       adsurl = {https://ui.adsabs.harvard.edu/abs/2022MNRAS.515L..23I}
}

@ARTICLE{2019ApJ...883...71C,
       author = {{Carrasco-Gonz{\'a}lez}, Carlos and {Sierra}, Anibal and {Flock}, Mario and {Zhu}, Zhaohuan and {Henning}, Thomas and {Chandler}, Claire and {Galv{\'a}n-Madrid}, Roberto and {Mac{\'\i}as}, Enrique and {Anglada}, Guillem and {Linz}, Hendrik and {Osorio}, Mayra and {Rodr{\'\i}guez}, Luis F. and {Testi}, Leonardo and {Torrelles}, Jos{\'e} M. and {P{\'e}rez}, Laura and {Liu}, Yao},
        title = "{The Radial Distribution of Dust Particles in the HL Tau Disk from ALMA and VLA Observations}",
      journal = {\apj},
         year = 2019,
        month = sep,
       volume = {883},
       number = {1},
          eid = {71},
        pages = {71},
          doi = {10.3847/1538-4357/ab3d33},
archivePrefix = {arXiv},
       eprint = {1908.07140},
 primaryClass = {astro-ph.EP},
       adsurl = {https://ui.adsabs.harvard.edu/abs/2019ApJ...883...71C}
}

@ARTICLE{2022A&A...664A.137G,
       author = {{Guidi}, G. and {Isella}, A. and {Testi}, L. and {Chandler}, C.~J. and {Liu}, H.~B. and {Schmid}, H.~M. and {Rosotti}, G. and {Meng}, C. and {Jennings}, J. and {Williams}, J.~P. and {Carpenter}, J.~M. and {de Gregorio-Monsalvo}, I. and {Li}, H. and {Liu}, S.~F. and {Ortolani}, S. and {Quanz}, S.~P. and {Ricci}, L. and {Tazzari}, M.},
        title = "{Distribution of solids in the rings of the HD 163296 disk: a multiwavelength study}",
      journal = {\aap},
         year = 2022,
        month = aug,
       volume = {664},
          eid = {A137},
        pages = {A137},
          doi = {10.1051/0004-6361/202142303},
archivePrefix = {arXiv},
       eprint = {2207.01496},
 primaryClass = {astro-ph.EP},
       adsurl = {https://ui.adsabs.harvard.edu/abs/2022A&A...664A.137G}
}

@ARTICLE{2011A&A...525A..11B,
       author = {{Birnstiel}, T. and {Ormel}, C.~W. and {Dullemond}, C.~P.},
        title = "{Dust size distributions in coagulation/fragmentation equilibrium: numerical solutions and analytical fits}",
      journal = {\aap},
         year = 2011,
        month = jan,
       volume = {525},
          eid = {A11},
        pages = {A11},
          doi = {10.1051/0004-6361/201015228},
archivePrefix = {arXiv},
       eprint = {1009.3011},
 primaryClass = {astro-ph.EP},
       adsurl = {https://ui.adsabs.harvard.edu/abs/2011A&A...525A..11B}
}

@ARTICLE{2019ApJ...877L..22L,
       author = {{Liu}, Hauyu Baobab},
        title = "{The Anomalously Low (Sub)Millimeter Spectral Indices of Some Protoplanetary Disks May Be Explained By Dust Self-scattering}",
      journal = {\apjl},
         year = 2019,
        month = jun,
       volume = {877},
       number = {2},
          eid = {L22},
        pages = {L22},
          doi = {10.3847/2041-8213/ab1f8e},
archivePrefix = {arXiv},
       eprint = {1904.00333},
 primaryClass = {astro-ph.SR},
       adsurl = {https://ui.adsabs.harvard.edu/abs/2019ApJ...877L..22L}
}

@ARTICLE{2020AJ....160..270F,
       author = {{Francis}, Logan and {Johnstone}, Doug and {Herczeg}, Gregory and {Hunter}, Todd R. and {Harsono}, Daniel},
        title = "{On the Accuracy of the ALMA Flux Calibration in the Time Domain and across Spectral Windows}",
      journal = {\aj},
         year = 2020,
        month = dec,
       volume = {160},
       number = {6},
          eid = {270},
        pages = {270},
          doi = {10.3847/1538-3881/abbe1a},
archivePrefix = {arXiv},
       eprint = {2010.02186},
 primaryClass = {astro-ph.IM},
       adsurl = {https://ui.adsabs.harvard.edu/abs/2020AJ....160..270F}
}

@ARTICLE{2019ApJ...881..159M,
       author = {{Mac{\'\i}as}, Enrique and {Espaillat}, Catherine C. and {Osorio}, Mayra and {Anglada}, Guillem and {Torrelles}, Jos{\'e} M. and {Carrasco-Gonz{\'a}lez}, Carlos and {Flock}, Mario and {Linz}, Hendrik and {Bertrang}, Gesa H.-M. and {Henning}, Thomas and {G{\'o}mez}, Jos{\'e} F. and {Calvet}, Nuria and {Dent}, William R.~F.},
        title = "{Characterization of Ring Substructures in the Protoplanetary Disk of HD 169142 from Multiwavelength Atacama Large Millimeter/submillimeter Array Observations}",
      journal = {\apj},
         year = 2019,
        month = aug,
       volume = {881},
       number = {2},
          eid = {159},
        pages = {159},
          doi = {10.3847/1538-4357/ab31a2},
archivePrefix = {arXiv},
       eprint = {1907.07277},
 primaryClass = {astro-ph.SR},
       adsurl = {https://ui.adsabs.harvard.edu/abs/2019ApJ...881..159M}
}

@ARTICLE{2019AJ....158...15P,
       author = {{P{\'e}rez}, Sebasti{\'a}n and {Casassus}, Simon and {Baruteau}, Cl{\'e}ment and {Dong}, Ruobing and {Hales}, Antonio and {Cieza}, Lucas},
        title = "{Dust Unveils the Formation of a Mini-Neptune Planet in a Protoplanetary Ring}",
      journal = {\aj},
         year = 2019,
        month = jul,
       volume = {158},
       number = {1},
          eid = {15},
        pages = {15},
          doi = {10.3847/1538-3881/ab1f88},
archivePrefix = {arXiv},
       eprint = {1902.05143},
 primaryClass = {astro-ph.EP},
       adsurl = {https://ui.adsabs.harvard.edu/abs/2019AJ....158...15P}
}

@ARTICLE{2023MNRAS.522L..51H,
       author = {{Hammond}, Iain and {Christiaens}, Valentin and {Price}, Daniel J. and {Toci}, Claudia and {Pinte}, Christophe and {Juillard}, Sandrine and {Garg}, Himanshi},
        title = "{Confirmation and Keplerian motion of the gap-carving protoplanet HD 169142 b}",
      journal = {\mnras},
         year = 2023,
        month = jun,
       volume = {522},
       number = {1},
        pages = {L51-L55},
          doi = {10.1093/mnrasl/slad027},
archivePrefix = {arXiv},
       eprint = {2302.11302},
 primaryClass = {astro-ph.EP},
       adsurl = {https://ui.adsabs.harvard.edu/abs/2023MNRAS.522L..51H}
}

@ARTICLE{2014ApJ...792L..22B,
       author = {{Biller}, Beth A. and {Males}, Jared and {Rodigas}, Timothy and {Morzinski}, Katie and {Close}, Laird M. and {Juh{\'a}sz}, Attila and {Follette}, Katherine B. and {Lacour}, Sylvestre and {Benisty}, Myriam and {Sicilia-Aguilar}, Aurora and {Hinz}, Philip M. and {Weinberger}, Alycia and {Henning}, Thomas and {Pott}, J{\"o}rg-Uwe and {Bonnefoy}, Micka{\"e}l and {K{\"o}hler}, Rainer},
        title = "{An Enigmatic Point-like Feature within the HD 169142 Transitional Disk}",
      journal = {\apjl},
         year = 2014,
        month = sep,
       volume = {792},
       number = {1},
          eid = {L22},
        pages = {L22},
          doi = {10.1088/2041-8205/792/1/L22},
archivePrefix = {arXiv},
       eprint = {1408.0794},
 primaryClass = {astro-ph.EP},
       adsurl = {https://ui.adsabs.harvard.edu/abs/2014ApJ...792L..22B}
}

@ARTICLE{2022MNRAS.517.5942G,
       author = {{Garg}, H. and {Pinte}, C. and {Hammond}, I. and {Teague}, R. and {Hilder}, T. and {Price}, D.~J. and {Calcino}, J. and {Christiaens}, V. and {Poblete}, P.~P.},
        title = "{A kinematic excess in the annular gap and gas-depleted cavity in the disc around HD 169142}",
      journal = {\mnras},
         year = 2022,
        month = dec,
       volume = {517},
       number = {4},
        pages = {5942-5958},
          doi = {10.1093/mnras/stac3039},
archivePrefix = {arXiv},
       eprint = {2210.10248},
 primaryClass = {astro-ph.EP},
       adsurl = {https://ui.adsabs.harvard.edu/abs/2022MNRAS.517.5942G}
}

@ARTICLE{2023ApJ...957...11D,
       author = {{Doi}, Kiyoaki and {Kataoka}, Akimasa},
        title = "{Constraints on the Dust Size Distributions in the HD 163296 Disk from the Difference of the Apparent Dust Ring Widths between Two ALMA Bands}",
      journal = {\apj},
         year = 2023,
        month = nov,
       volume = {957},
       number = {1},
          eid = {11},
        pages = {11},
          doi = {10.3847/1538-4357/acf5df},
archivePrefix = {arXiv},
       eprint = {2308.16574},
 primaryClass = {astro-ph.EP},
       adsurl = {https://ui.adsabs.harvard.edu/abs/2023ApJ...957...11D}
}

@ARTICLE{2024ApJ...974L..25D,
       author = {{Doi}, Kiyoaki and {Kataoka}, Akimasa and {Liu}, Hauyu Baobab and {Yoshida}, Tomohiro C. and {Benisty}, Myriam and {Dong}, Ruobing and {Yamato}, Yoshihide and {Hashimoto}, Jun},
        title = "{Asymmetric Dust Accumulation of the PDS 70 Disk Revealed by ALMA Band 3 Observations}",
      journal = {\apjl},
         year = 2024,
        month = oct,
       volume = {974},
       number = {2},
          eid = {L25},
        pages = {L25},
          doi = {10.3847/2041-8213/ad7f51},
archivePrefix = {arXiv},
       eprint = {2408.09216},
 primaryClass = {astro-ph.EP},
       adsurl = {https://ui.adsabs.harvard.edu/abs/2024ApJ...974L..25D}
}

@ARTICLE{2020MNRAS.495.3209J,
       author = {{Jennings}, Jeff and {Booth}, Richard A. and {Tazzari}, Marco and {Rosotti}, Giovanni P. and {Clarke}, Cathie J.},
        title = "{frankenstein: protoplanetary disc brightness profile reconstruction at sub-beam resolution with a rapid Gaussian process}",
      journal = {\mnras},
         year = 2020,
        month = jul,
       volume = {495},
       number = {3},
        pages = {3209-3232},
          doi = {10.1093/mnras/staa1365},
archivePrefix = {arXiv},
       eprint = {2005.07709},
 primaryClass = {astro-ph.EP},
       adsurl = {https://ui.adsabs.harvard.edu/abs/2020MNRAS.495.3209J}
}

@ARTICLE{2024A&A...684A.134R,
       author = {{Rota}, A.~A. and {Meijerhof}, J.~D. and {van der Marel}, N. and {Francis}, L. and {van der Tak}, F.~F.~S. and {Sellek}, A.~D.},
        title = "{Correlation between accretion rate and free-free emission in protoplanetary disks. A multiwavelength analysis of central mm/cm emission in transition disks}",
      journal = {\aap},
         year = 2024,
        month = apr,
       volume = {684},
          eid = {A134},
        pages = {A134},
          doi = {10.1051/0004-6361/202348387},
archivePrefix = {arXiv},
       eprint = {2401.05798},
 primaryClass = {astro-ph.EP},
       adsurl = {https://ui.adsabs.harvard.edu/abs/2024A&A...684A.134R}
}

@ARTICLE{2025ApJ...980...50Y,
       author = {{Yoshida}, Tomohiro C. and {Nomura}, Hideko and {Tsukagoshi}, Takashi and {Doi}, Kiyoaki and {Furuya}, Kenji and {Kataoka}, Akimasa},
        title = "{Dust Scattering Albedo at Millimeter Wavelengths in the TW Hya Disk}",
      journal = {\apj},
         year = 2025,
        month = feb,
       volume = {980},
       number = {1},
          eid = {50},
        pages = {50},
          doi = {10.3847/1538-4357/ad9f31},
archivePrefix = {arXiv},
       eprint = {2412.10731},
 primaryClass = {astro-ph.EP},
       adsurl = {https://ui.adsabs.harvard.edu/abs/2025ApJ...980...50Y}
}

@ARTICLE{2017ApJ...838...97M,
       author = {{Mac{\'\i}as}, Enrique and {Anglada}, Guillem and {Osorio}, Mayra and {Torrelles}, Jos{\'e} M. and {Carrasco-Gonz{\'a}lez}, Carlos and {G{\'o}mez}, Jos{\'e} F. and {Rodr{\'\i}guez}, Luis F. and {Sierra}, Anibal},
        title = "{Imaging a Central Ionized Component, a Narrow Ring, and the CO Snowline in the Multigapped Disk of HD 169142}",
      journal = {\apj},
         year = 2017,
        month = apr,
       volume = {838},
       number = {2},
          eid = {97},
        pages = {97},
          doi = {10.3847/1538-4357/aa6620},
archivePrefix = {arXiv},
       eprint = {1703.02957},
 primaryClass = {astro-ph.SR},
       adsurl = {https://ui.adsabs.harvard.edu/abs/2017ApJ...838...97M}
}

@ARTICLE{2014ApJ...791L..36O,
       author = {{Osorio}, Mayra and {Anglada}, Guillem and {Carrasco-Gonz{\'a}lez}, Carlos and {Torrelles}, Jos{\'e} M. and {Mac{\'\i}as}, Enrique and {Rodr{\'\i}guez}, Luis F. and {G{\'o}mez}, Jos{\'e} F. and {D'Alessio}, Paola and {Calvet}, Nuria and {Nagel}, Erick and {Dent}, William R.~F. and {Quanz}, Sascha P. and {Reggiani}, Maddalena and {Mayen-Gijon}, Juan M.},
        title = "{Imaging the Inner and Outer Gaps of the Pre-transitional Disk of HD 169142 at 7 mm}",
      journal = {\apjl},
         year = 2014,
        month = aug,
       volume = {791},
       number = {2},
          eid = {L36},
        pages = {L36},
          doi = {10.1088/2041-8205/791/2/L36},
archivePrefix = {arXiv},
       eprint = {1407.6549},
 primaryClass = {astro-ph.SR},
       adsurl = {https://ui.adsabs.harvard.edu/abs/2014ApJ...791L..36O}
}

@ARTICLE{2026ApJ...996L..17R,
       author = {{Rampinelli}, Luna and {Facchini}, Stefano and {Leemker}, Margot and {Isella}, Andrea and {Curone}, Pietro and {Benisty}, Myriam and {Humphreys}, Elizabeth and {Testi}, Leonardo},
        title = "{Water Vapor Emission at the Warm Cavity Wall of the HD 100546 Disk as Revealed by ALMA}",
      journal = {\apjl},
         year = 2026,
        month = jan,
       volume = {996},
       number = {1},
          eid = {L17},
        pages = {L17},
          doi = {10.3847/2041-8213/ae2868},
archivePrefix = {arXiv},
       eprint = {2512.06439},
 primaryClass = {astro-ph.EP},
       adsurl = {https://ui.adsabs.harvard.edu/abs/2026ApJ...996L..17R}
}

@ARTICLE{2010A&A...512A..15R,
       author = {{Ricci}, L. and {Testi}, L. and {Natta}, A. and {Neri}, R. and {Cabrit}, S. and {Herczeg}, G.~J.},
        title = "{Dust properties of protoplanetary disks in the Taurus-Auriga star forming region from millimeter wavelengths}",
      journal = {\aap},
         year = 2010,
        month = mar,
       volume = {512},
          eid = {A15},
        pages = {A15},
          doi = {10.1051/0004-6361/200913403},
archivePrefix = {arXiv},
       eprint = {0912.3356},
 primaryClass = {astro-ph.EP},
       adsurl = {https://ui.adsabs.harvard.edu/abs/2010A&A...512A..15R}
}

@ARTICLE{2019JOSS....4.1632T,
       author = {{Teague}, Richard},
        title = "{GoFish: Fishing for Line Observations in Protoplanetary Disks}",
      journal = {The Journal of Open Source Software},
         year = 2019,
        month = sep,
       volume = {4},
       number = {41},
          eid = {1632},
        pages = {1632},
          doi = {10.21105/joss.01632},
       adsurl = {https://ui.adsabs.harvard.edu/abs/2019JOSS....4.1632T}
}

@ARTICLE{2020NatMe..17..261V,
       author = {{Virtanen}, Pauli and {Gommers}, Ralf and {Oliphant}, Travis E. and {Haberland}, Matt and {Reddy}, Tyler and {Cournapeau}, David and {Burovski}, Evgeni and {Peterson}, Pearu and {Weckesser}, Warren and {Bright}, Jonathan and {van der Walt}, St{\'e}fan J. and {Brett}, Matthew and {Wilson}, Joshua and {Millman}, K. Jarrod and {Mayorov}, Nikolay and {Nelson}, Andrew R.~J. and {Jones}, Eric and {Kern}, Robert and {Larson}, Eric and {Carey}, C.~J. and {Polat}, {\.I}lhan and {Feng}, Yu and {Moore}, Eric W. and {VanderPlas}, Jake and {Laxalde}, Denis and {Perktold}, Josef and {Cimrman}, Robert and {Henriksen}, Ian and {Quintero}, E.~A. and {Harris}, Charles R. and {Archibald}, Anne M. and {Ribeiro}, Ant{\^o}nio H. and {Pedregosa}, Fabian and {van Mulbregt}, Paul and {SciPy 1.  0 Contributors}},
        title = "{SciPy 1.0: fundamental algorithms for scientific computing in Python}",
      journal = {Nature Medicine},
         year = 2020,
        month = feb,
       volume = {17},
        pages = {261-272},
          doi = {10.1038/s41592-019-0686-2},
archivePrefix = {arXiv},
       eprint = {1907.10121},
 primaryClass = {cs.MS},
       adsurl = {https://ui.adsabs.harvard.edu/abs/2020NatMe..17..261V}
}

@ARTICLE{2015JOSAA..32..611B,
       author = {{Baddour}, Natalie and {Chouinard}, Ugo},
        title = "{Theory and operational rules for the discrete Hankel transform}",
      journal = {Journal of the Optical Society of America A},
         year = 2015,
        month = apr,
       volume = {32},
       number = {4},
        pages = {611},
          doi = {10.1364/JOSAA.32.000611},
       adsurl = {https://ui.adsabs.harvard.edu/abs/2015JOSAA..32..611B}
}

@ARTICLE{2019arXiv191211554P,
       author = {{Phan}, Du and {Pradhan}, Neeraj and {Jankowiak}, Martin},
        title = "{Composable Effects for Flexible and Accelerated Probabilistic Programming in NumPyro}",
      journal = {arXiv e-prints},
         year = 2019,
        month = dec,
          eid = {arXiv:1912.11554},
        pages = {arXiv:1912.11554},
          doi = {10.48550/arXiv.1912.11554},
archivePrefix = {arXiv},
       eprint = {1912.11554},
 primaryClass = {stat.ML},
       adsurl = {https://ui.adsabs.harvard.edu/abs/2019arXiv191211554P}
}

@ARTICLE{2025ApJ...984L...7L,
       author = {{Loomis}, Ryan A. and {Facchini}, Stefano and {Benisty}, Myriam and {Curone}, Pietro and {Ilee}, John D. and {Cataldi}, Gianni and {Yen}, Hsi-Wei and {Teague}, Richard and {Pinte}, Christophe and {Huang}, Jane and {Garg}, Himanshi and {Orihara}, Ryuta and {Czekala}, Ian and {Zawadzki}, Brianna and {Andrews}, Sean M. and {Wilner}, David J. and {Bae}, Jaehan and {Barraza-Alfaro}, Marcelo and {Fasano}, Daniele and {Flock}, Mario and {Fukagawa}, Misato and {Galloway-Sprietsma}, Maria and {Izquierdo}, Andr{\'e}s F. and {Kanagawa}, Kazuhiro and {Lesur}, Geoffroy and {Longarini}, Cristiano and {Menard}, Francois and {Price}, Daniel J. and {Rosotti}, Giovanni and {Stadler}, Jochen and {Wafflard-Fernandez}, Gaylor and {W{\"o}lfer}, Lisa and {Yoshida}, Tomohiro C.},
        title = "{exoALMA. II. Data Calibration and Imaging Pipeline}",
      journal = {\apjl},
         year = 2025,
        month = may,
       volume = {984},
       number = {1},
          eid = {L7},
        pages = {L7},
          doi = {10.3847/2041-8213/adc43a},
archivePrefix = {arXiv},
       eprint = {2504.19870},
 primaryClass = {astro-ph.EP},
       adsurl = {https://ui.adsabs.harvard.edu/abs/2025ApJ...984L...7L}
}
\bibliographystyle{aasjournalv7}



\end{document}